\documentclass{bmcart}

\usepackage{amsmath}
\usepackage{amsfonts}
\usepackage{amssymb}
\usepackage{graphicx}
\usepackage{psfrag}
\usepackage{array}
\usepackage{epsfig}
\usepackage{color}
\usepackage{ dsfont }
\usepackage{moreverb,url}
\usepackage{lipsum}
\usepackage[colorlinks,bookmarksopen,bookmarksnumbered,citecolor=red,urlcolor=red]{hyperref}
\usepackage[latin1]{inputenx}
\usepackage[T1]{fontenc}
\usepackage{bm}
\usepackage{epstopdf}
\usepackage{subfigure}
\usepackage{xcolor}
\usepackage{blindtext}
\usepackage{graphicx}
\usepackage{caption}
\usepackage{float}
\definecolor{cream}{RGB}{222,217,201}

\newcommand{\matacc}[2]{\left\{ \begin{array}{#1} #2 \end{array} \right. }

\usepackage{afterpage}
\startlocaldefs
\endlocaldefs

\UseRawInputEncoding

\begin{document}

\begin{frontmatter}

\begin{fmbox}
\dochead{Research}

\title{Hybrid Machine Learning and Physical Modeling of Feedstock Deformation During Robotic 3D Printing of Continuous Fiber Thermoplastic Composites}

\author[
  addressref={unf},
  email={chady.ghnatios@unf.edu}
]{\inits{C.G.}\fnm{Chady} \snm{Ghnatios}}
\author[
  addressref={tmu},                   
  email={kazem@torontomu.ca}   
]{\inits{K.F.}\fnm{Kazem} \snm{Fayazbakhsh}}


\address[id=unf]{
  \orgdiv{Department of Mechanical Engineering},
  \orgname{University of North Florida},
\city{Jacksonville, FL},
\cny{United States}
}

\address[id=tmu]{
  \orgdiv{Department of Aerospace Engineering},
  \orgname{Toronto Metropolitan University},
\city{Toronto, ON},
\cny{Canada}
}
\end{fmbox}

\begin{abstractbox}
\begin{abstract}
Feedstock deformation during 3D printing of continuous fiber composites is a critical challenge in path planning and a main driver in the generation of manufacturing defects. The proposed work addressed the feedstock deformation during the deposition through several experimental and numerical pathways. The experimental setups and numerical simulations are used to identify the main driving phenomena in the deformation of feedstock through residual stress relief and drying, crystallization, and thermal stresses. A hybrid physics-based and data-driven modeling effort is performed, using Kelvin-Voigt viscoelastic modeling of the composite prepregs and a stabilized neural ODE for the modeling of drying and crystallization. The identified hybrid models from DMA and DSC experiments are used in robotic 3D printing to validate the deposition of a composite prepreg in real printing settings. The results show the ability of the model to reproduce the prepreg behavior far above the temperature used in the training, showcasing its robustness and generalizability.
\end{abstract}

\begin{keyword}
\kwd{Proper generalized decomposition} 
\kwd{Model order reduction} 
\kwd{Stabilized Neural Ordinary Differential Equation}
\kwd{Continuous fiber-reinforced composite}
\kwd{Crystallization} 
\kwd{Feedstock Deformation}
\end{keyword}
\end{abstractbox}

\end{frontmatter}

\section{Introduction}
Feedstock deformation during 3D printing of continuous fiber-reinforced composites is observed experimentally and reported on multiple occasions \cite{Zheng2025-jf,https://doi.org/10.1002/pc.29895,KABIR2020111476}. Different phenomena come into play when 3D printing continuous fiber-reinforced composites. Multiple works analyzed different phenomena happening on the manufactured part level. For instance, thermal gradients and non-uniform heating/cooling effects generate residual thermal stresses and warping when printing several tapes \cite{OUYANG2025111209,GHNATIOS2022102796,DIAZ2025104749,El_Moumen2019-pl}. Even though thermal gradients are a major driving phenomenon in the warping of 3D printed tape and the estimation of the final shape, other effects exist and are reported in the literature, such as the squeeze flow effect \cite{FEREIDOUNI2025108519,Miri2025-iu}, initial residual stresses \cite{Jiang02012023}, and crystallization shrinkage \cite{SAMY2021389}. Multiple experimental studies have highlighted the importance of the crystalline structure in tailoring the material properties in 3D-printed polymers and have analyzed crystallization as a function of the thermal field using classical crystallization models \cite{Pourali2024-ny,MALAGUTTI2023117961,LEE2022100085}, but not the instant volume change of the feedstock under the effect of crystallization. Sreejith et al.  \cite{SREEJITH2023103789} proposed a mechanics-based constitutive framework where crystallization is a dominant factor in generating residual stresses and macroscopic distortion in polymers and thermoplastics. The through-thickness variation of crystallization gradients is analyzed in \cite{WANG2026109310}.

Most process-induced deformation mechanisms in continuous fiber-reinforced additive manufacturing, such as cumulative residual stress buildup and global warpage, predominantly manifest when multi-layer and multi-path parts are fabricated. In contrast, deformation of the feedstock, driven by crystallization and the release of initial thermo-mechanical stresses, can already be observed at the scale of a single deposited tape. This single-path deformation is of particular importance for path definition and optimization in robotic arm-based 3D printing, as it directly affects intrinsic process variables, including fiber alignment \cite{ZHANG2021101775,KUMAR2026119819}, as well as robotic deposition path and control challenges such as gaps and overlaps \cite{Lu31122024,Afanasev2026-jo,YADAV2023107654}. In fact, feedstock deformation can play a major role in defect generation during 3D printing of continuous fiber composites when not accounted for. The present work therefore focuses on isolating and analyzing single-tape deformations induced by thermal effects, residual stress release, and crystallization-induced deformation in robotic 3D printing. This problem is highly complex due to the coupled multiphysics involved in the crystallization formation and properties changing on the microscopic level \cite{SAMY2021389,OUYANG2025111209}, hindering correct modeling and multiscale simulation of large-scale parts \cite{Ghnatios2021-zq}. 


With the recent developments in machine learning technologies, multiscale modeling and complex multiphysics can now be modeled with data models, inheriting established mechanical properties by construction \cite{GHNATIOS2024105542}. Moreover, data-driven modeling in 3D printing of composites is conquering many of the previously unaddressed challenges, such as the process variability \cite{Ghnatios2021-bd}, online defect detection \cite{TUO2026113114} and material properties \cite{DAGHIGH2024100600,polym17182557}. However, the data-driven approach is criticized for being unreliable beyond the tested regions and for engineering experimental results being too expensive to produce and therefore often being too scarce \cite{reviewmlinengineering}. Moreover, many processes and systems are previously modeled, and their modeling is reliable on many levels. These considerations paved the way for the creation of hybrid modeling frameworks, where reliable engineering models are used when available, coupled to data-driven approaches, to close the gap between the experimental results and the modeling ones \cite{hybridfedatamodel2019,mydigitaltwin,FERDOUSI2023110958}. This work will use a hybrid modeling approach to investigate the dynamic evolution of the strains generated by residual stress release, drying, crystallization, and thermal stresses of the resin.

When it comes to advanced dynamic machine learning technologies, Neural Ordinary Differential Equations (neural ODEs) are a family of dynamic, time-dependent modeling technologies of data \cite{neuralodeoriginal}. Unlike more classical models like ResNet or Long-Short Term Memory (LSTM) \cite{7780459,10.1162/neco.1997.9.8.1735}, they tend to preserve stability and predictability in time far beyond the training region, allowing reliable forecasting of the quantities of interest \cite{SHAWLY20253033}. This work uses a novel form of stabilized neural ODE to learn a reliable form of the crystallization and initial strain release derivatives. This form is later integrated using a numerical integration scheme \cite{en16155790}. The hybrid modeling effort starts through numerical simulation using a custom finite element and Proper Generalized Decomposition (PGD) \cite{Archives-PGD} code to evaluate the evolution and deformation shape of a deposited tape. Later, we propose to integrate the established mechanical viscoelastic models with the stabilized neural ODE formulation for temperature- and time-dependent modeling of the crystallization and initial strains.

The work starts with a review of the robotic 3D printing setup, feedstock material, and experimental results, including tape deformation after deposition and material characterization using Differential Scanning Calorimetry (DSC) and Dynamic Mechanical Analyzer (DMA) machines in Section \ref{experimentalsec}. The numerical modeling based on the established physical models and data-driven approaches is described in Section \ref{physicalmodelingsec}. Section \ref{thermalsec} reviews the thermal modeling and simulation of the prepreg tape obtained from the DMA experiments. The deformation observed in the DMA is modeled using a Kelvin-Voigt viscoelastic model in Section \ref{mechanicalsec}. The discrepancy between the viscoelastic simulation and the experimental measurements is used to quantify the initial strain release during prepreg heating in Section \ref{dryingsec}, and the crystallization-induced strains in Section \ref{crystallssec}. A discussion of the results is presented in section \ref{discussionsec}. Section \ref{exterapolationtorealcomposites} shows the integration of the derived models from the DMA experiments into a holistic simulation framework, with a comparison to the tape deformation after deposition. Finally, Section \ref{comclusionsec} concludes the article with some review of the results, their implications for 3D printing continuous carbon fiber composites with minimum defects, and directions for future works.

\section{Experimental setup and results}\label{experimentalsec}
\begin{figure}[!ht]
    \centering
  \subfigure[Robotic 3D printing setup.\label{fig:fig1akazem}]
    {
        \includegraphics[width=0.56\textwidth]{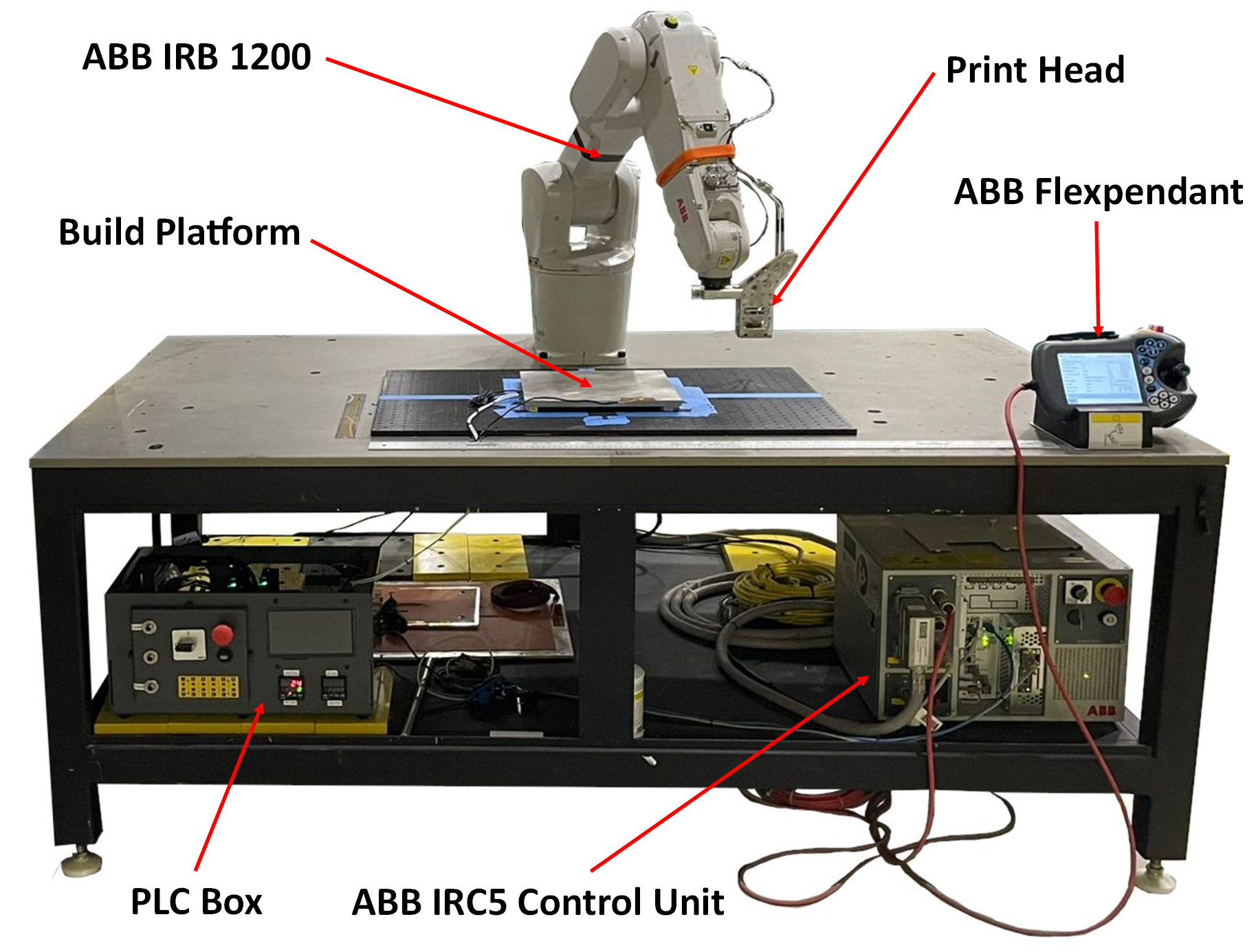}
    }
   \subfigure[The used feedstock.\label{fig:fig1bkazem}]
    {
        \includegraphics[width=0.36\textwidth]{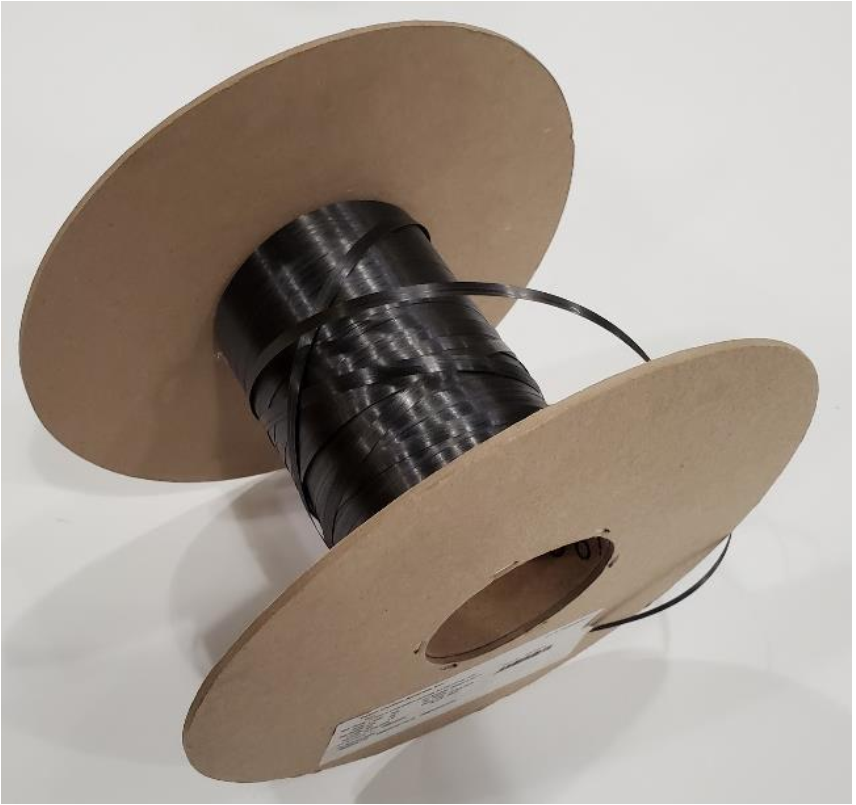}
        }
\caption{Robotic 3D printing setup used in this work and the feedstock.\label{fig:fig1kazem}}
\end{figure}

The robotic 3D printing setup includes an ABB IRB 1200 robotic arm with a custom-built 3D printing head as an end effector, including a slotted nozzle. A 304.8 $\times$ 304.8 mm (12 in. $\times$ 12 in.) steel plate with a thickness of 12.7 mm (0.5 in.) is used as the build platform. A Programmable Logic Controller (PLC) box regulates the nozzle and build platform temperatures while the robot movement is optimized through the ABB IRC5 Industrial Robot Controller, illustrated in Figure \ref{fig:fig1akazem}.

The feedstock, illustrated in Figure \ref{fig:fig1bkazem}, is a continuous carbon fiber (CCF) prepreg tape with a nominal width of 6.35 mm (0.25 in.) and a thickness of about 0.180 mm on a spool with an inner diameter of 95 mm (Figure 1, right). It combines high-strength carbon fiber (HTS45) with low-melt poly-aryl ether ketone (LM-PAEK). 

\begin{figure}[!ht]
    \centering
  \subfigure[Representation of one LM PAK-CCF tape deposited on the build platform.\label{fig:fig2akazem}]
    {
        \includegraphics[width=0.6\textwidth]{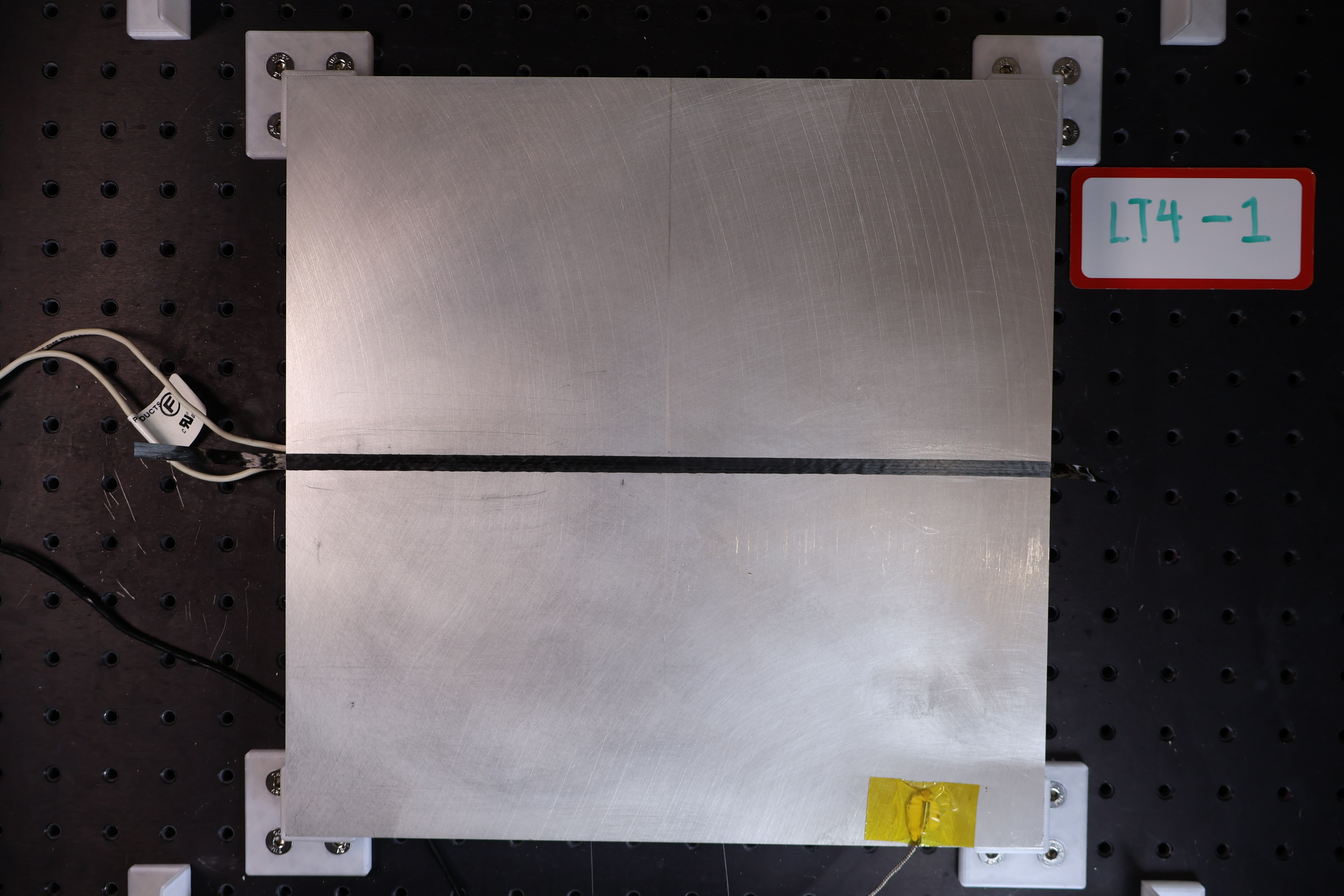}
    }
   \subfigure[Tape width measurement along its length after deposition.\label{fig:fig2bkazem}]
    {
        \includegraphics[width=0.9\textwidth]{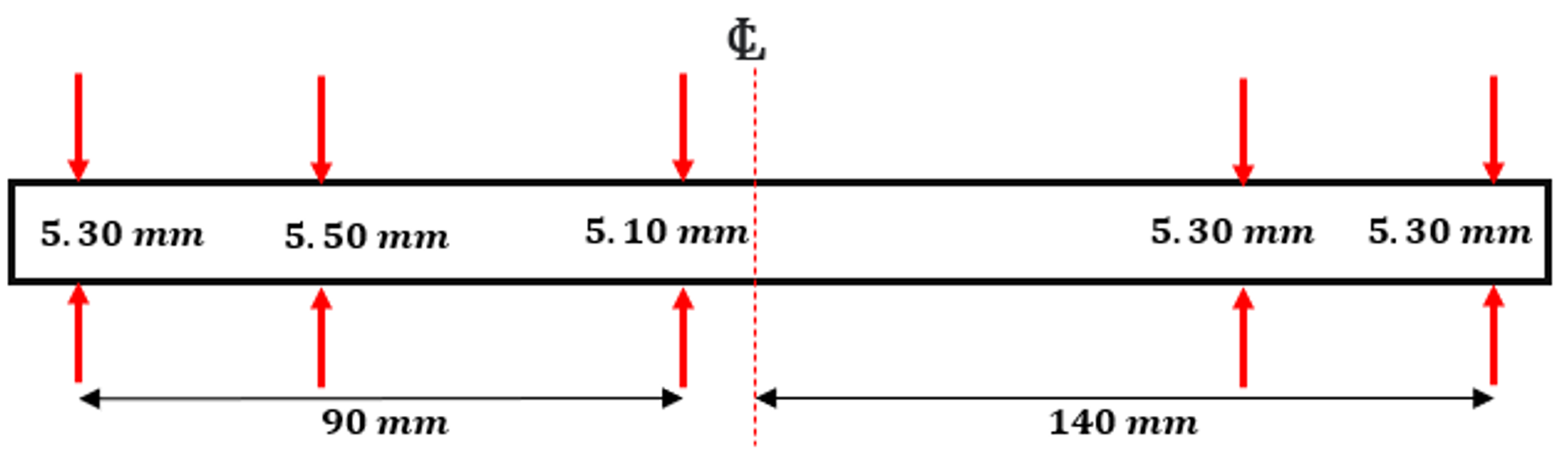}
    }
\caption{Robotic 3D printing of a single tape.\label{fig:fig2kazem}}
\end{figure}

To manufacture a part, multiple tapes need to be deposited side by side to create one layer, and then several layers are stacked on top of each other to provide sufficient structural performance to the final component. During tape deposition within one ply, no gaps or overlaps are generally desirable. Since the tape dimensions change after deposition, the raw feedstock width cannot be used as the distance between the centerlines of two adjacent tapes, i.e., the Tape Raster Offset (TRO). Figure \ref{fig:fig2akazem} shows one tape deposited on the build platform per the following 3D printing process parameters: nozzle temperature of 380$^\textrm{o}$C, build platform temperature of 160$^\textrm{o}$C, 3D printing speed of 10 mm/s, and layer height of 0.200 mm. Tape width was measured at multiple points using a caliper (Figure \ref{fig:fig2bkazem}).

The tape width along its length changes between 5.10 mm and 5.50 mm, and it is mainly around 5.30 mm, which is significantly lower than the pre-processed tape width of 6.35 mm. While the nominal layer thickness for this feedstock is 0.14 mm, a thicker layer of 0.200 mm was used for the first layer. This is the general practice in 3D printing to have a high-quality first layer and better bonding to the subsequent one and have a direct impact on the tape width. Furthermore, residual stresses from spooling, moisture release, and crystallization impact the tape width after deposition. 

One way to produce high-quality parts with no gaps and overlaps is to perform many experimental trials and find the tape width after deposition for each specific case. Instead, numerical modeling can predict the tape width for all possible cases and significantly reduce time and cost. For this purpose, two thermal tests were performed to characterize the feedstock material and provide input for numerical simulation: Differential Scanning Calorimetry (DSC) and Dynamic Mechanical Analyzer (DMA). 

\begin{figure}[!ht]
    \centering
    \includegraphics[width=0.85\linewidth]{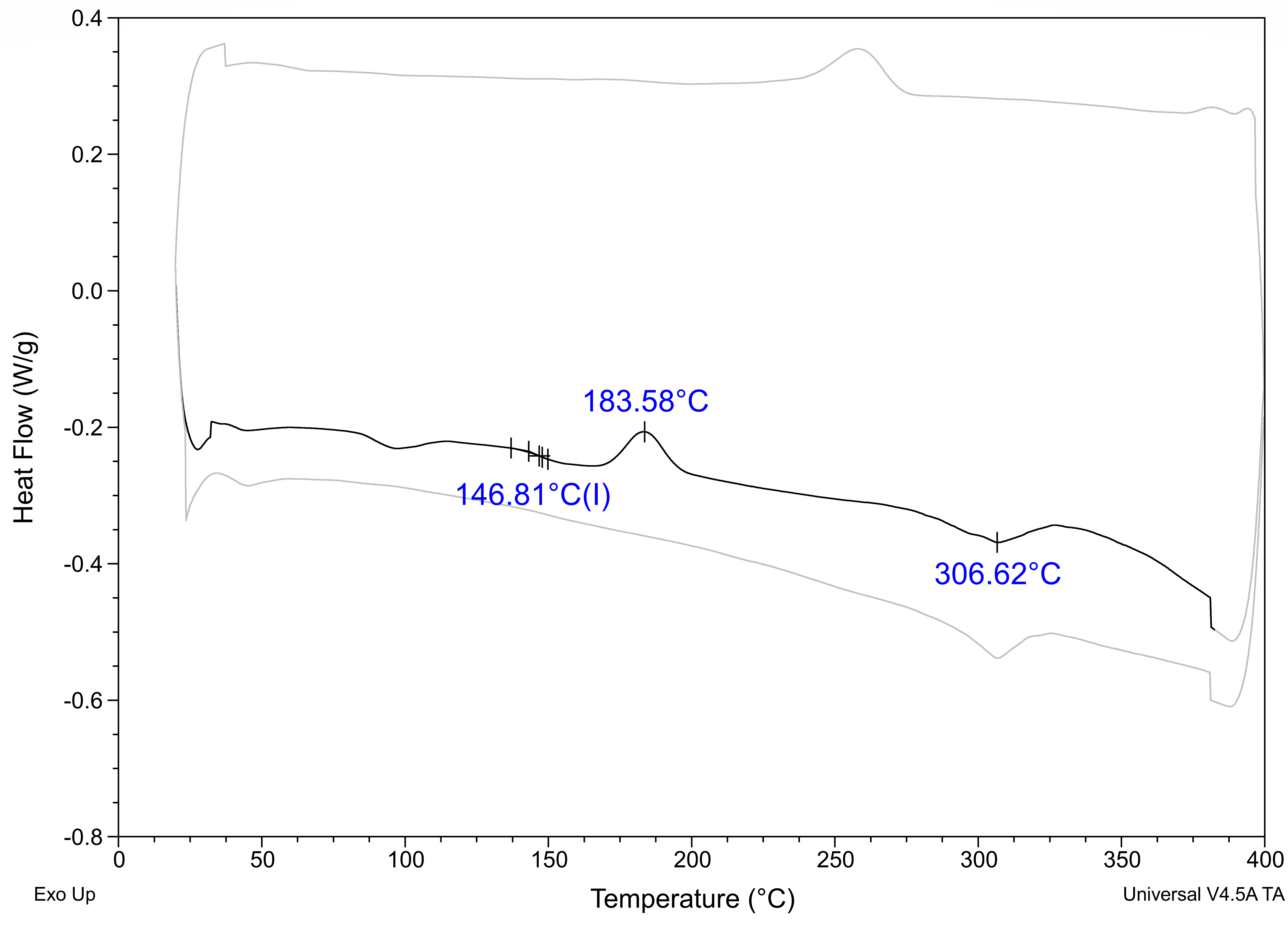}
    \caption{DSC thermograms for LM PAEK-CCF pre-processed feedstock.}
    \label{dscpic}
\end{figure}

A DSC Q2000 (TA Instruments, New Castle, Delaware) was used for thermal testing of the feedstock and to find its glass transition temperature ($T_g$) and crystallization peak temperature ($T_{pc}$) per ASTM D3418-21. The test profile included a heating ramp from 20$^\textrm{o}$C to 400$^\textrm{o}$C at 10$^\textrm{o}$C/min with an isothermal hold for 5 min, followed by a cooling ramp to 20$^\textrm{o}$C at the same rate and isothermal hold time at the end. The second heating ramp followed the same test parameters as the first one, and the specimen mass was 2.89 mg. Figure \ref{dscpic} shows the DSC curves with the following values obtained from the first heating ramp: inflection temperature of 147$^\textrm{o}$C ($T_{i}$) considered as $T_g$, $T_{pc}$ of 184$^\textrm{o}$C, and enthalpy of cold crystallization of 5.208 J/g.

\begin{figure}[!ht]
    \centering
  \subfigure[Specimen dimensions.\label{fig:fig4akazem}]
    {
        \includegraphics[width=0.46\textwidth]{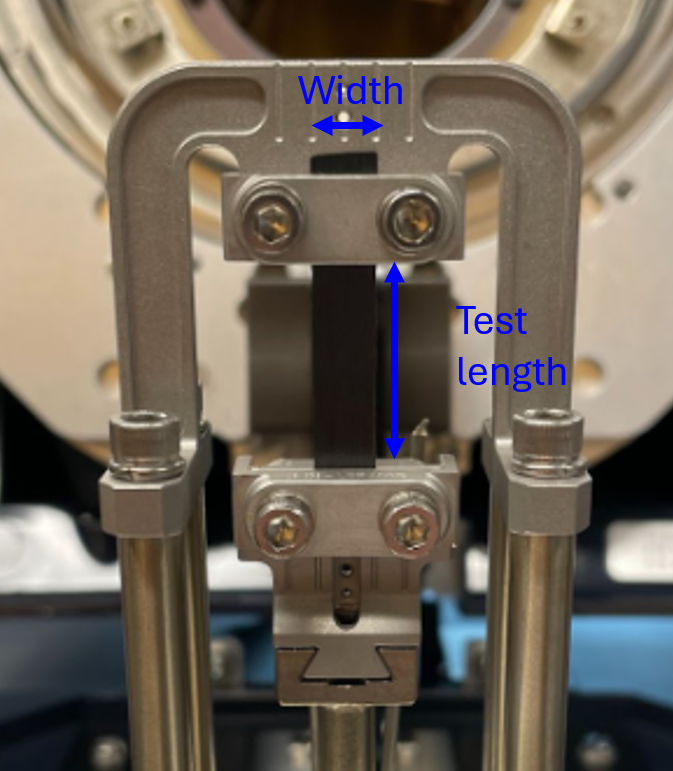}
    }
   \subfigure[Reference system used in defining the boundary conditions.\label{fig:fig4bkazem}]
    {
        \includegraphics[width=0.46\textwidth]{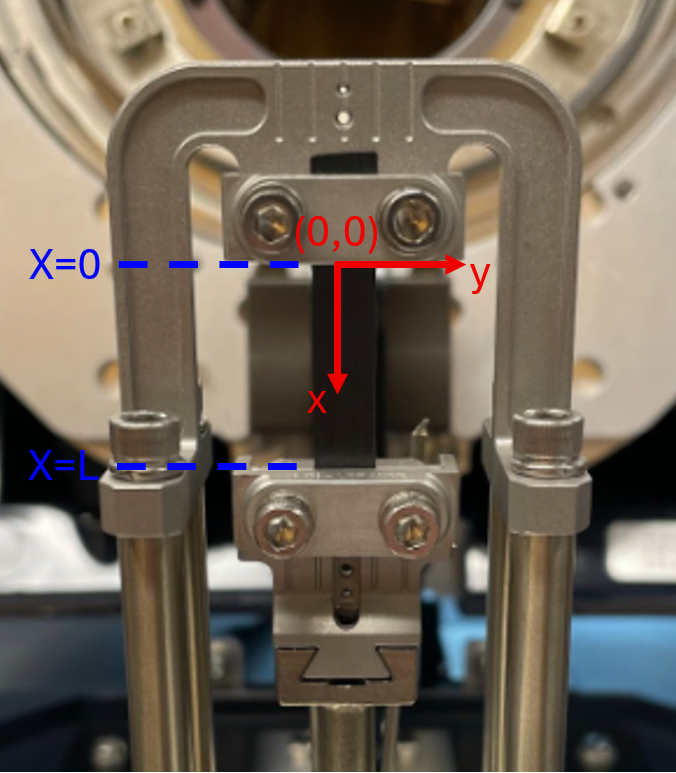}
    }
\caption{Prepreg tape inside the film tensile fixture inside the DMA testing apparatus.\label{fig:fig4kazem}}
\end{figure}

A DMA 850 (TA Instruments, New Castle, Delaware) was used to measure the feedstock dimensional change for two temperatures, one below and one above the Tg, respectively, 120$^\textrm{o}$C and 180$^\textrm{o}$C. A film tension clamp was utilized, and pieces of the prepreg tape about 40 mm in length were used for testing. Figure \ref{fig:fig4akazem} shows one specimen placed inside the DMA 850 machine with the black LM PAEK-CCF preprocessed tape piece mounted in the clamp. For both experiments, four measurements of specimen width and thickness were made. The average width and thickness, respectively, were 6.29 mm and 0.172 mm (DMA-1) and 6.30 mm and 0.181 mm (DMA-2). The DMA machine measures the test length and records its change during the experiment. Figure \ref{fig:fig4bkazem} shows the coordinates for describing the boundary conditions. The two top and the two bottom screws were tightened to 3 lbs$ \cdot $in. The top half of the clamp is fixed; therefore, at $x=0$, the edge does not have any translation or rotation. The bottom half of the clamp sits on an actuator and is free to move along the x-axis. Therefore, at $x=L$, translation along x is free, while y is constrained, as well as the rotation. The other two edges at $y = \pm$width/2 are completely free.

The actuator was left free to move to see dimensional change in the tape; therefore, a negligible force ($0.01 N$) was set in the machine. Multiple trials were completed to find the heating and cooling rates and the soak time for the two tests that produce a steady-state change in the tape dimensions. It means that the tape's dimensional change had stabilized by the time a subsequent heating or cooling was about to start. DMA-1 (test temperature of 120$^\textrm{o}$C) and DMA-2 (test temperature of 180$^\textrm{o}$C) included four iso-stress steps at $0.01 N$ summarized below:
\begin{itemize}
    \item First: (Heating 1) room temperature to test temperature at 0.5$^\textrm{o}$C/min, soak time 60 min.
    \item Second: (Cooling 1) test temperature to 30$^\textrm{o}$C at -0.5$^\textrm{o}$C/min, soak time 180 min.
    \item Third: (Heating 2) 30$^\textrm{o}$C to test temperature at 0.5$^\textrm{o}$C/min, soak time 60 min.
    \item Fourth: (Cooling 2) test temperature to 30$^\textrm{o}$C at -0.5$^\textrm{o}$C/min, soak time 180 min.
\end{itemize}

\begin{figure}[!ht]
    \centering
  \subfigure[Strain VS temperature evolution in the experiment DMA-1.\label{fig:fig5akazem}]
    {
        \includegraphics[width=0.46\textwidth]{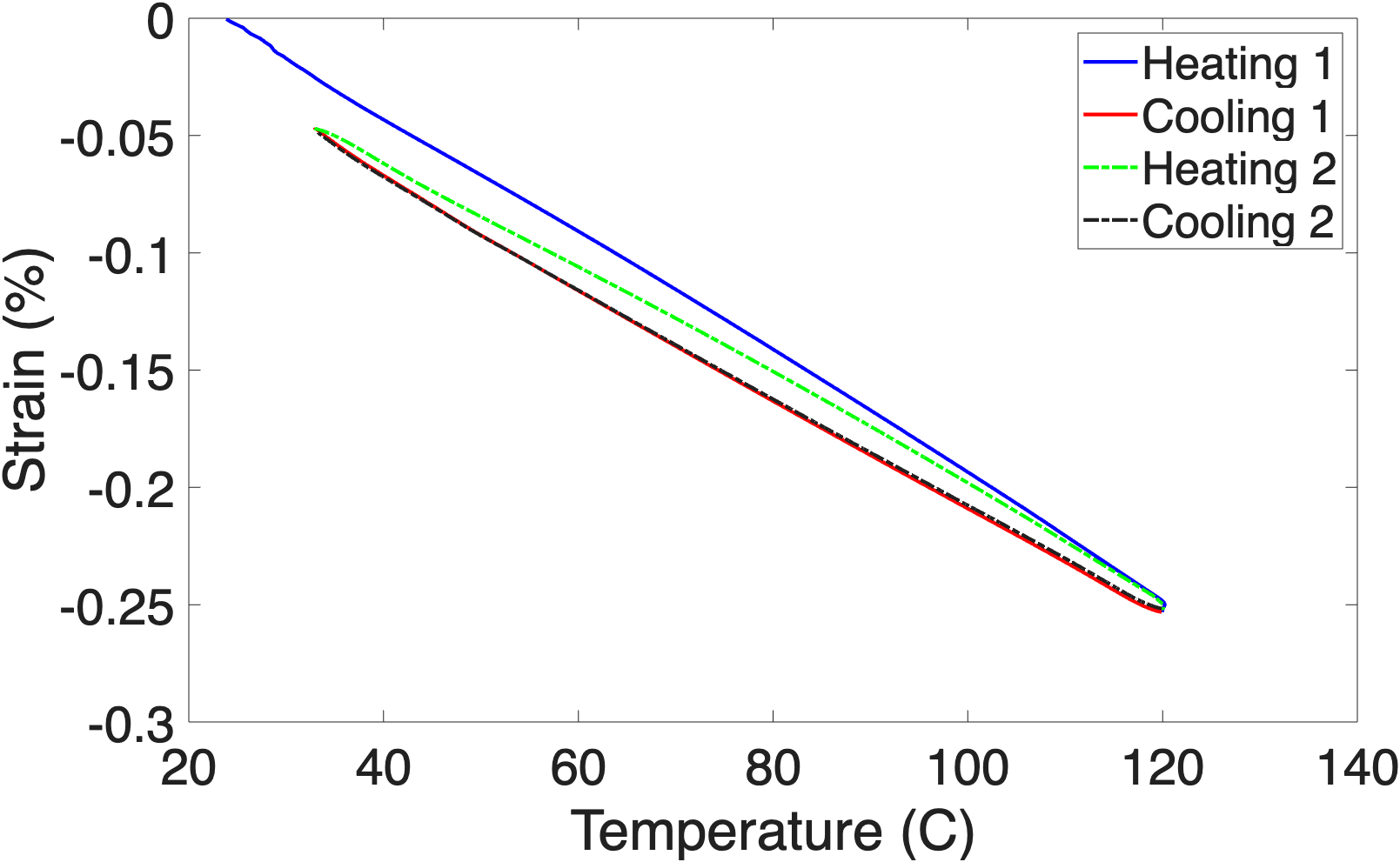}
    }
   \subfigure[Strain VS temperature evolution in the experiment DMA-2.\label{fig:fig5bkazem}]
    {
        \includegraphics[width=0.46\textwidth]{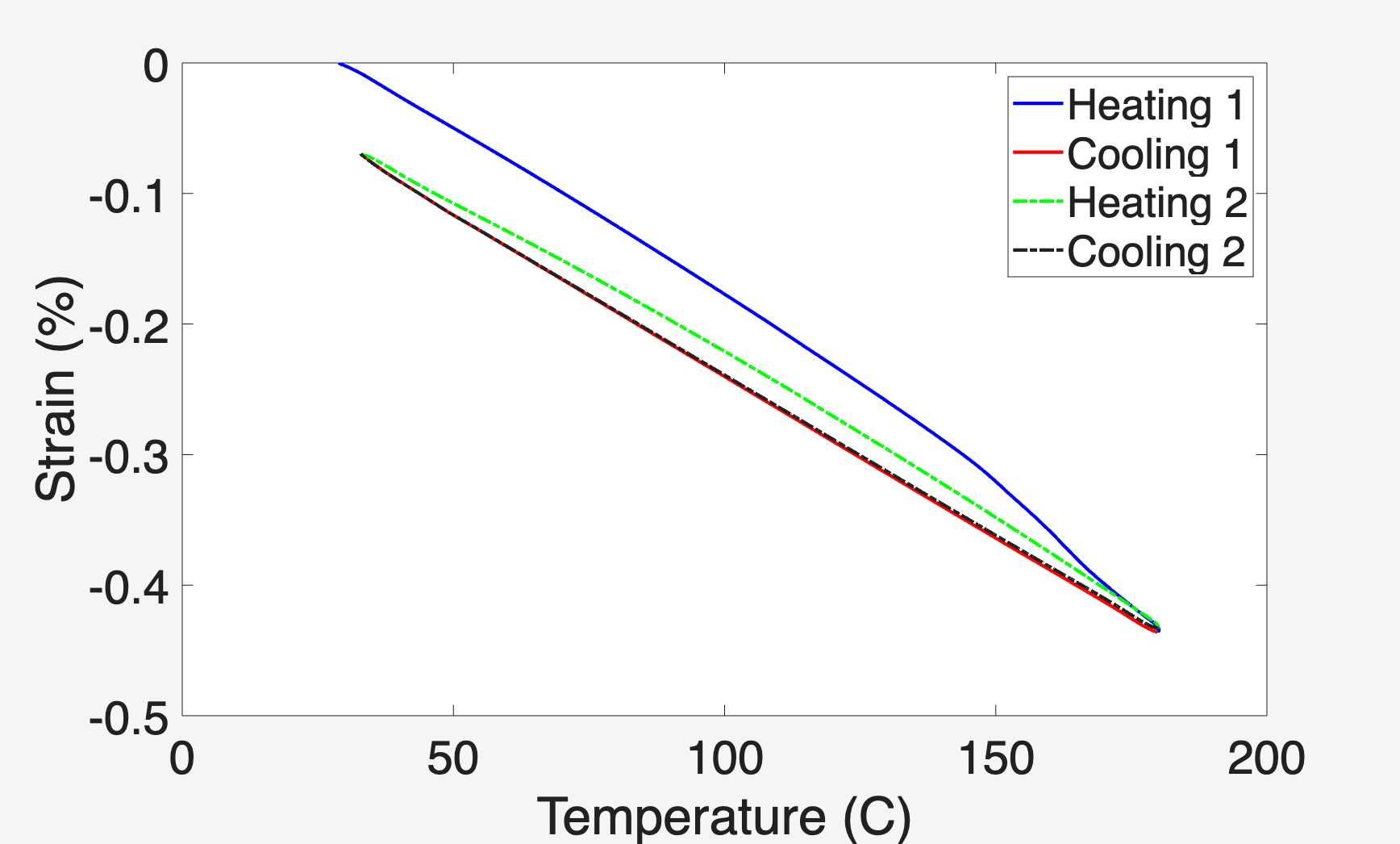}
    }
\caption{Strain VS temperature evolution in the performed DMA experiments.\label{fig:fig5kazem}}
\end{figure}

Figure \ref{fig:fig5kazem} shows the change in the test length in terms of strain percentage versus temperature for both experiments (DMA-1 and DMA-2).

\section{Numerical analysis and modeling}\label{physicalmodelingsec}
In this part, we compare numerical results with the experimental findings to explore the physical phenomenon occurring in the feedstock material. The analysis is decomposed into several parts, starting with the thermal simulation, followed by the mechanical one; then modeling of the deformations induced by the initial conditions, like drying and release of residual stress; and, finally, the deformation induced by crystallization.

\subsection{Thermal simulation}\label{thermalsec}
The thermal modeling uses the heat transfer equation written by:
\begin{equation}\label{strongformthermal}
\rho c_p\frac{\partial T}{\partial t}-\nabla\left(\mathbf{K}\cdot \nabla T\right)=Q,
\end{equation}

with $\nabla$ the gradient operator, $\rho$ being the homogenized density; $c_p$ the homogenized thermal capacitance of the prepreg tape at atmospheric pressure; $T$ the thermal field; $t$ the time; and $\mathbf{K}$ the orthotropic thermal conductivity tensor defined as:
\begin{equation}
\mathbf{K}=\left[\begin{array}{ccc}
K_{\mathbin{/\!/}} & 0 & 0\\
0 & K_{\perp} & 0\\
0 & 0 & K_{\perp}
\end{array}\right]
\end{equation}

The reaction term $Q$ is neglected in this simulation as the temperature is increased to a maximum of $120^{\textrm{o}}C$ (or $393.15K$) in experiment $1$, and $180^{\textrm{o}}C$ (or $453.15K$)in experiment $2$. The used boundary conditions are convection with the ambient temperature throughout the domain boundary, written as:
\begin{equation}\label{bcthermalequation}
\matacc{l}{
T(\mathbf{x},0)=T_0\\
\left. -\mathbf{K}\cdot \nabla T\right|_\Gamma=h\left(T(\Gamma,t)-T_{\infty}(t)\right)\\
T_{\infty}(t)=T_{TMA}(t)
}
\end{equation}

\begin{figure}[!ht]
\centering
  \subfigure[Imposed temperature in the first $DMA$ experiment, reaching $120^{\textrm{o}}C$.]
    {
        \includegraphics[width=0.46\textwidth]{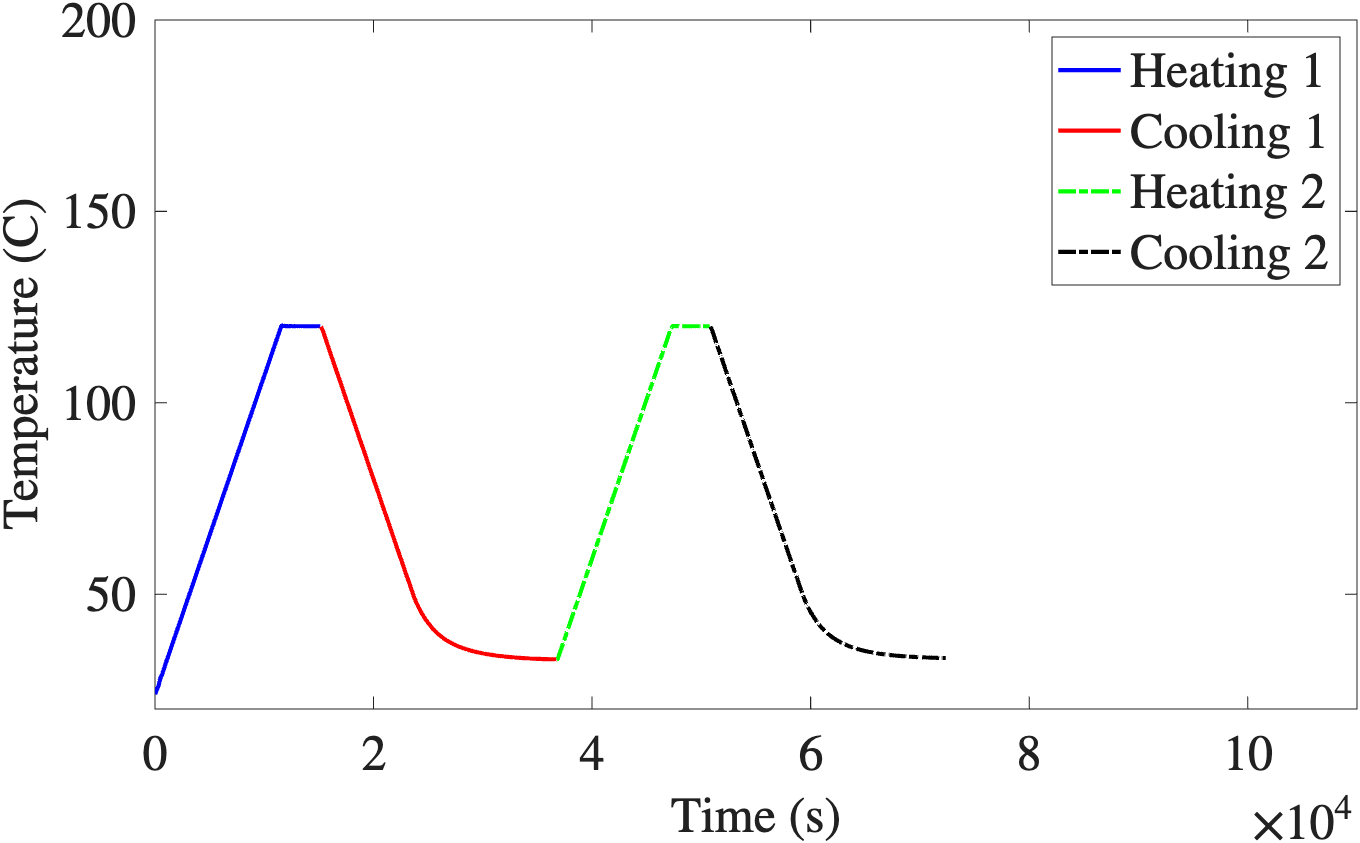}
    }
   \subfigure[Imposed temperature in the second $DMA$ experiment, reaching $180^{\textrm{o}}C$.]
    {
        \includegraphics[width=0.46\textwidth]{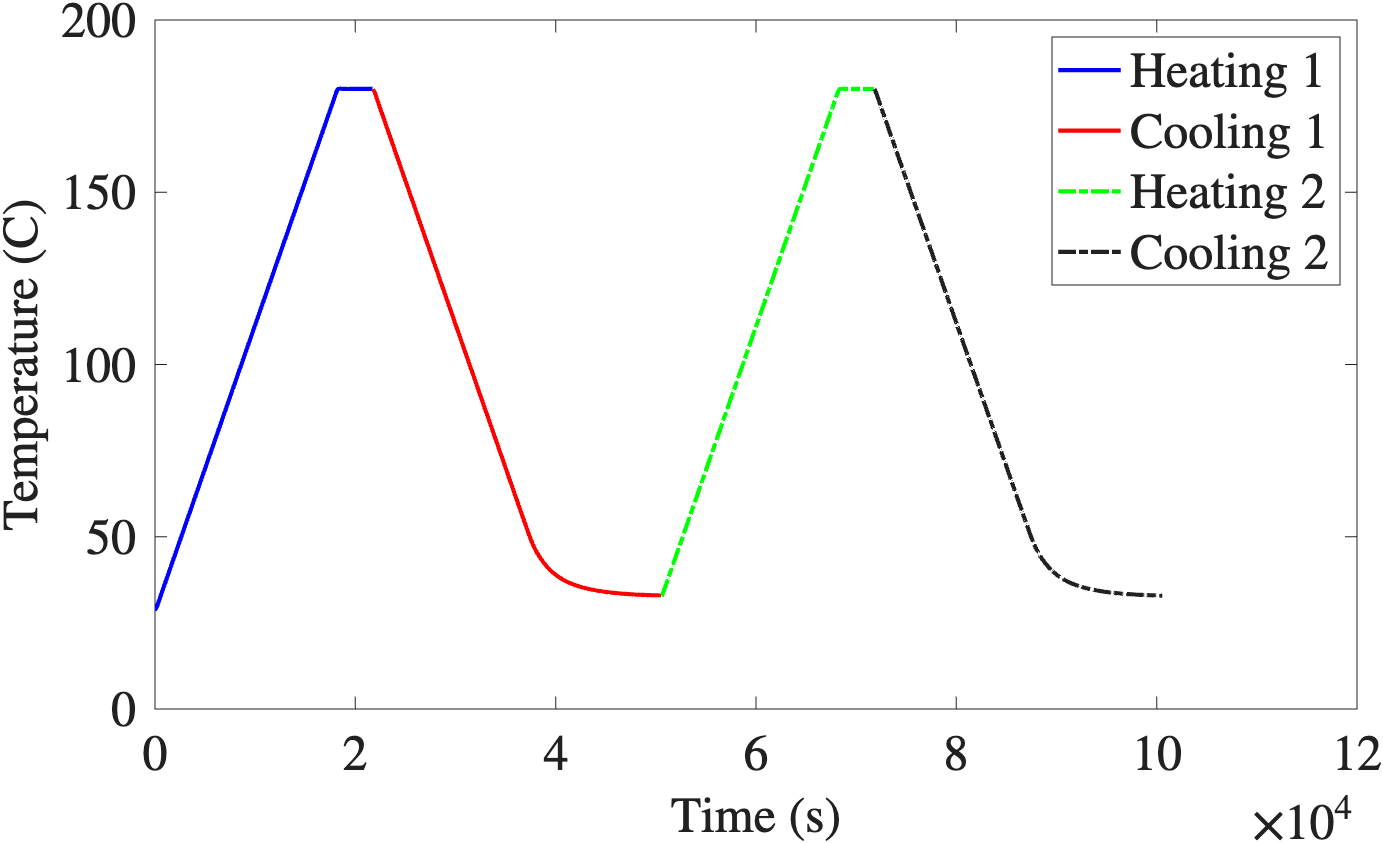}
    }
\caption{Imposed air temperature $T_\infty$ in the $DMA$ experiments as a function of time.\label{tinftyvaluesbothexp}}
\end{figure}

In equation (\ref{bcthermalequation}), $\mathbf{x}=(x,y,z), x\in[0,L], y\in[0,W], z\in[0,H]$ is the vector nodal 3D coordinates, $t_0$ is the initial temperature, $\Gamma$ the domain boundary, $h$ the coefficient of thermal convection between the air and the prepreg tape, and $T_\infty$ is the air temperature imposed by the time-dependent $DMA$ cycle $T_{DMA}$. For the simulation, the thermal properties are obtained using the manufacturer's data sheet or are homogenized using the law of mixture. The prepreg tape has a fiber volume fraction $v_f=0.6$. The thermal properties for the material are summarized in Table \ref{thermalproperties}. The imposed temperature $T_\infty$ is illustrated in Figure \ref{tinftyvaluesbothexp}.\\

\begin{table}
\centering
\begin{tabular}{|c|c|}
\hline
Thermal properties & Values (SI units)\\
\hline
$K_{\mathbin{/\!/}}$ & $14.48 W/m.K$\\
$K_{\perp}$ & $1.52 W/m.K$\\
$\rho$ & $1078 kg/m^3$\\
$c_p$ & $1054 J/K$\\
$h$ & $25 W/m^2.K$\\
$T_0$ & $25^{\textrm{o}}C$\\
$L$ & $20 mm$\\
$W$ & $6.35 mm$\\
$H$ & $0.177 mm$\\
\hline
\end{tabular}
\caption{Used thermal properties in the $DMA$ experiment simulation \cite{datasheettejin}.\label{thermalproperties}}
\end{table}

The thermal simulation uses the Proper Generalized Decomposition (PGD) method to compute the solution of the differential equation provided in equation (\ref{strongformthermal}) \cite{mynpmpgd,GHNATIOS201229}. The thermal solution is therefore computed in a separate form using space-time and in-plane-out-of-plane decomposition:
\begin{equation}
T(\mathbf{x},t)=\sum\limits_{i=1}^{n} F_i(x,y) \cdot Z_i(z) \cdot G_i(t)
\end{equation}

\begin{figure}[!ht]
\centering
\includegraphics[width=0.99\textwidth]{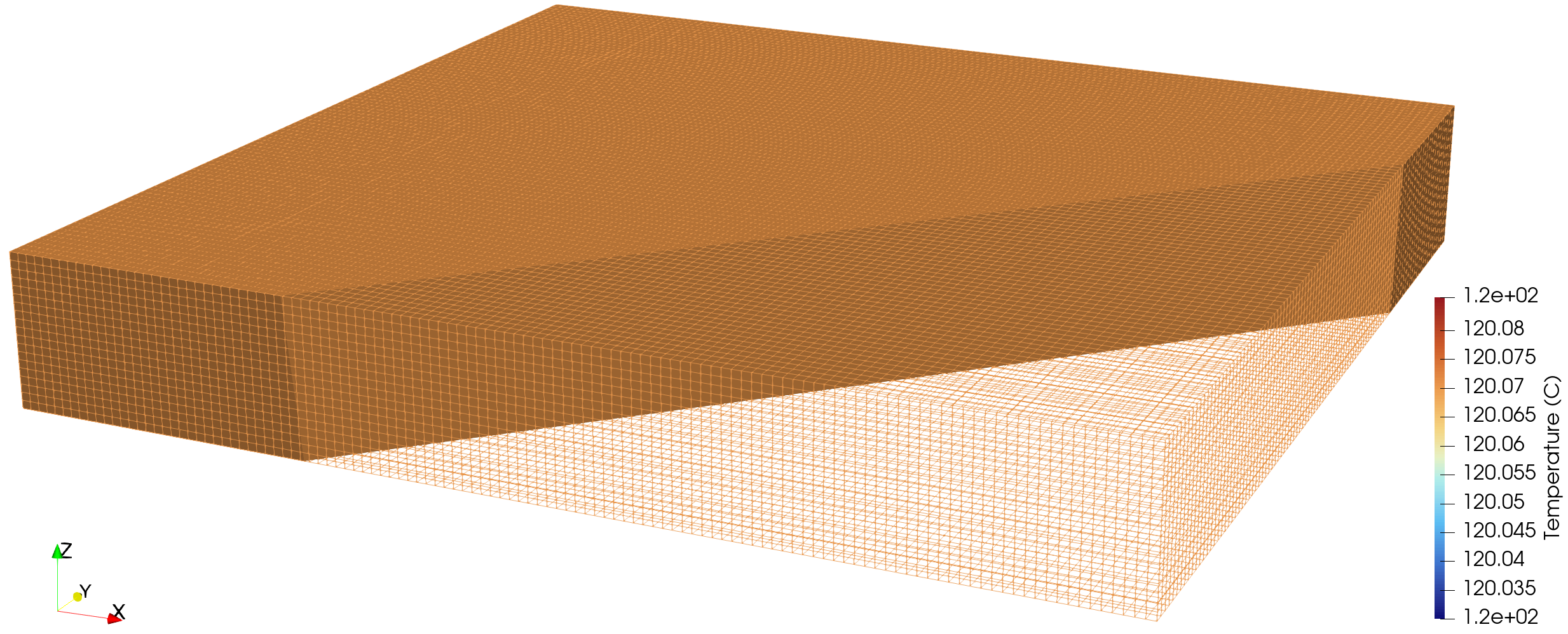}
\caption{Solution of the thermal fields (in K) as a function of time for DMA-1. A cut in the domain at $t=12,000s$ shows the mesh and the homogeneous thermal fields. The thickness is amplified 10 times.\label{3dcutfig}}
\end{figure}

\begin{figure}[!ht]
\centering
  \subfigure[Thermal fields as a function of the time.]
    {
        \includegraphics[width=0.46\textwidth]{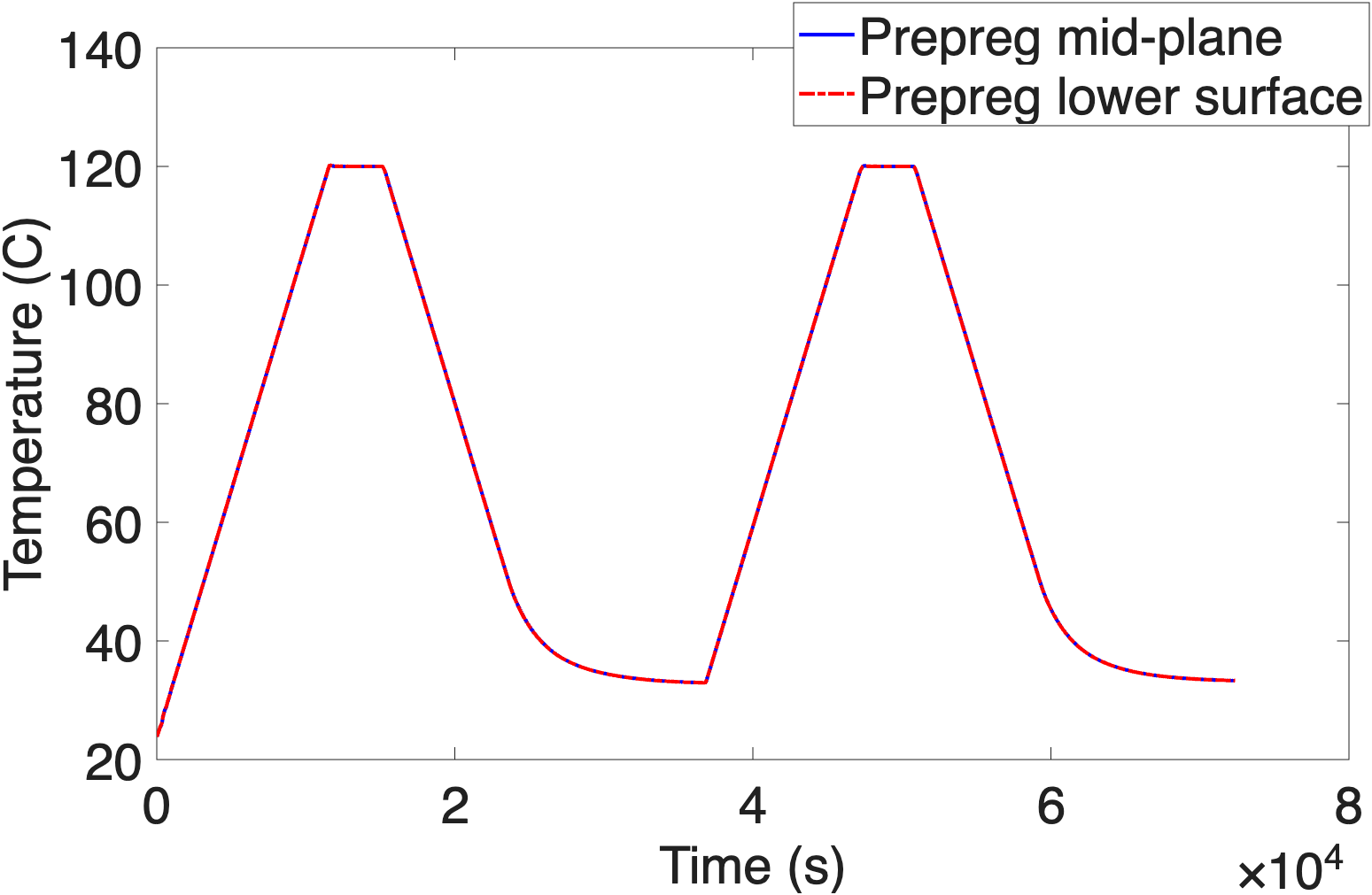}
    }
   \subfigure[Thermal fields as a function of the time, zoom at the peak in temperature.]
    {
        \includegraphics[width=0.46\textwidth]{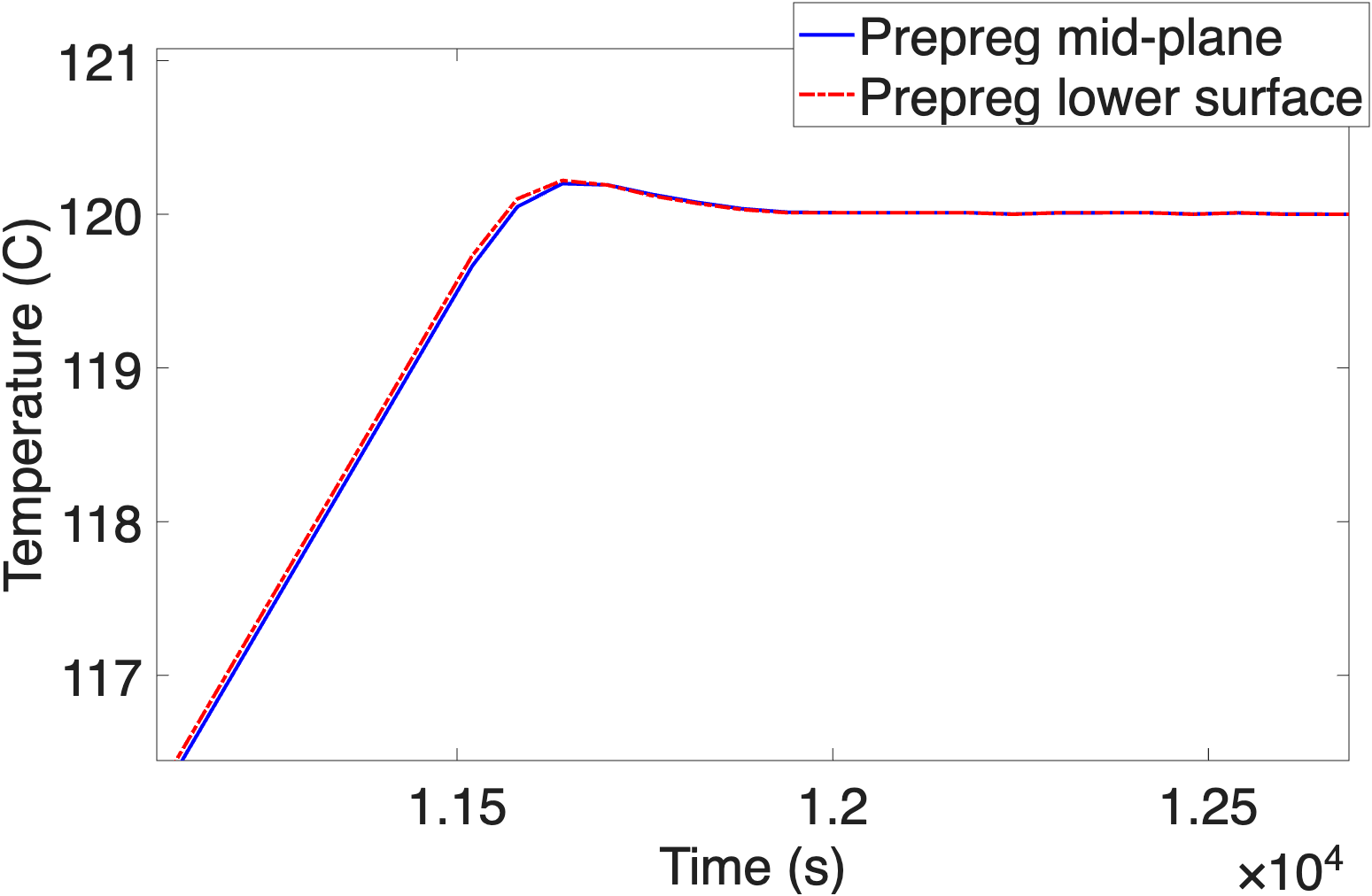}
    }
\caption{Solution of the thermal fields as a function of time for DMA-1, with $T_\infty$ increased to $120^{\textrm{o}}C$, for $z=0$ and $z=H/2$.\label{exp28solutionzoom}}
\end{figure}

The PGD solution is computed using a fixed-point iterative algorithm. The user unfamiliar with the process is referred to \cite{Archives-PGD}, and the references therein. The solution uses a fine out-of-plane mesh, along the z-direction, with 21 nodes used through the thickness dimension $z$. The obtained solution for DMA-1 with a maximum temperature reaching $120^{\textrm{o}}C$, is illustrated in Figure \ref{3dcutfig}. A plot of the thermal variations in the central in-plane node ($x=L/2$ and $y=W/2$), and for $z=0$, the bottom layer, as well as for $z=H/2$, is illustrated in Figure \ref{exp28solutionzoom}, using the simulated data for DMA-1.\\

The thermal simulation shows a homogeneous thermal distribution in the part during the experiments; therefore, the thermal fields can be treated as being a function of time only. The same remark applies for DMA-2.

\subsection{Mechanical simulation}\label{mechanicalsec}
Based on the experimental results, the prepreg shows a visco-elastic behavior since there are differences in deformation between the heating and cooling cycles. Therefore, a Kelvin-Voigt viscoelastic model is used to model the prepreg behavior \cite{ghnatios2017computational}. The Kelvin-Voigt viscoelastic model writes the stress tensor $\bm\sigma$ as a sum of an elastic part and a viscous part:
\begin{equation}
\bm\sigma=\bm\sigma_{elastic}+\bm\sigma_{viscous}=\mathbf{C}\bm\epsilon + \bm\eta \dot{\bm\epsilon},
\end{equation}

with $\mathbf{C}$ the stiffness matrix, $\bm\epsilon$ the strain tensor, $\bm\eta$ the viscous tensor, and $\dot{\bm\epsilon}$ the strain rate tensor. For simplicity, the Voigt notations are used in the remainder of this work.\\

The stiffness tensor is computed as being the inverse of the compliance tensor $\mathbf{S}$, for instance $\mathbf{C}=\mathbf{S}^{-1}$, where the compliance tensor is as follows:
\begin{equation}
\mathbf{S}=\begin{bmatrix}
\displaystyle \frac{1}{E_{\mathbin{/\!/}}} & 
\displaystyle -\frac{\nu_{21}}{E_\perp} &
\displaystyle -\frac{\nu_{31}}{E_\perp} &
0 & 0 & 0 \\[10pt]

\displaystyle -\frac{\nu_{12}}{E_{\mathbin{/\!/}}} &
\displaystyle \frac{1}{E_\perp} &
\displaystyle -\frac{\nu_{32}}{E_\perp} &
0 & 0 & 0 \\[10pt]

\displaystyle -\frac{\nu_{13}}{E_{\mathbin{/\!/}}} &
\displaystyle -\frac{\nu_{23}}{E_\perp} &
\displaystyle \frac{1}{E_\perp} &
0 & 0 & 0 \\[10pt]

0 & 0 & 0 &
\displaystyle \frac{1}{G_{tt}} & 0 & 0 \\[10pt]

0 & 0 & 0 &
0 & \displaystyle \frac{1}{G_{nt}} & 0 \\[10pt]

0 & 0 & 0 &
0 & 0 & \displaystyle \frac{1}{G_{nt}}
\end{bmatrix}
\end{equation}

The mechanical properties appearing in $\mathbf{S}$ and used in this work are taken from the manufacturer's datasheet and the Halpin-Tsai homogenization equations \cite{Jones1999,doi:10.1177/002199836900300419}. The mechanical properties are provided in Table \ref{mechproperties}.\\

\begin{table}
\centering
\begin{tabular}{|c|c|}
\hline
Elastic properties & Values (SI units)\\
\hline
$E_{\mathbin{/\!/}}$ & $140 GPa$\\
$E_\perp$ & $7.78 GPa$\\
$\nu_{12}$ & $0.293$\\
$\nu_{21}$ & $\nu_{21}=\nu_{12}\frac{E_\perp}{E_{\mathbin{/\!/}}}$\\
$\nu_{13}$ & $0.293$\\
$\nu_{31}$ & $\nu_{31}=\nu_{13}\frac{E_\perp}{E_{\mathbin{/\!/}}}$\\
$\nu_{23}$ & $0.2794$\\
$\nu_{32}$ & $\nu_{32}=\nu_{23}$\\
$G_{nt}$ & $4.4GPa$\\
$G_{tt}$ & $4.4GPa$\\
\hline
\end{tabular}
\caption{Mechanical properties in the simulation of $DMA$ experiments.\label{mechproperties}}
\end{table}

The viscoelastic properties of uncured prepregs have been studied in multiple research studies in the literature \cite{Kulkarni2021,Xiong2019-bd,Wang2024,Chan2018,Wei2023}, but no work is found for the LM PAEK-carbon fiber material. For the material at hand, no direct characterization of $\bm\eta$ is performed. Therefore, in this work, we assume the following form of the orthotropic viscelastic tensor:
\begin{equation}\label{etaform}
\bm\eta=\begin{bmatrix}
\displaystyle 0 & 
\displaystyle \eta_1 &
\displaystyle \eta_1 &
0 & 0 & 0 \\[10pt]

\displaystyle \eta_1 &
\displaystyle \eta_2 &
\displaystyle \eta_3 &
0 & 0 & 0 \\[10pt]

\displaystyle \eta_1 &
\displaystyle \eta_3 &
\displaystyle \eta_2 &
0 & 0 & 0 \\[10pt]

0 & 0 & 0 &
\displaystyle 0 & 0 & 0 \\[10pt]

0 & 0 & 0 &
0 & \displaystyle 0 & 0 \\[10pt]

0 & 0 & 0 &
0 & 0 & \displaystyle 0
\end{bmatrix}
\end{equation}

The rationale behind the assumption in equation (\ref{etaform}) comes from the fact that the $x$ direction parallel to the fiber is dominated by the fiber behavior; therefore, it is considered purely elastic. On the other hand, the coupling $(x,y)$ and $(x,z)$ have similar behavior, modeled by $\eta_1$. The direction normal to the fiber is viscoelastic with a viscous term $\eta_2$, and the coupling $(y,z)$ is modeled by $\eta_3$. The three parameters modeling $\bm\eta$ will be fitted to DMA-1, where the temperature increases $120^{\textrm{o}}C$. It is below the glass transition temperature; therefore, no micro-structural changes are observed.\\

The equilibrium equation with temperature variation is written as:
\begin{equation}\label{stronformmechanical}
\nabla \cdot \left(\bm\sigma - \bm\sigma_0\right)=\mathbf{f}.
\end{equation}

In this work, acceleration is neglected, and a quasi-static approach is adopted. This choice is motivated by the reduced rate of temperature evolution, fixed to a low value of $0.5K/min$. Moreover, the gravity body force $\mathbf{f}$ is neglected and set to $\mathbf{0}$. The thermal stresses are written as:
\begin{equation}\label{sigmazeroeq}
\bm\sigma_0=\mathbf{E}\bm\epsilon_0+\bm\eta\dot{\bm\epsilon}_0,
\end{equation}

with the thermal strains written as:
\begin{equation}
\bm\epsilon_0=\bm\alpha \Delta T.
\end{equation}

The thermal variation $\Delta T$ is computed as the simulated temperature $T$ minus the initial prepreg temperature. The coefficient of thermal expansion $\bm\alpha=(\alpha_{\mathbin{/\!/}},\alpha_\perp,\alpha_\perp)$ is unknown a priori and will be fitted using the simulation data from DMA-1. In the proposed approach, ${\bm\epsilon}_0$ are computed using the forward finite differences method and smoothed using a Savitzky-Golay filter with a polynomial order of 3 and a moving window of size 9.\\

The parameters to estimate from DMA-1 are therefore noted as $\mathbf{p}=(\alpha_{\mathbin{/\!/}},\alpha_\perp,\eta_1,\eta_2,\eta_3)$. The identification of $\mathbf{p}$ is performed after solving the finite element simulation of the problem described in equation (\ref{stronformmechanical}) with the following boundary conditions:
\begin{equation}
\matacc{l}{
u(x=0,y,z,t)=0\\
v(x=0,y,z,t)=0\\
w(x=0,y,z,t)=0\\
u(x=L,y,z,t)=u_L\\
v(x=L,y,z,t)=0\\
w(x=L,y,z,t)=0\\
}
\end{equation}
The boundary conditions are imposed on the displacements $\mathbf{u}=(u,v,w)$, where the part is cantilevered at $x=0$, and a single degree of freedom, $u=u_L$, along $x$-direction is allowed at $x=L$. The value of $u_L$ is obtained from solving the discrete, quasi-static, finite element problem at every time step, and the average normal strain along $x$ is computed using:
\begin{equation}
\epsilon_{xx}^{sim}=\frac{u_L(t)}{L}
\end{equation}

The optimization problem is solved by fitting the numerical strains obtained from the simulation to the experimental ones, obtained from DMA-1. This is performed by aligning the end of the second cooling with the point from the simulation. Therefore, the experimental results are subjected to the following vertical translation:
\begin{equation}
\bar{\epsilon}_{xx}=\epsilon_{xx}^{exp}-\epsilon_{xx}^{exp}(t=t_{end})+\epsilon_{xx}^{sim}(t=t_{end})
\end{equation}

For simplicity, the subscript $xx$ will be omitted in the following equation. The parameters $\mathbf{p}$ are fitted by solving the following optimization problem:
\begin{equation}
\begin{array}{r}
\left(\mathbf{p} \right)=\underset{\tiny{\begin{array}{c}
\mathbf{p}\in(\mathds{R}^+)^{k}
\end{array}}}{arg\min}\left\{ \sum\limits_{i=2}^{4}\left(\sum\limits_{j=1}^{N_i}\left(\bar{\epsilon}_{ij}-\epsilon^{sim}_{ij}\right)^2\right)\right\},
\end{array}
\end{equation}

where $k=5$, the number of parameters in $\mathbf{p}$, and $j$ indicates the minimization is performed over the stages 2, 3, and 4, which refers to the first cooling, second heating, and second cooling. The first heating stage is omitted due to the irreversible behavior observed in this stage, because of the drying and release of residual stresses from the prepreg tape.\\

A custom finite element code in MATLAB is used to solve the deformation problem and is coupled to the $fmincon$ optimizer subroutine in MATLAB to identify the optimal properties. Starting from an initial guess $mathbf{p}_0$, first the interior-point algorithm is used to identify the optimal parameters. After convergence, the converged results are set as the initial guess to the active-set algorithm, which runs again until convergence. The optimization converged to $\mathbf{p}_{converged}=(-0.2\times 10^{-6}, 26.97\times 10^{-6},0.876\times 10^{13}, 4.62\times 10^{13}, 0.76\times 10^{13})$. All values are in SI units.\\

\begin{figure}[!ht]
\centering
  \subfigure[Deformations along x direction (m).]
    {
        \includegraphics[width=0.99\textwidth]{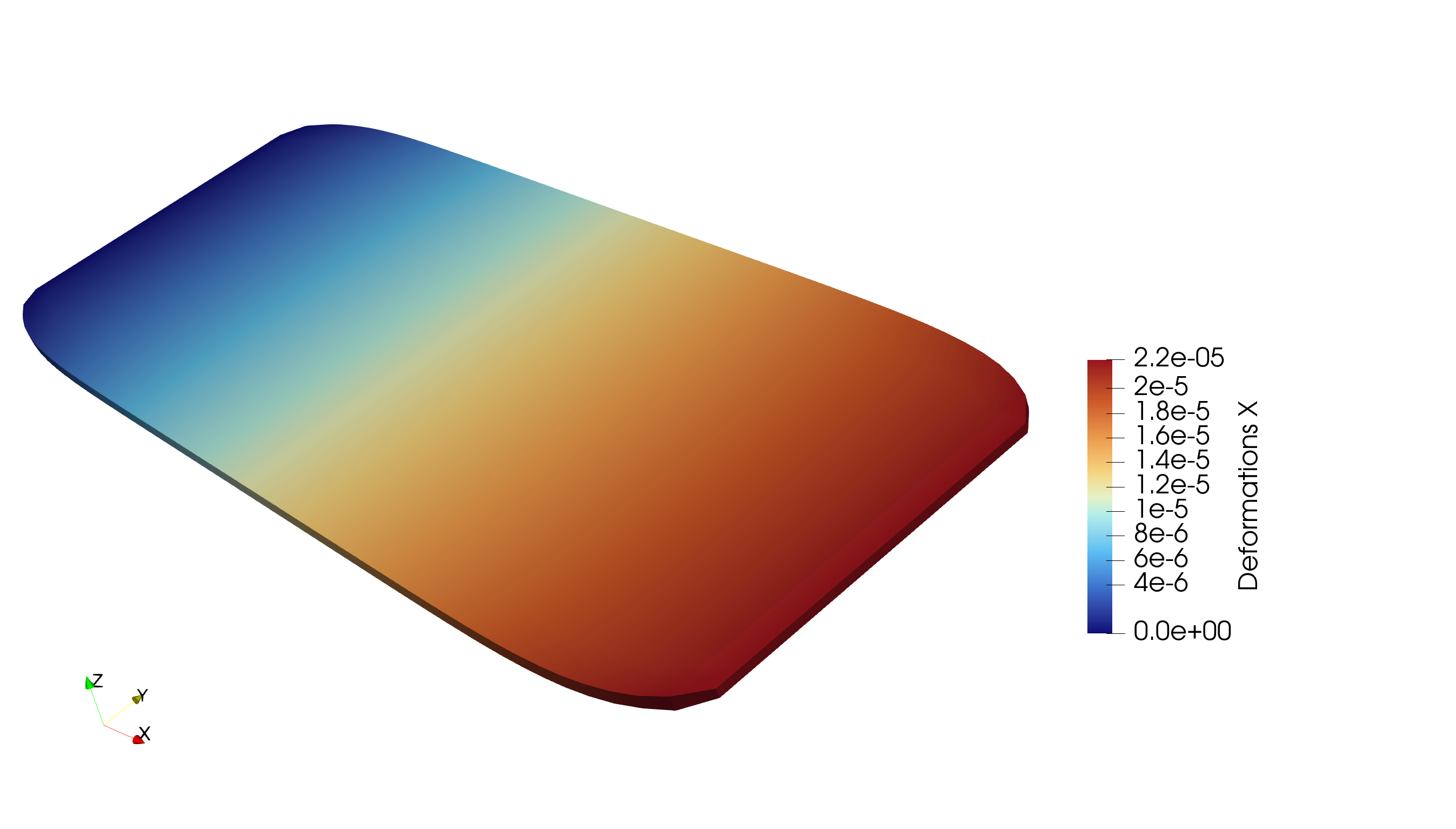}
    }
   \subfigure[Deformations along y direction (m).]
    {
        \includegraphics[width=0.99\textwidth]{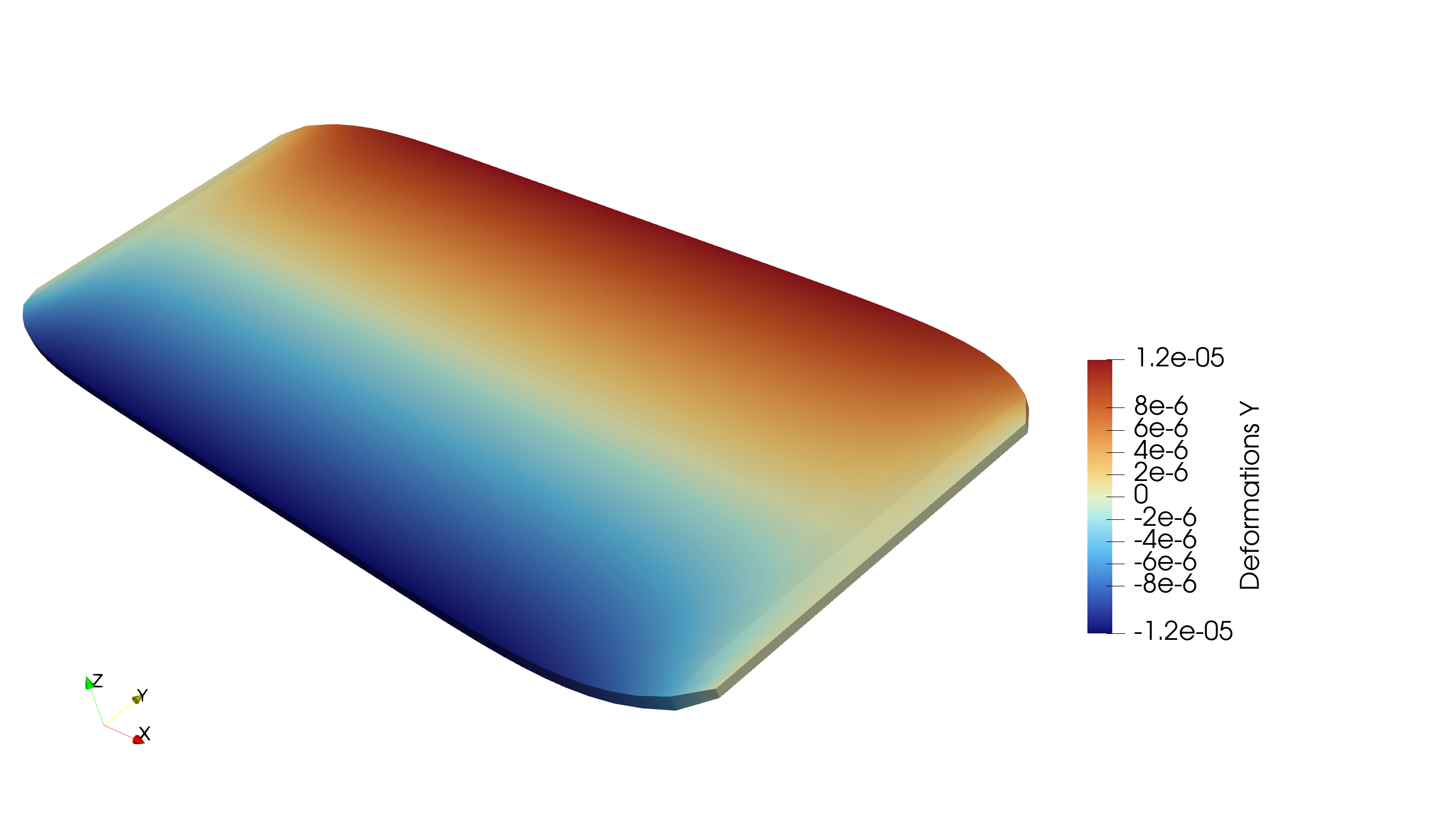}
    }
\caption{Solution of the mechanical deformation fields at the end of the first heating stage, with $T_\infty$ increased to $120^{\textrm{o}}C$.\label{deformationsolution}}
\end{figure}

\begin{figure}[!ht]
\centering
\includegraphics[width=0.99\textwidth]{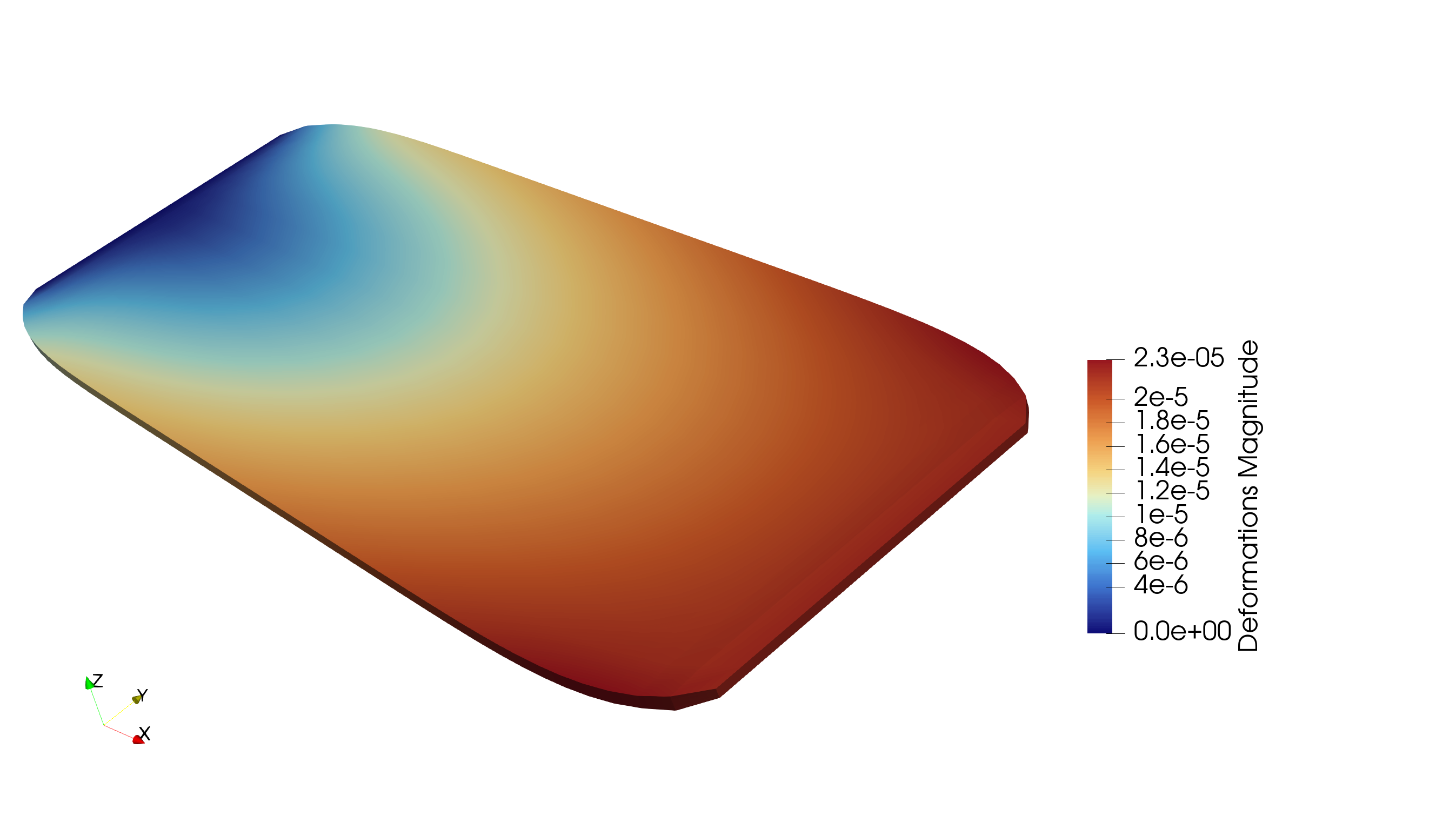}  
\caption{Deformations' magnitude (m) of the mechanical deformation fields at the end of the first heating stage, with $T_\infty$ increased to $120^{\textrm{o}}C$.\label{deformationsolutionmagnitude}}
\end{figure}

Figure \ref{deformationsolution} shows the deformation results along the axial $x$ direction and the lateral $y$ direction at the peak temperature after the first heating stage using the converged parameters. The deformations are amplified 100 times in the figures for better visualization. The deformation magnitude $\|\mathbf{u}\|$ is illustrated in Figure \ref{deformationsolutionmagnitude}.\\

\begin{figure}[!ht]
\centering
\includegraphics[width=0.9\textwidth]{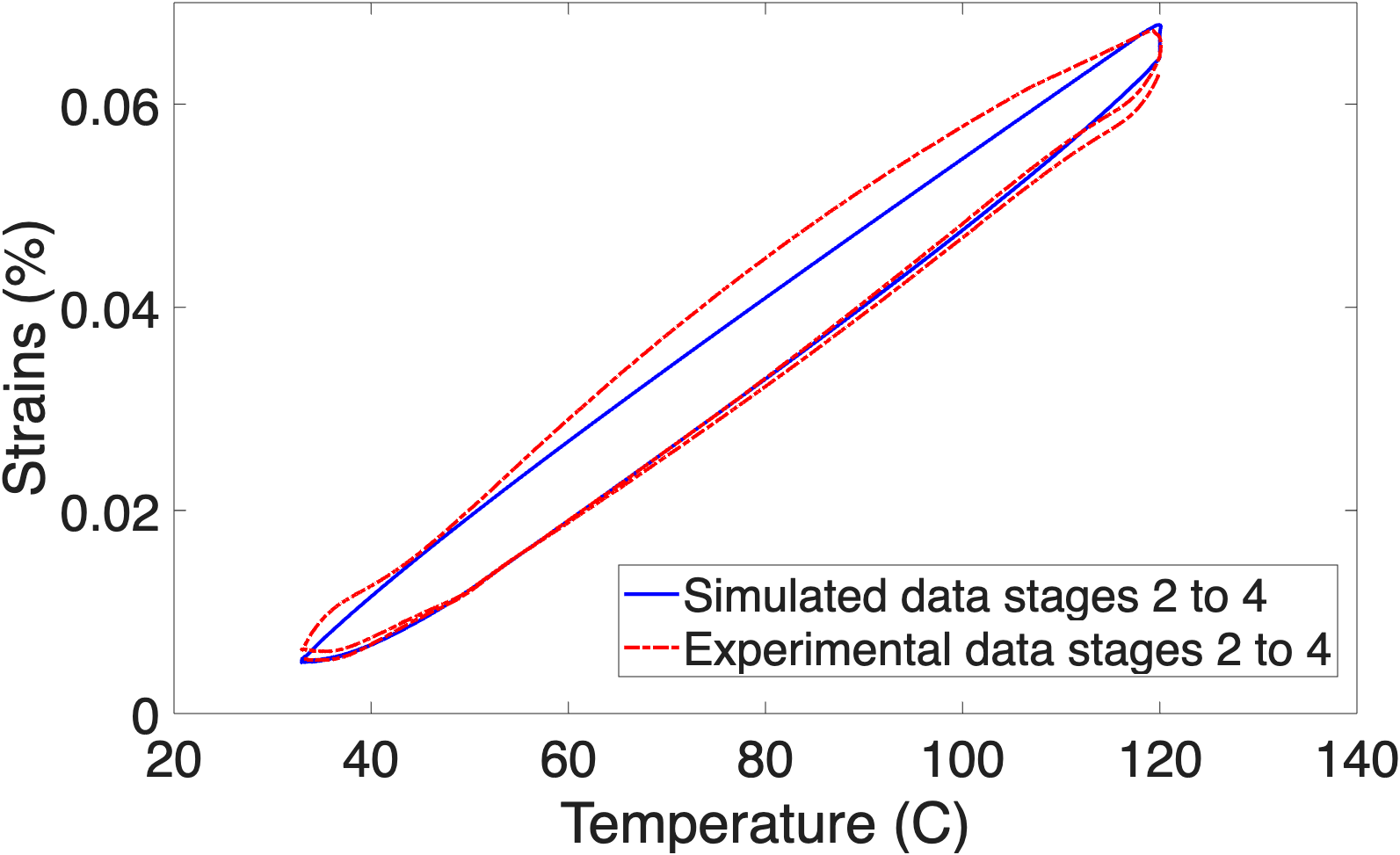}  
\caption{Fitted strain in DMA-1 with a maximum temperature of $120^{\textrm{o}}C$.\label{fittedstrainexp28}}
\end{figure}

To perform reliable identification of the parameters, first the expansion of the steel bars existing in the DMA experiments is subtracted from the strains, leading to a positive strain, not a negative one, as expected for the setup at hand. The coefficient of thermal expansion of the steel is also identified in the optimization process and found to be equal to $17.4^\textrm{o}C^{-1}$, in the expected range for stainless steel. The identified values are within the normal range for the prepreg used in this experiment. The resulting normal strain is illustrated in Figure \ref{fittedstrainexp28}.\\

\begin{figure}[!ht]
\centering
  \subfigure[Deformations along x direction (m).]
    {
        \includegraphics[width=0.99\textwidth]{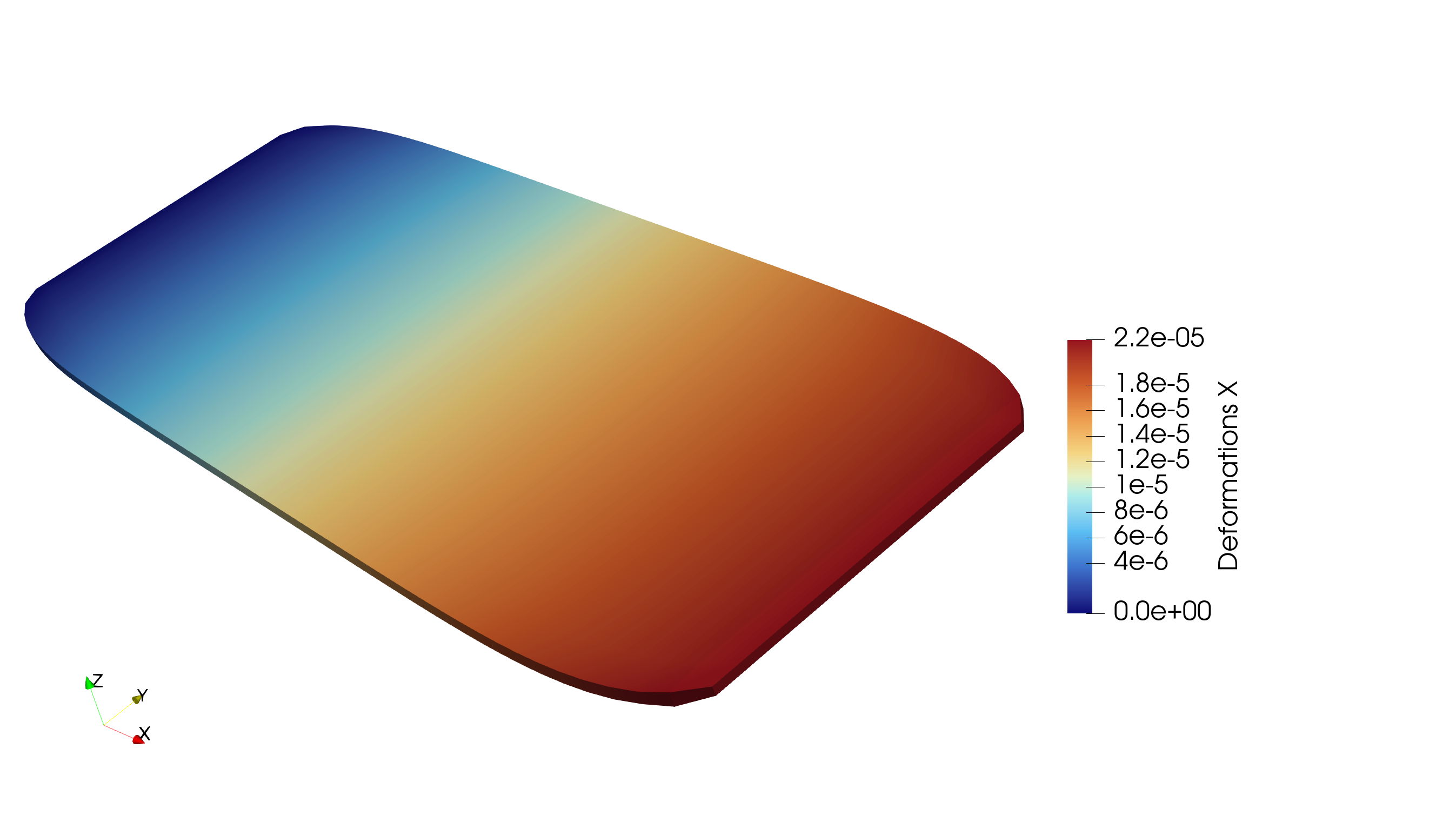}
    }
   \subfigure[Deformations along y direction (m).]
    {
        \includegraphics[width=0.99\textwidth]{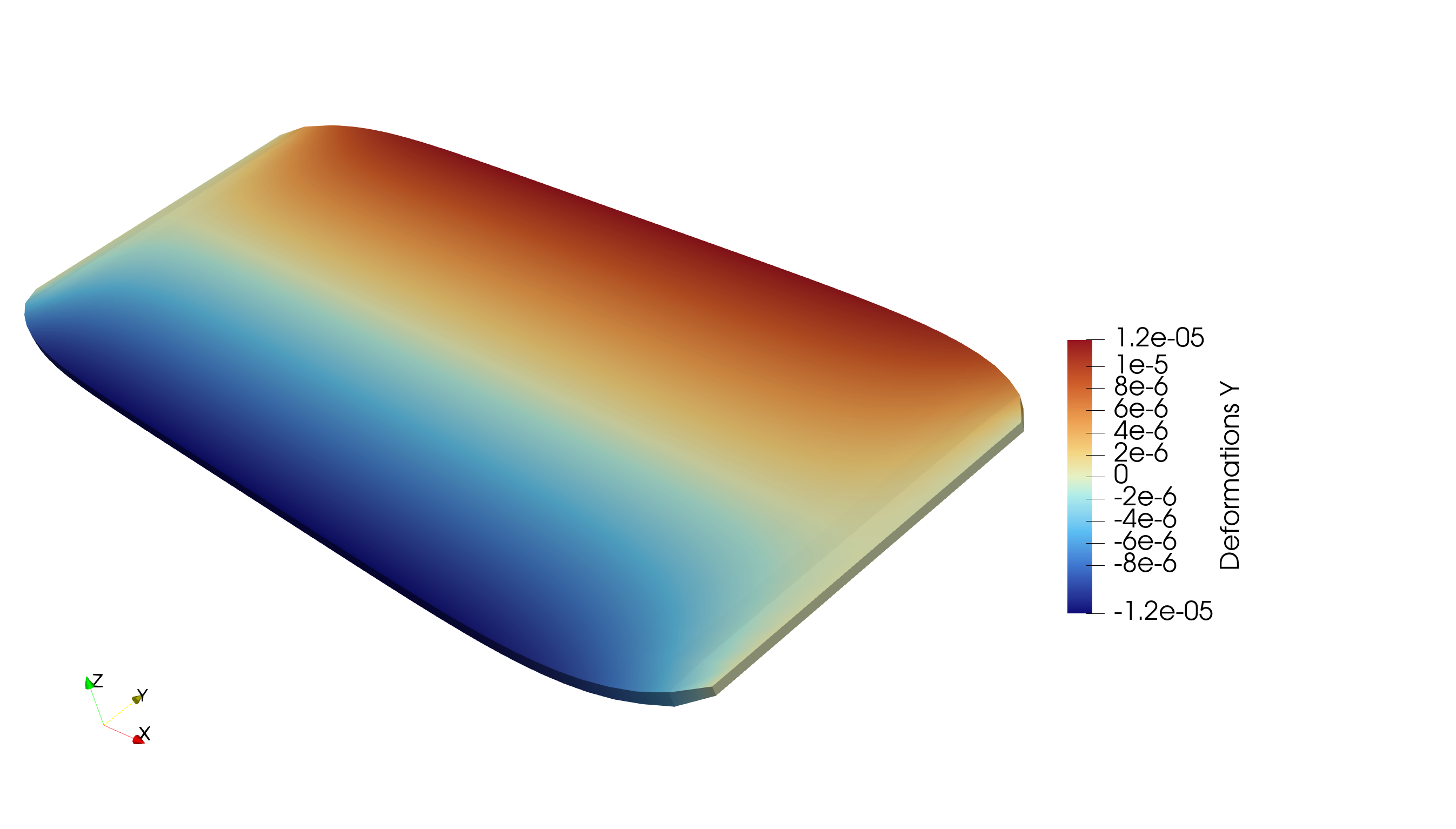}
    }
\caption{Solution of the mechanical deformation fields at the end of the first heating stage, with $T_\infty$ increased to $180^{\textrm{o}}C$.\label{deformationsolution27}}
\end{figure}

\begin{figure}[!ht]
\centering
\includegraphics[width=0.99\textwidth]{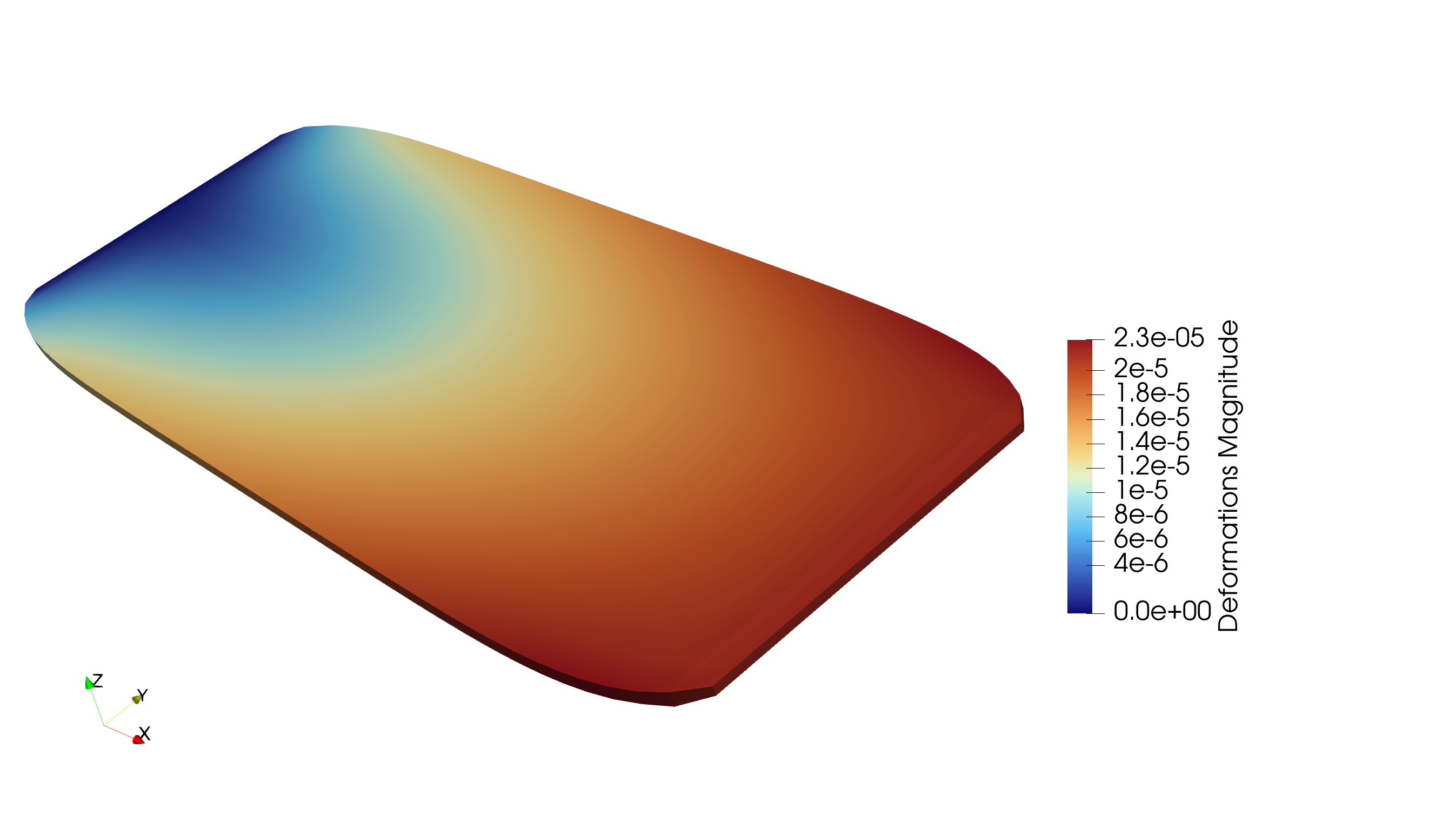}  
\caption{Deformations' magnitude (m) of the mechanical deformation fields at the end of the first heating stage, with $T_\infty$ increased to $180^{\textrm{o}}C$.\label{deformationsolution27magnitude}}
\end{figure}

\begin{figure}[!ht]
\centering
\includegraphics[width=0.9\textwidth]{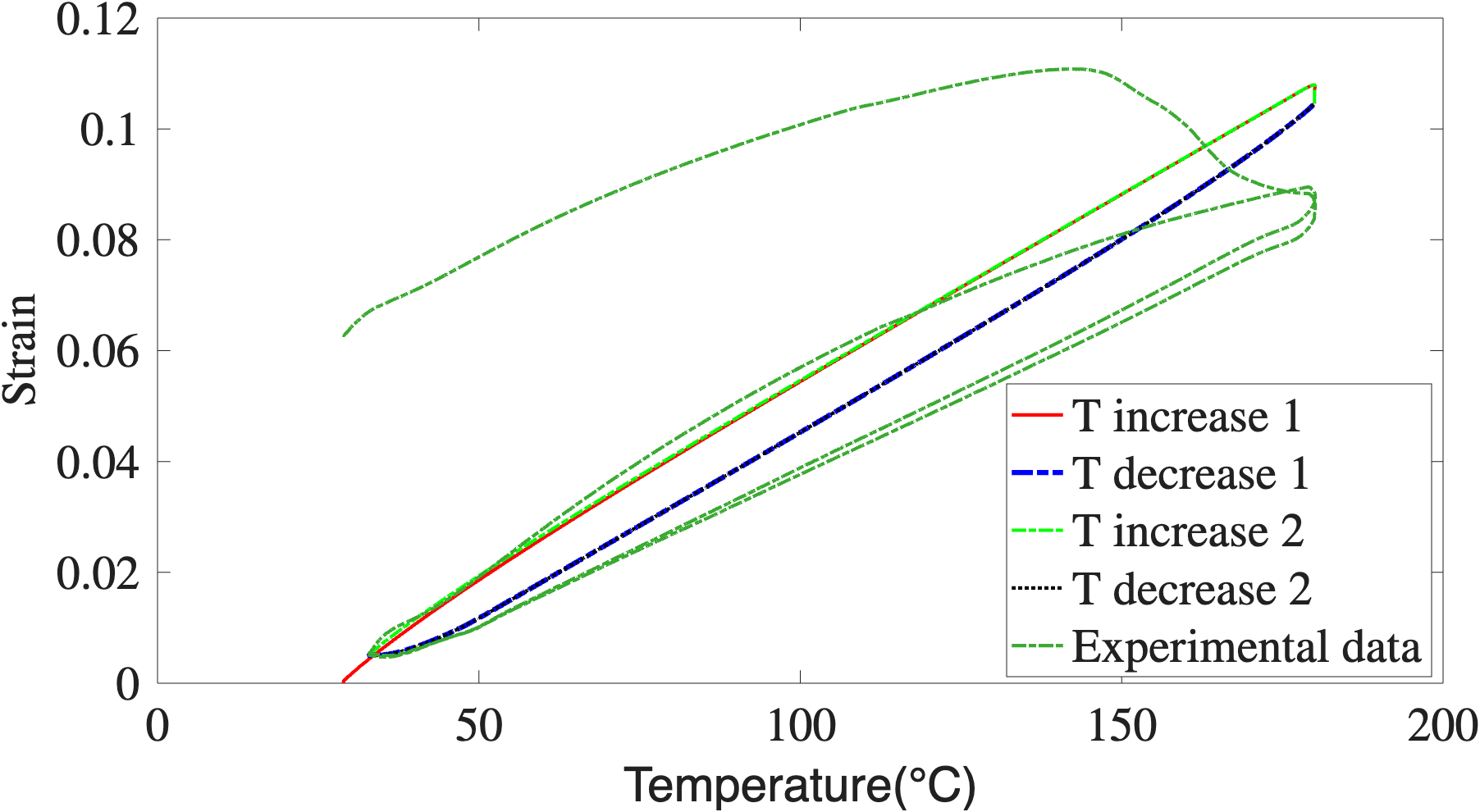}  
\caption{Predicted strains in DMA-2 with a maximum heating temperature of $180^{\textrm{o}}C$, no refitting of parameters is performed. The discrepancies will be used to identify the crystallization contributions.\label{fittedstrainexp27}}
\end{figure}

Once the parameters $\mathbf{p}_{converged}$ are identified from DMA-1, they are used for simulating DMA-2, where the temperature is increased to $180^{\textrm{o}}C$. The resulting deformations are illustrated in Figure \ref{deformationsolution27} for the $x$ and $y$ components. Figure \ref{deformationsolution27magnitude} illustrates the deformation magnitude, all at the maximum temperature at the end of the first heating cycle. Note that the deformations are amplified 100 times in the figures. Figure \ref{fittedstrainexp27} shows the comparison between the numerical simulation and the experimental measurements in DMA-2.
 
The initial shrinkage observed in both experiments 1 and 2 is the result of an initial irreversible shrinkage from drying and residual stress release. This effect will be quantified and modeled in Section \ref{dryingsec} using stabilized Neural Ordinary Differential Equations. The shift in the experimental strain measurements in DMA-2 is due to crystallization shrinkage and will be analyzed in Section \ref{crystallssec}.

\subsection{Drying and residual stresses modeling}\label{dryingsec}
In this section, we consider the first phase of the process, where the strains were not fitted in Section \ref{mechanicalsec}. The discrepancy in phase 1 of the experiments is used to compute the cumulative strain release as a function of time. This initial strain release is attributed to the drying of the part and the release of the existing residual stresses. From this point on, we refer by initial strain release to the sum of both effects, drying and residual stress release. The first heating stage discrepancy in DMA-1, where the temperature is increased to $120^{\textrm{o}}C$, is used to create an AI-driven model of the drying/residual stress release.

A stabilized Neural ODE model is used in this effort \cite{9959914,LINOT2023111838,en16155790}. Therefore, we define the initial strains to be released as:
\begin{equation}
\left\{
\begin{array}{l}
\epsilon^{ini}_{1}=\bar{\epsilon}_1(T<393.15K)-\epsilon^{s}_1(T<393.15K)\\
\epsilon^{ini}_{2}=\bar{\epsilon}_1(T<453.15K)-\epsilon^{s}_2(T<453.15K),\\
\end{array}
\right.
\end{equation} 
with the subscripts $1$ and $2$ indicates the experiment number and $\epsilon^{s}$ are the simulation results after parametric fitting. The training is performed on experiment $1$ only, by defining \cite{en16155790}:
\begin{equation}
\frac{\partial \epsilon^{ini}}{\partial t}=\mathcal{G}(\epsilon^{ini},T,t,\epsilon^{s})\cdot \epsilon^{ini} + \mathcal{H}(T,t,\epsilon^{s})
\end{equation}

The fitting is performed using an Euler integration scheme, leading to:
\begin{equation}
\epsilon^{ini}_{i+1}=\frac{\partial \epsilon^{ini}}{\partial t}|_i \cdot \Delta t+\epsilon^{ini}_{i}
\end{equation}

To train the neural ODE, the time sequence is divided into $n$ chains, and the integration is performed over $m$ time steps, and the loss function $\mathcal{L}$ is computed using:
\begin{equation}
\mathcal{L}=\sum\limits_{j=1}^{n}\left[ \sum\limits_{i=1}^{m} \left( \epsilon^{ini}_{i,j} -\hat{\epsilon}^{ini}_{i,j} \right)^2 \right],
\end{equation}

\begin{figure}[!ht]
\centering
\includegraphics[width=\textwidth]{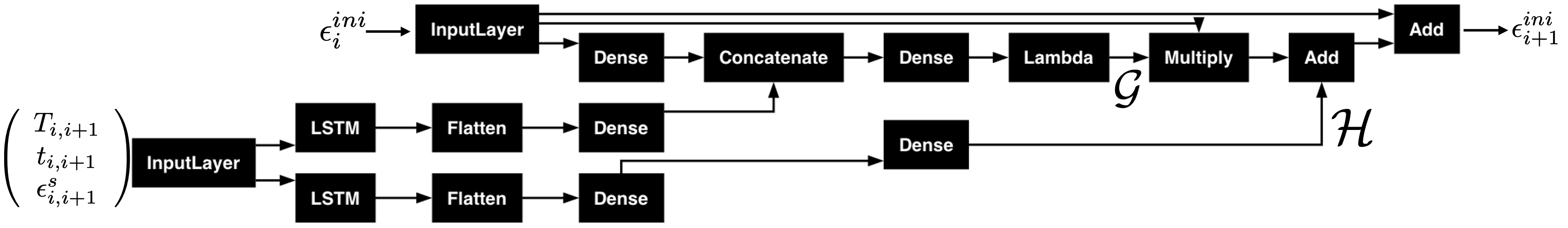}  
\caption{Neural networks are used to predict $\epsilon^{ini}$.\label{residualneuralode}}
\end{figure}

Where $\hat{\epsilon}^{ini}_{i,j}$ is the prediction of the residual strain released at time step $i$ from chain $j$. The gradient descent can be performed by batches using several chains from the $n$ available ones at a time. The used neural networks to train the surrogates $\mathcal{G}$ and $\mathcal{H}$ are illustrated in Figure \ref{residualneuralode}. The neural network employs two LSTM layers, taking the last two known values of $(T,t,\epsilon^{s})$ as input. The second input layer receives as input the previously predicted value $\hat{\epsilon}^{ini}_i$ to perform the integration for the next step and predict $\epsilon^{ini}_{i+1}$. Note that the initial value of the released stresses at ambient temperature is always set to zero, that is $\epsilon^{ini}_{0}=0$, allowing the use of this model in any prepreg manufacturing or handling process.

\begin{figure}[!ht]
\centering
\includegraphics[width=0.7\textwidth]{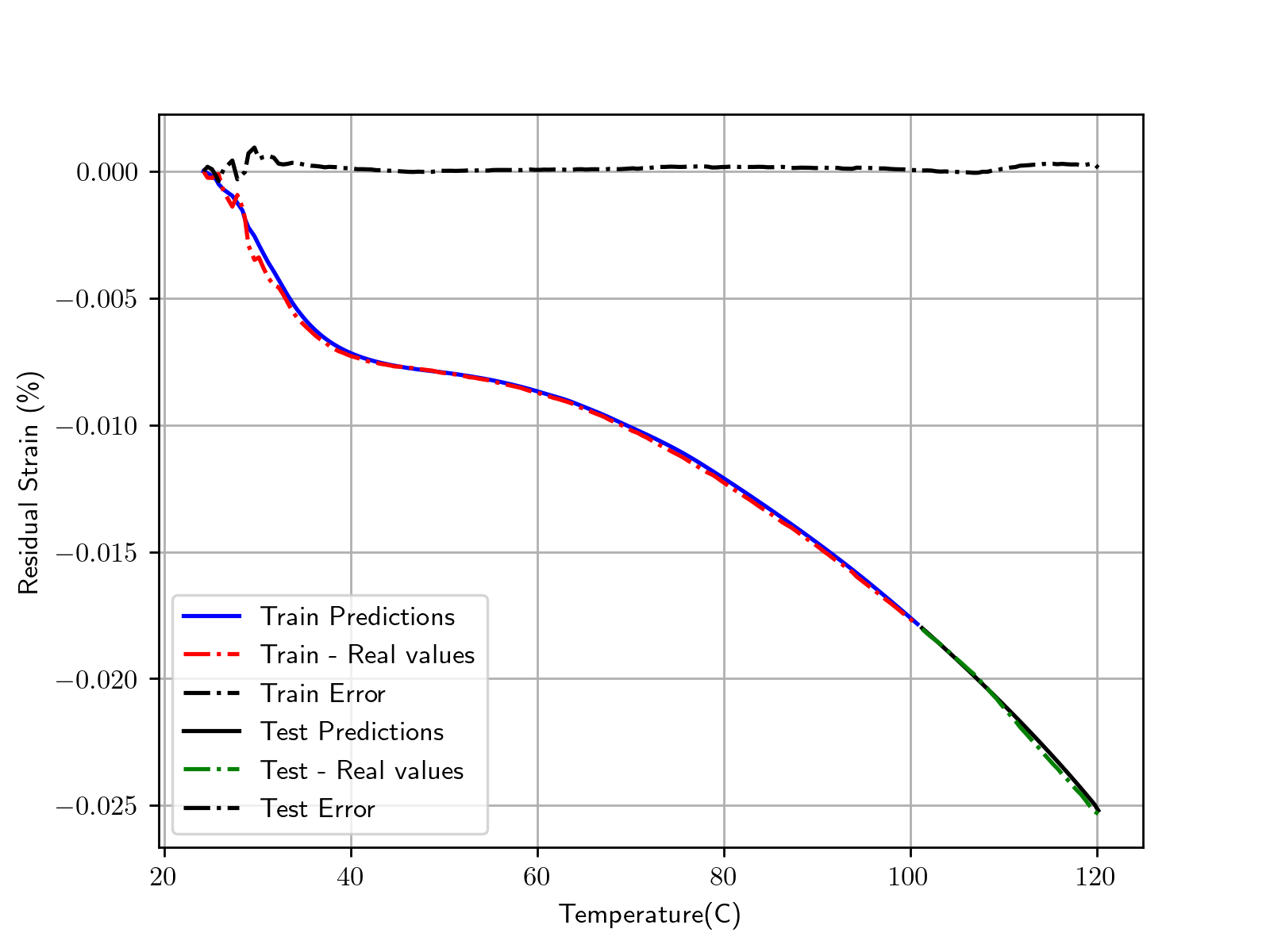}  
\caption{Results of the predicted $\epsilon^{ini}$ in DMA-1.\label{residualneuraloderesults1}}
\end{figure}

The training of the neural network is performed over the first 80\% of the time sequence data available from DMA-1. The remaining 20\% are used to validate the results. The network prediction of the initial strain release is illustrated in Figure \ref{residualneuraloderesults1}. The results show the ability of the network to predict the residual strains in high fidelity even beyond the training region.

\begin{figure}[!ht]
\centering
\subfigure[Random permutation features importance method outputs.\label{rpfimoutall}]{
\includegraphics[width=0.46\textwidth]{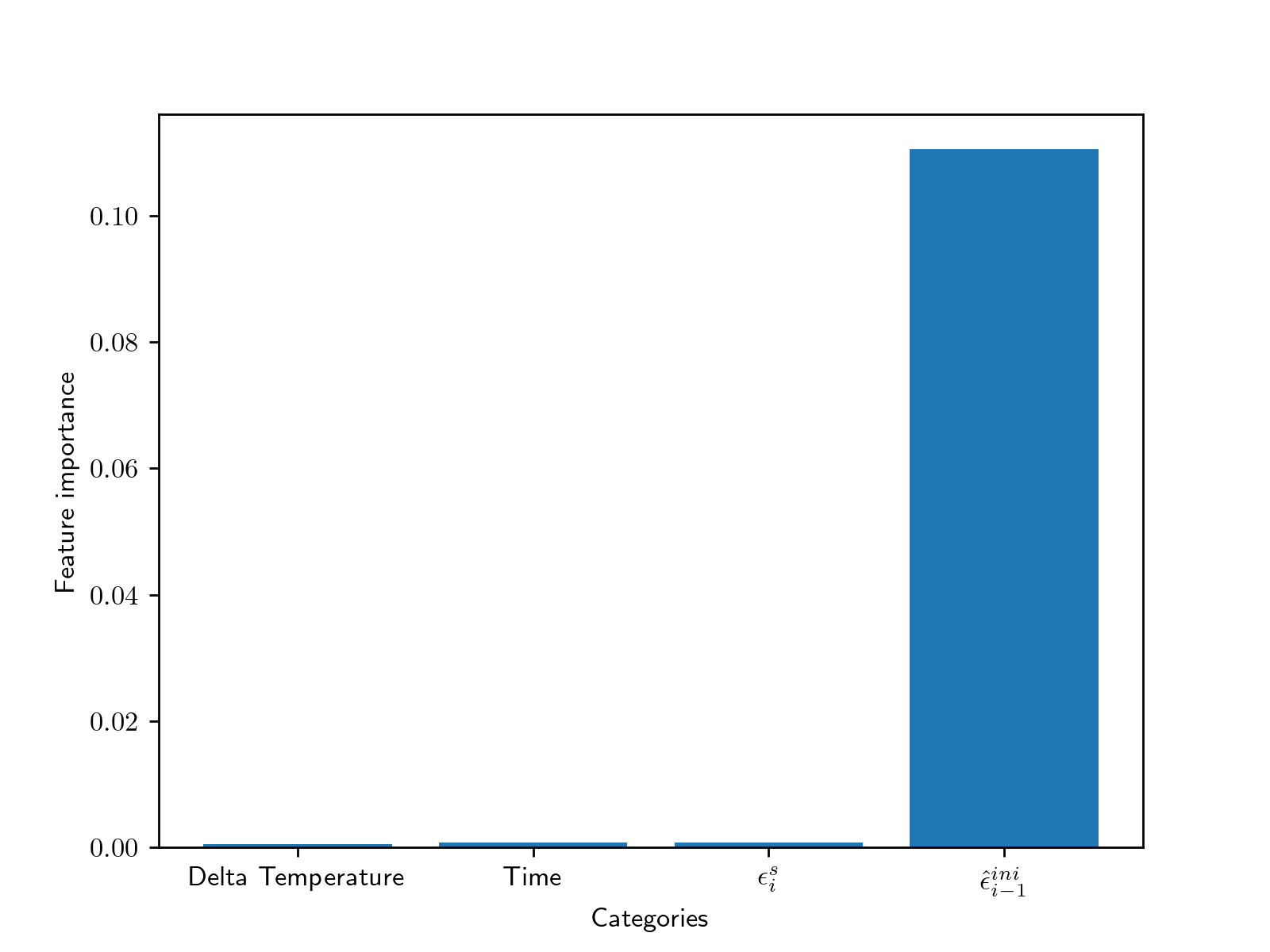}}
\subfigure[Random permutation features importance method outputs without the last predicted $\epsilon^{ini}$ value, normalized and shown as a fraction value. \label{rpfimoutonlylstm}]{
\includegraphics[width=0.46\textwidth]{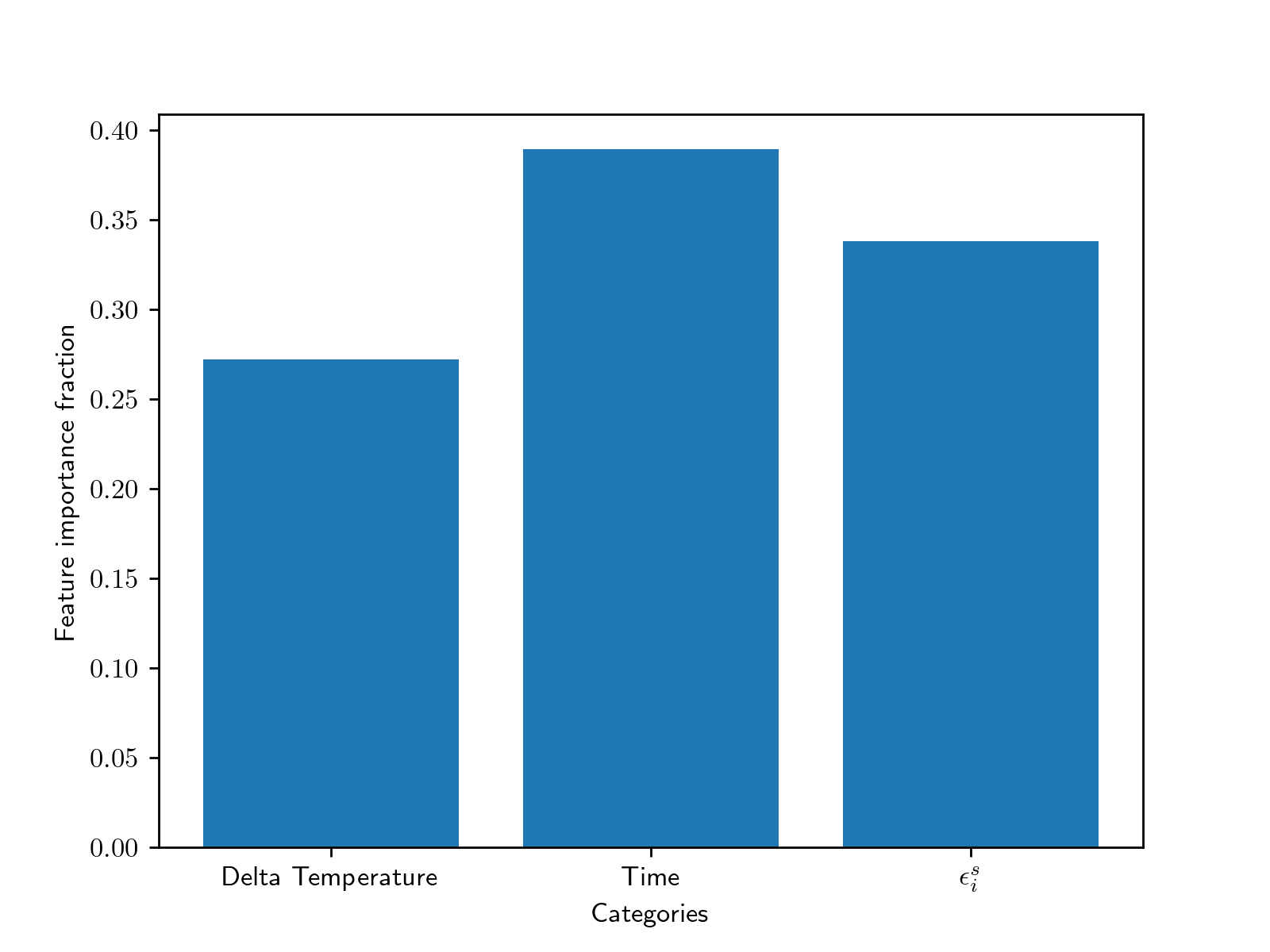}}
\caption{Random permutation features importance method on the surrogate model of $\frac{\partial \epsilon^{ini}}{\partial t}$ estimating the importance of the selected inputs.}
\end{figure}

To gain further insights on the way the network is using the inputs to predict the drying and residual stress release, the random permutation feature importance method is used to predict the effect of the inputs on the network's predictions. The results are illustrated in Figure \ref{rpfimoutall}. The results show that the previous prediction $\hat{\epsilon}^{ini}_{i-1}$ has the highest effect on the next value prediction $\hat{\epsilon}^{ini}_{i}$, which is obvious as it appears in the nonlinear form $\mathcal{G}$ and induces error accumulations. To further refine the effect of the inputs. This feature is eliminated in Figure \ref{rpfimoutonlylstm}, and the three remaining features are evaluated. We can see all the three features have similar contributions. This is expected since $\Delta T$ is set to a constant in our dataset, with a heating rate that is controlled at $0.5 ^{\textrm{o}}C/min$.

The relative errors $e_a^{ini}$ of the predictions are shown to be 0.76\% for the train set and 0.57\% for the test set, computed using the following equation:
\begin{equation}
e_a^{ini}=100\times \sum\limits_{i=1}^{N} \frac{\left|\epsilon^{ini}_i-\hat{\epsilon}^{ini}_i \right|}{\left\| \epsilon^{ini}\right\|_{\infty}}
\end{equation}

A division by the norm infinity $\left\| \cdot \right\|_{\infty}$ is used since the strain release is close to zero in the beginning of the predictions.

\begin{figure}[!ht]
\centering
\subfigure[Predictions $\hat{\epsilon}^{ini}_2$ as a function of the temperature \label{exp2temp}]{\includegraphics[width=0.47\textwidth]{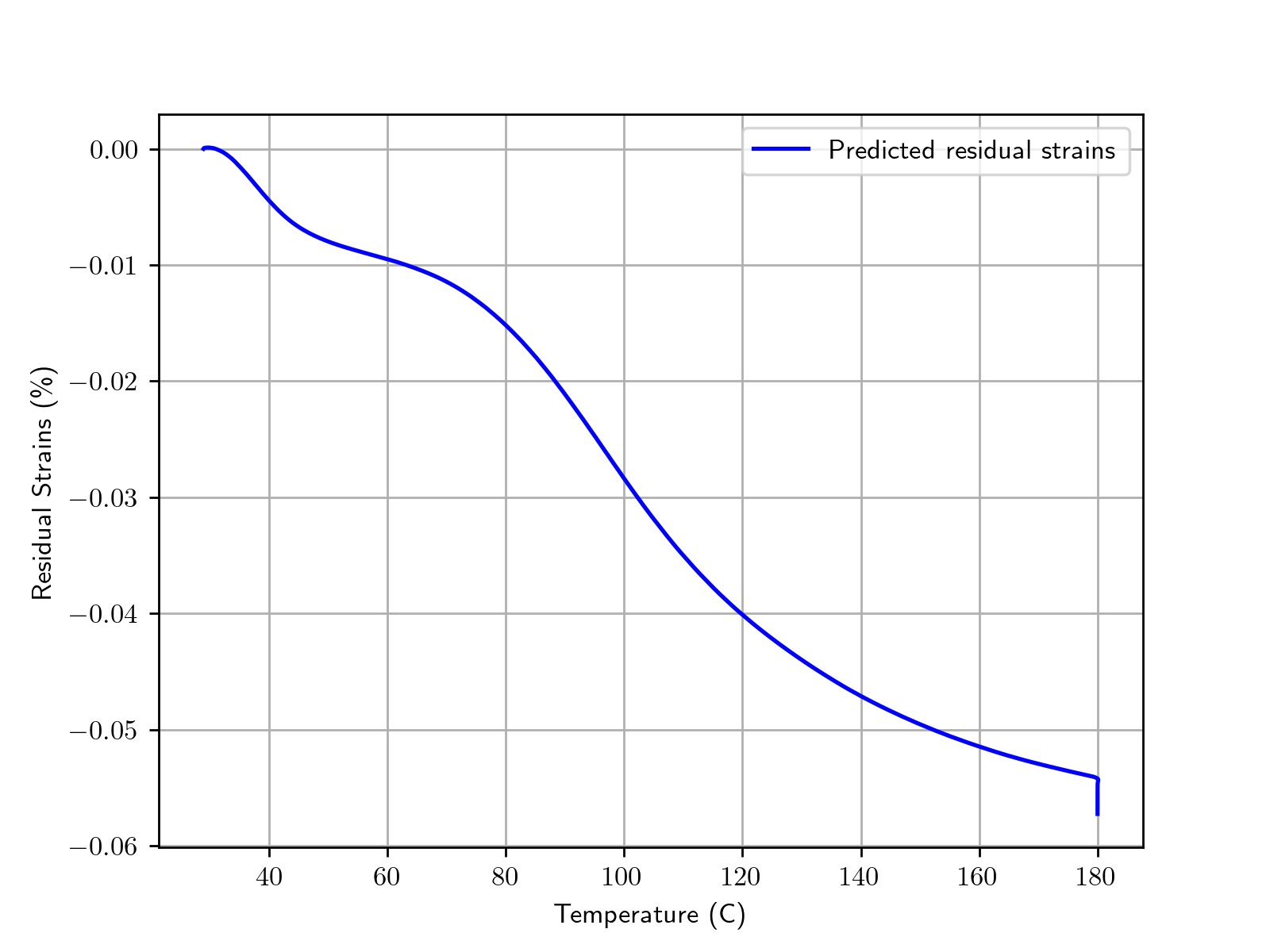}  
}
\subfigure[Predictions $\hat{\epsilon}^{ini}_2$ as a function of the time \label{exp2time}]{\includegraphics[width=0.43\textwidth]{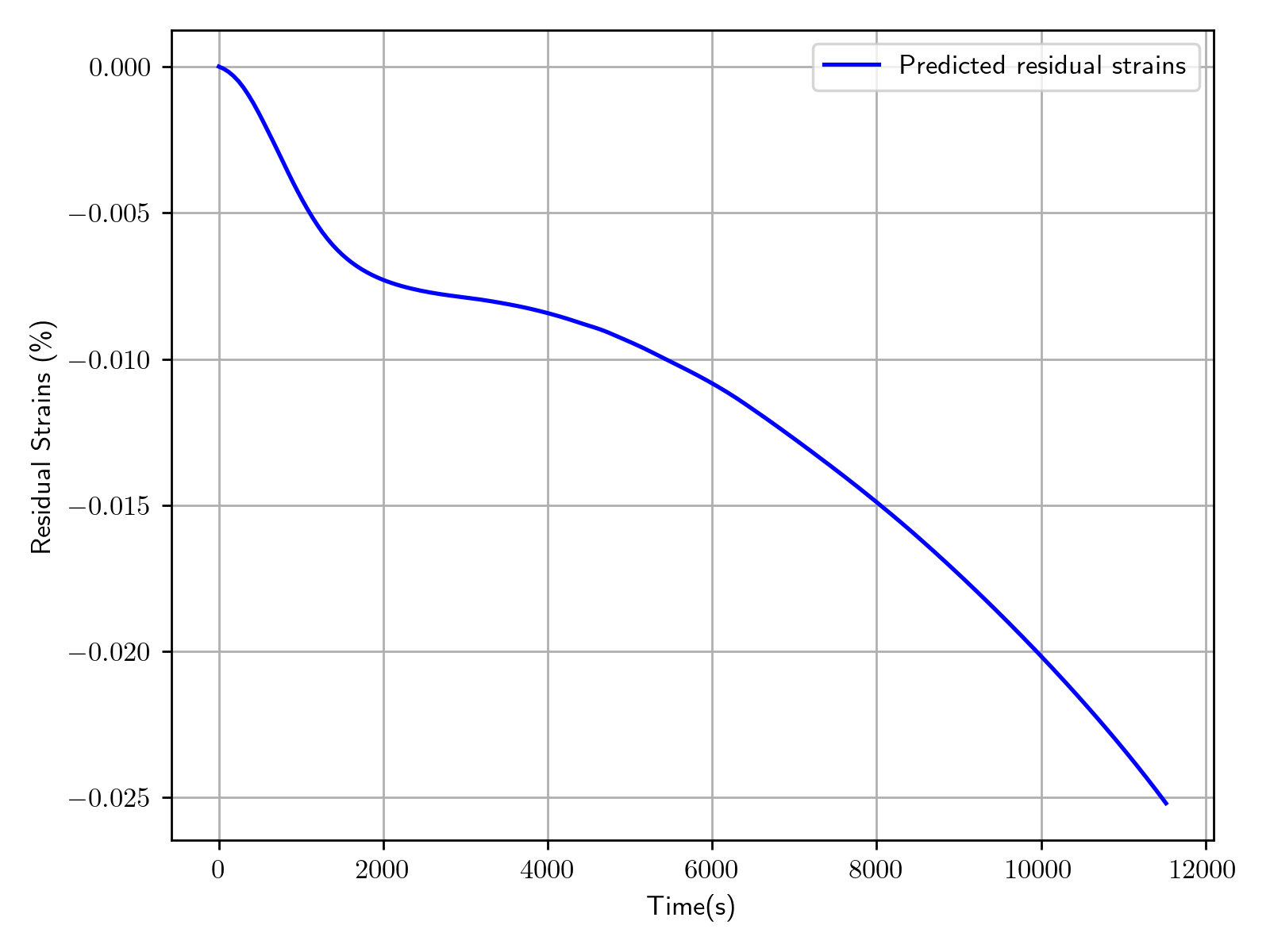}}
\caption{Results of the $\hat{\epsilon}^{ini}_2$ prediction, with the experimental inputs taken from DMA-2.\label{residualneuraloderesults2}}
\end{figure}

The same model, without further refinement, is used to predict the initial strain release in the second experiment, with a maximum temperature being $60^{\textrm{o}}C$ higher. The results are illustrated in Figure \ref{residualneuraloderesults2}. We can note that the neural network learned to predict the saturation automatically for higher temperatures, appearing better in Figure \ref{exp2time}, despite never seeing temperatures beyond $120^{\textrm{o}}C$ during training. This is a plausible effect induced by the stable form of the neural ODE \cite{9959914,en16155790}. Moreover, we note that the shrinking accelerates just around $T=100^{\textrm{o}}C$, the evaporation temperature of the water, before reaching saturation later on.

\subsection{Crystallization-induced deformation}\label{crystallssec}
In this section, we used DMA-2 with a maximum temperature reaching $180^{\textrm{o}}C$, to model the crystallization effect on the strains. It is commonly accepted that crystallization onsets start to germinate just above the crystallization temperature in similar thermoplastics \cite{KEMMISH1985905,ATKINSON2002731,KURTZ20074845}. It is also shown that and a measure maximum in the heat of crystallization, of $22J/g$, immediately above the glass transition region is emitted in PEEK \cite{KEMMISH1985905}. For our LM PAEK-CF material, the glass transition temperature is around $T_g=147^{\textrm{o}}C$; therefore, we aim to explore the effect of crystallization on the strains by exploring DMA-2 between $146^{\textrm{o}}C$ to $180^{\textrm{o}}C$.

We define the crystallization strains $\epsilon^{crys}$ as follows:
\begin{equation}\label{crystallizationstraineq}
\epsilon^{crys}=\bar{\epsilon}-\epsilon^{s}-\epsilon^{ini}
\end{equation}

\begin{figure}[!ht]
\centering
\includegraphics[width=\textwidth]{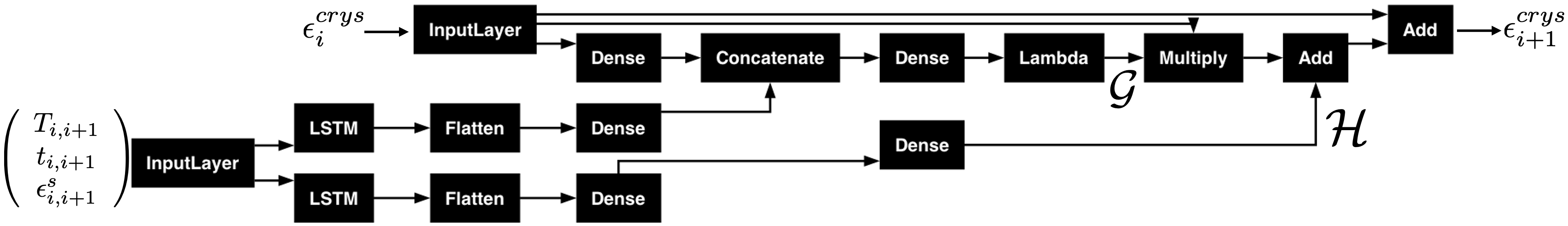}  
\caption{Neural network used for the prediction of the crystallization effect on strains\label{nncrystals}}
\end{figure}

The crystallization strains given in equation (\ref{crystallizationstraineq}) are defined throughout  DMA-2 for all temperatures. However, the initial strain release $\epsilon^{ini}$ is computed on the first heating stage only and remains constant afterwards. The crystallization strains are modeled using the same form as for the initial strain release, using a stabilized neural ODE form, defined by:

\begin{equation}
\frac{\partial \epsilon^{crys}}{\partial t}=\mathcal{G}_c(\epsilon^{crys},T,t,\epsilon^{s})\cdot \epsilon^{crys} + \mathcal{H}_c(T,t,\epsilon^{s}),
\end{equation}

where  $\mathcal{G}_c$ and $\mathcal{H}_c$ are two new deep neural networks with LSTM layers, as illustrated in Figure \ref{nncrystals}. The same structure is used for the prediction of the crystallization strains, starting with $\epsilon^{crys}_0=0$, and integrating through the network using an Euler integration scheme:
\begin{equation}
\epsilon^{crys}_{i+1}=\left.\frac{\partial \epsilon^{crys}}{\partial t}\right|_{i} \Delta t+\epsilon^{crys}_{i}
\end{equation}

The loss function $\mathcal{L}_c$ also employs an integration over several time steps and batches before computing the loss value and the gradient descent:
\begin{equation}
\mathcal{L}_c=\sum\limits_{j=1}^{n}\left[ \sum\limits_{i=1}^{m} \left( \epsilon^{crys}_{i,j} -\hat{\epsilon}^{crys}_{i,j} \right)^2 \right],
\end{equation}

\begin{figure}[!ht]
\centering
\subfigure[Predictions $\hat{\epsilon}^{crys}_2$ as a function of temperature \label{exp2onset1temp}]{\includegraphics[width=0.47\textwidth]{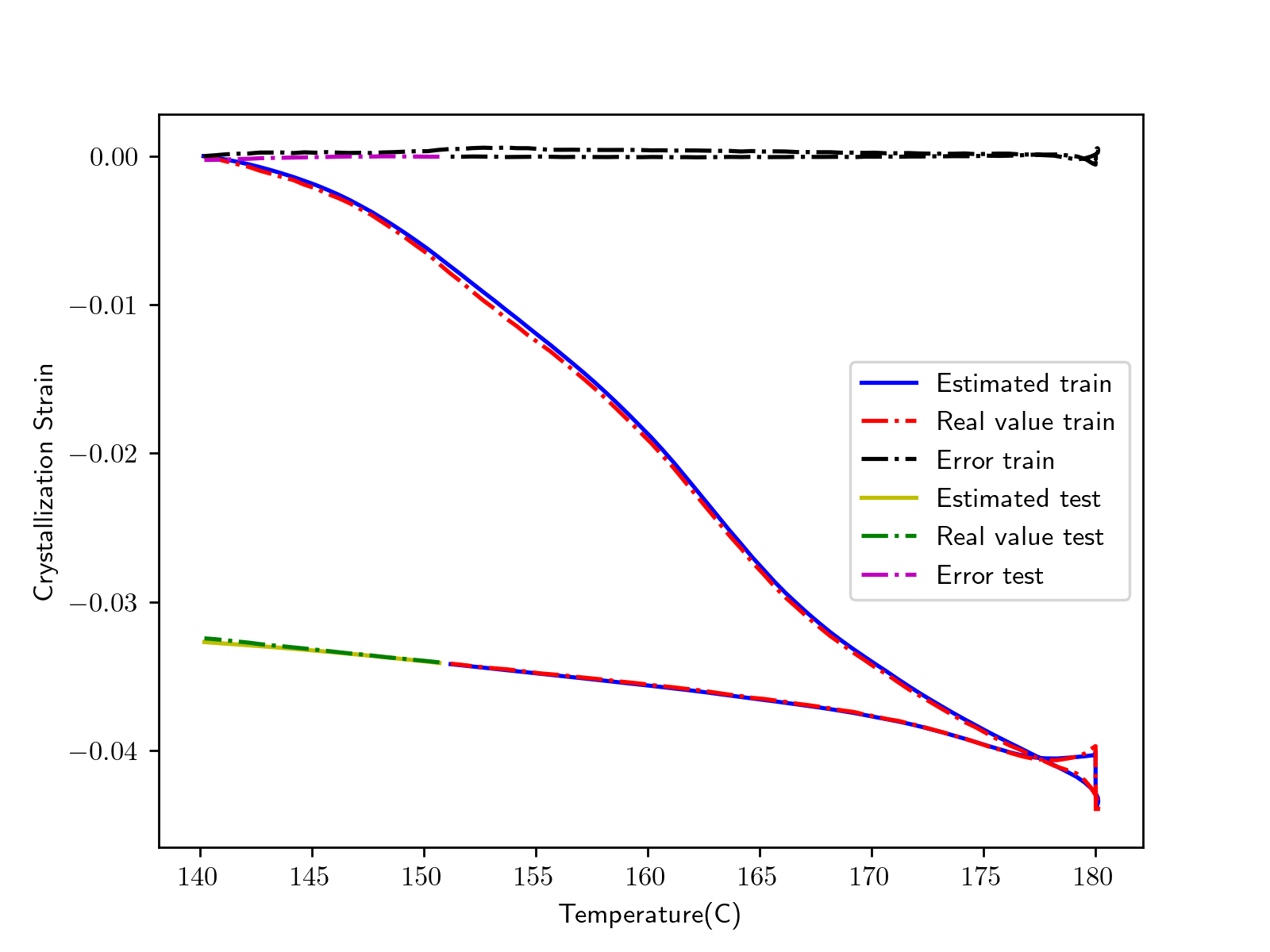}  
}
\subfigure[Predictions $\hat{\epsilon}^{crys}_2$ as a function of time \label{exp2onset1time}]{\includegraphics[width=0.47\textwidth]{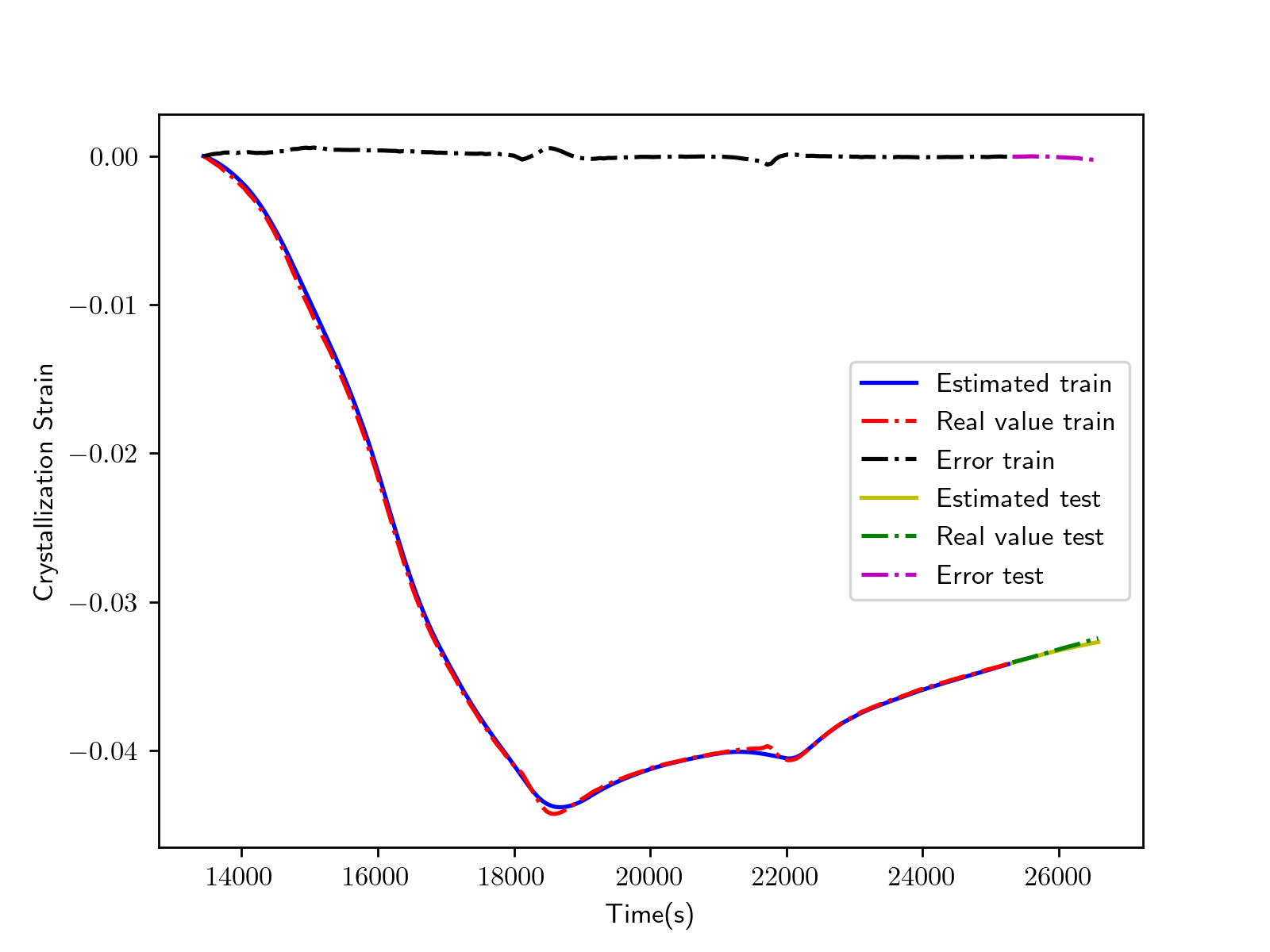}}
\caption{Results for the prediction $\epsilon^{crys}_2$ during the first onset of crystallization, with the inputs $(\Delta T,t,\epsilon^s)$ from DMA-2 when reaching a temperature $T>T_g$ and remaining above $T_g$ during the first heating cycle.\label{crystallizationonset1results}}
\end{figure}

ADAM gradient descent algorithm is used with a custom-built learning rate adaptation function. Initially, the problem is trained only on the first onset of crystallization, when the sample remains at a temperature above $T_g$ in the first heating and then cooling cycle. The modeling effort stopped at the point the temperature goes below $T_g$. The results of the neural network training for the first onset of crystallization are illustrated in figure \ref{crystallizationonset1results}. The results show the ability of the proposed regularized neural ODE to predict the results accurately and to generalize beyond the trained dataset.

\begin{figure}[!ht]
\centering
\subfigure[Predictions of $\hat{\epsilon}^{crys}_2$ as a function of temperature \label{exp2alltemp}]{\includegraphics[width=0.47\textwidth]{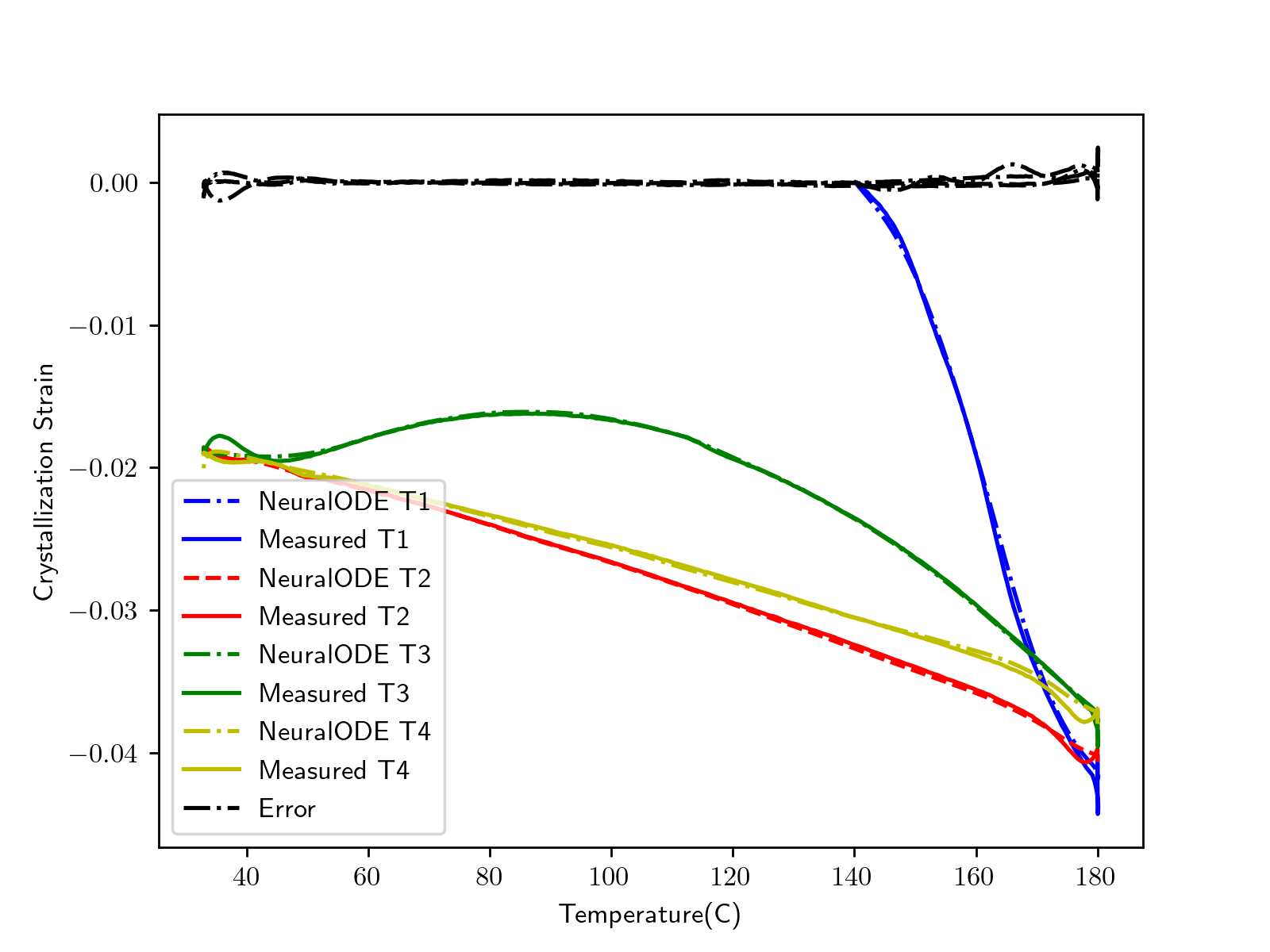}  
}
\subfigure[Predictions of $\hat{\epsilon}^{crys}_2$ as a function of time \label{exp2alltime}]{\includegraphics[width=0.47\textwidth]{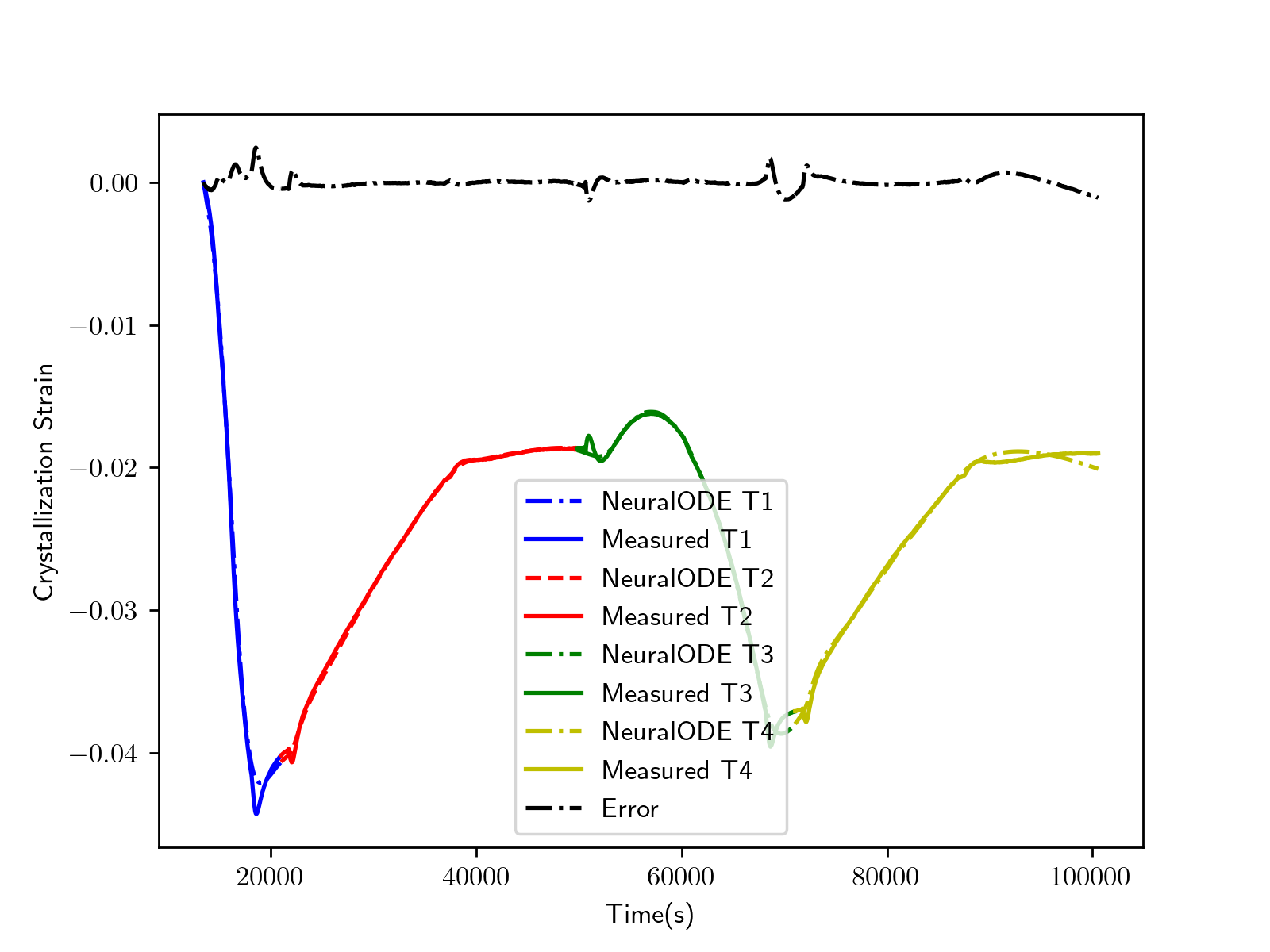}}
\caption{Predictions of $\epsilon^{crys}_2$ for the complete experiment beyond $T_g$, with the inputs $(\Delta T,t,\epsilon^s)$. $T1$ is the first heating stage, $T2$ the first cooling one, $T3$ the second heating stage and $T4$ the second cooling one.\label{crystallizationonallresults2}}
\end{figure}

Beyond the first onset of crystallization, the mechanical and thermal properties of the composite prepreg change. Therefore, the surrogates $\mathcal{G}_c$ and $\mathcal{H}_c$ are trained beyond the first onset of crystallization until the end of the process to model the deformation induced by the crystals themselves and the change in material properties. The same ADAM algorithm with the custom-built learning rate adaptation function is used for the training. The results are illustrated in Figure \ref{crystallizationonallresults2}, where $T1$ and $T3$ are the first and the second heating stages, and $T2$ and $T4$ are the first and the second cooling steps. The last cooling stage $T4$ is kept for testing and is never shown to the neural network during training.

The relative errors $e_a^{crys}$ of the predictions are shown to be 1.65\% for the train set and 2.43\% for the test set, computed using the following equation:
\begin{equation}
e_a^{crys}=100\times \sum\limits_{i=1}^{N} \frac{\left| \epsilon^{crys}_i-\hat{\epsilon}^{crys}_i\right|}{\left| \epsilon^{crys}_i\right|}
\end{equation}

\subsection{Discussion}\label{discussionsec}
\begin{figure}[!ht]
\centering
\includegraphics[width=0.7\textwidth]{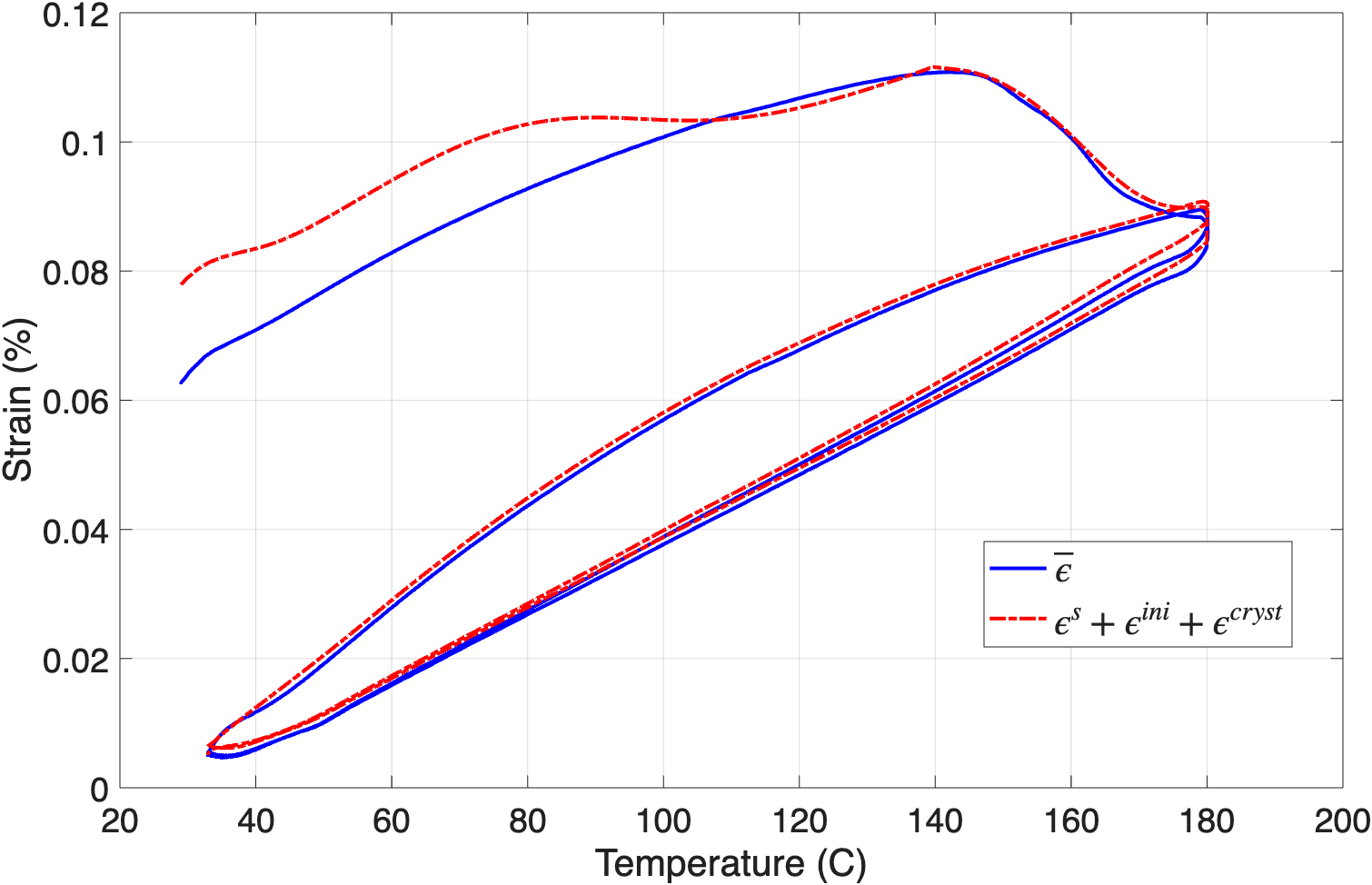}  
\caption{Comparison of strain as a summation of the three built models with the strain measured in DMA-2. \label{exp27_all_included_no_steel_final}}
\end{figure}

To validate the numerical models developed in this work, we sum the contribution of all strain components and compare the total strain with the strain measured in DMA-2. The results are illustrated in Figure \ref{exp27_all_included_no_steel_final}. It illustrates the prediction of the release of initial strains and/or drying is slightly overestimated in this model, as it was trained only on the experiment reaching a maximum temperature of $120^\textrm{o}C$, while the predictions are performed up to $180^\textrm{o}C$. The model did reach saturation thanks to its stabilized form but slightly overestimated the strain release. 

On the other hand, the results in Figure \ref{exp27_all_included_no_steel_final} show the ability of the network to predict the contribution of the crystallization to strain generation as soon as the temperature increases beyond $T_g$. 

These models will be used to generalize the characterization results obtained from the DMA machine and to predict the dimensional change of a prepreg tape after deposition by a robotic 3D printing in the next section.

\section{Prepreg tape dimensional change after robotic 3D printing}\label{exterapolationtorealcomposites}
In this part, we develop a complete simulation of a prepreg tape dimensional change after deposition by the robotic 3D printer combining the previously derived models. Here, the identified kelvin-voight viscoelastic model, the initial strain release, and the crystallization strain models are all taken into consideration.

\subsection{Thermal and mechanical simulation of the robotic 3D printing process}
\begin{figure}[!ht]
\centering
\includegraphics[width=\textwidth]{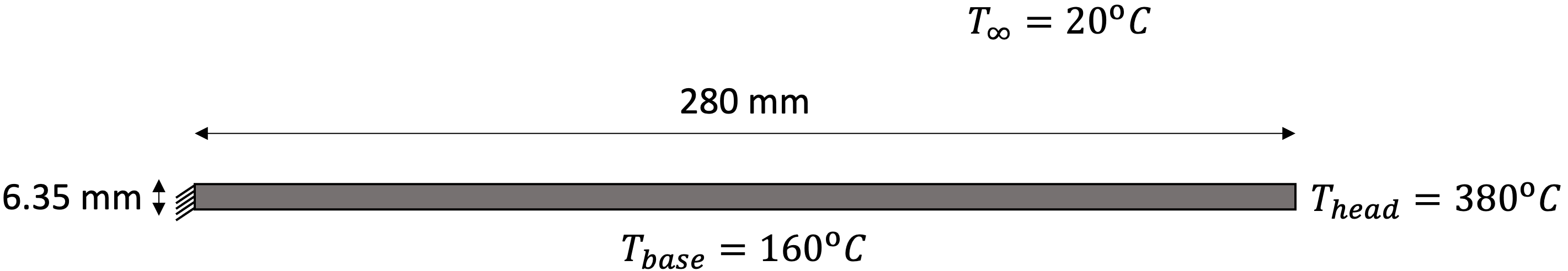}  
\caption{Top view of the 3D printed tape with the simulation boundary conditions. \label{3dprintingshape}}
\end{figure}

The simulated tape has a length of $L=280$ mm, a width of $W=6.35$ mm, as illustrated in Figure \ref{3dprintingshape}, and a thickness of $H=0.177$ mm. The boundary conditions are summarized as an imposed temperature $T_{base}=160^\textrm{o}C$ at the built platform, the incoming tape temperature set to $T_{head}=380^\textrm{o}C$, and the surrounding air temperature set to $T_{\infty}=20^\textrm{o}C$. The coefficient of convection with the air $h_{air}=25W/m^2 K$, and the one between the base support and the printed part is set to $h_{base}=60W/m^2 K$.

The thermal simulation solves the convection-diffusion heat transfer equation:
\begin{equation}
\rho C_p \mathbf{v}_{deposition} \frac{\partial T}{\partial x}-\nabla \left(\mathbf{K}\cdot \nabla T\right)=0
\end{equation}

with the tape 3D printing velocity $\mathbf{v}_{deposition}=10 mm/s$. The heat transfer equation is solved using the PGD with the Streamline Upwind Petrov-Galerkin (SUPG) stabilization method to ensure the stability of the convective term \cite{ghnatios:tel-00867281,Donea2003-tn,GHNATIOS2019769}.

\begin{figure}[!ht]
\centering
\includegraphics[width=0.7\textwidth]{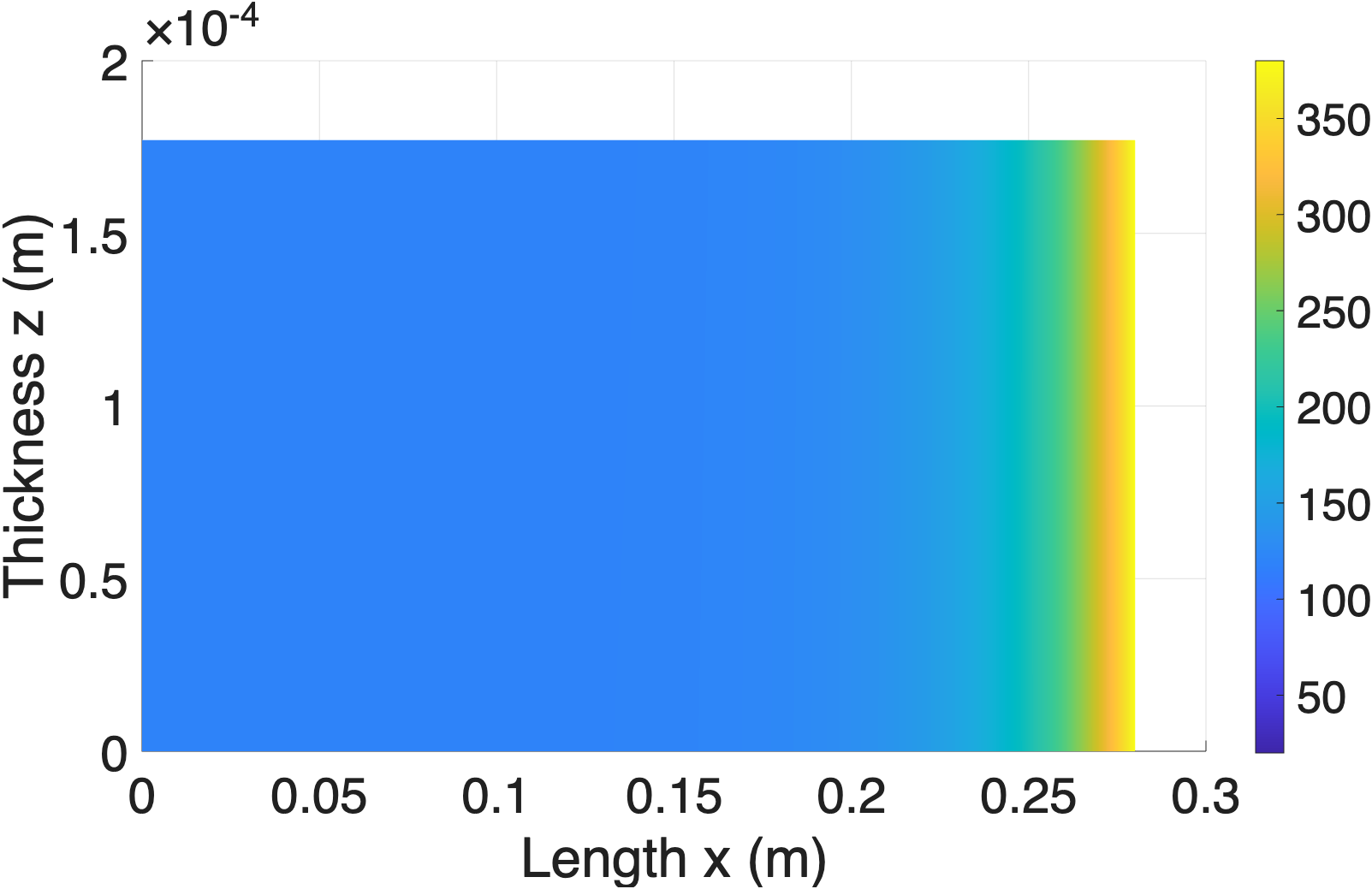}  
\caption{Through-thickness thermal fields in the 3D printed tape at the instant the head reaches the end of the deposition path of  $280mm$ length. \label{thermalfieldspic}}
\end{figure}

\begin{figure}[!ht]
\centering
\includegraphics[width=0.7\textwidth]{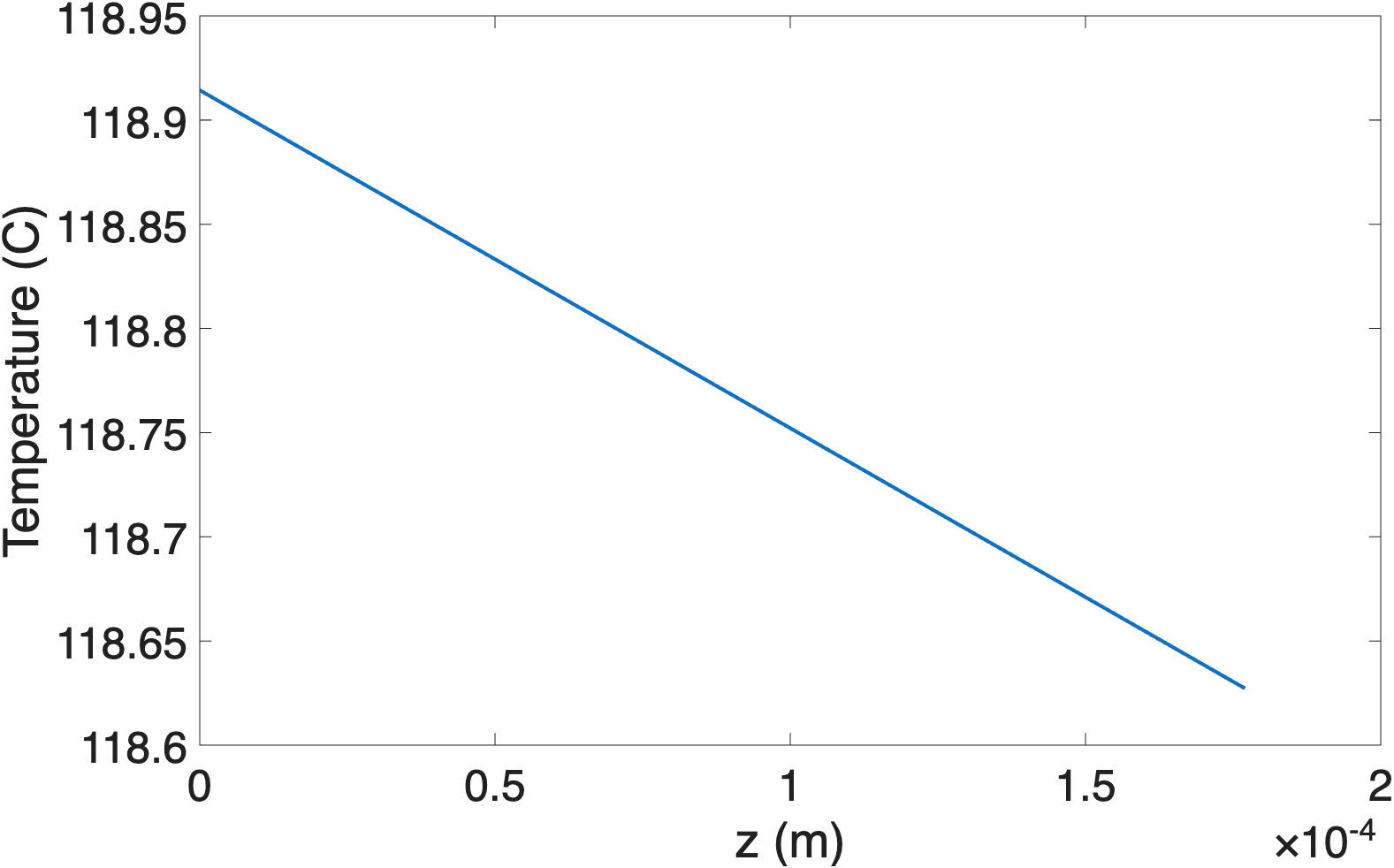}  
\caption{Thermal fields through the thickness of the 3D printed tape at an equilibrium node. \label{throughthicknesst}}
\end{figure}

The thermal simulation results are illustrated in Figure \ref{thermalfieldspic} for a section through the thickness of the 3D printed part. We can identify a deposition temperature of about $380^\textrm{o}C$ leading to an equilibrium temperature of about $120^\textrm{o}C$ for the given base and air temperature. The temperature through the thickness at $x=0$ is illustrated in Figure \ref{throughthicknesst}, showcasing a through-thickness temperature variation of less than $0.3^\textrm{o}C$.

The mechanical simulation leverages the thermal fields and the identified deformation models. The residual strain release model saturated at $\epsilon_{xx}^{ini}=0.06$\%. The crystallization strains along the $x$ direction leverage the DSC experiment results, with a crystallization factor dependent on the strain release. The DSC experiment of the prepreg tape is illustrated in Figure \ref{dscpic}. We can note a first onset of the crystallization around $T_g=147^{\textrm{o}}C$. The crystallization generation is proportional to the energy \cite{https://doi.org/10.1002/pen.26813}; therefore, we can write:
\begin{equation}
\mathcal{X}_{c}=\frac{\Delta H_{c}}{\lambda \cdot \Delta H_{final}},
\end{equation}


$\mathcal{X}_{c}$ being the crystallization fraction at the $\Delta H_{c}$ energy release obtained by integrating the enthalpy of energy release in the DSC plots up to 180$^\textrm{o}C$, $\Delta H_{final}$ the total energy release during the first recrystallization (after complete melting of the material).  

Considering the deformation induced by the crystallization is proportional to the crystallization fraction, one can write the following:
 \begin{equation}
{\epsilon_{xx}^{cryst}}_{final}=\frac{{\epsilon_{xx}^{cryst}}_{c}}{\mathcal{X}_{c}},
\end{equation}

with ${\epsilon_{xx}^{cryst}}_{final}$ being the final deformation generated along the $x$ direction after deposition and ${\epsilon_{xx}^{cryst}}_{c}$ is the fraction of strains generated during the DMA experiment. To compute the lateral strains, micromechanics relations are derived using the mechanics of materials assumptions. The micromechanics derivation is provided in \ref{micromechanicsappendix}.

The boundary conditions for the mechanical simulation are the following:
\begin{equation}
\begin{array}{l}
u(x=0)=v(x=0)=w(x=0)=0\\
w(z=0)=0\\
u(x=L_{tape})=u_{tape}\\
v(x=L_{tape})=w(x=L_{tape})=0,\\
\end{array}
\end{equation}

with $u_{tape}$ being the deformation at the tip of the tape, allowed to deform only along the $x$ direction, and considering a fixed edge at $x=0$. No external forces are applied on the modeled part. The simulation performed uses the governing equations illustrated in (\ref{stronformmechanical}) and (\ref{sigmazeroeq}), along with:
\begin{equation}\label{epsilonzerouse}
\bm\epsilon_0=\bm\alpha \Delta T+\bm\epsilon^{ini}+\bm\epsilon^{cryst}
\end{equation}

\begin{figure}[!ht]
\centering
\includegraphics[width=0.9\textwidth]{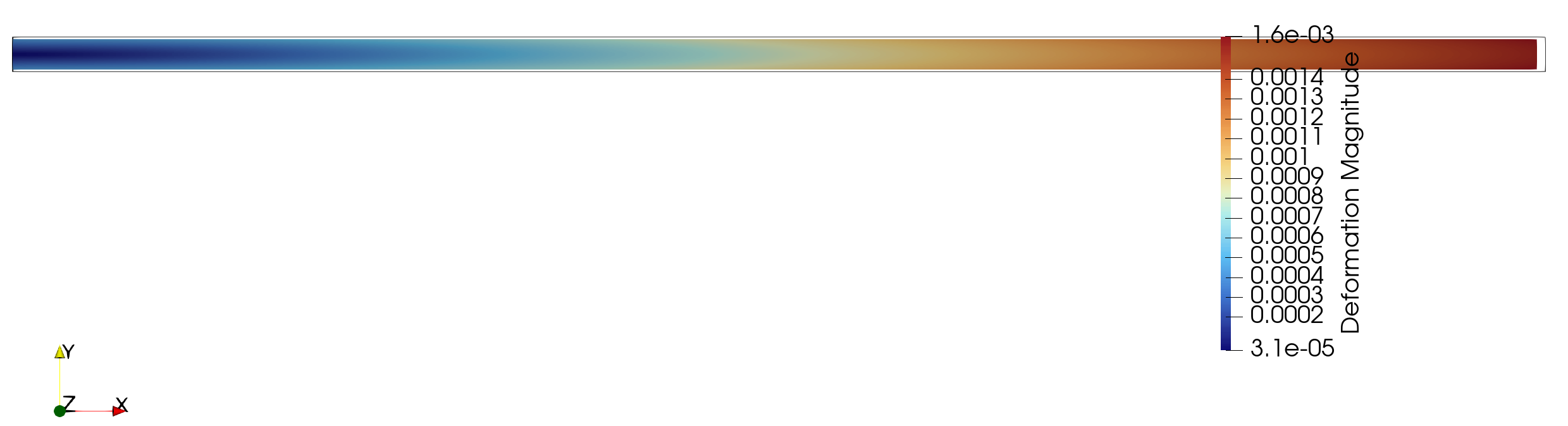}  
\caption{Deformed shape of the 3D printed tape, with the outline showing the original shape (displacements are in meters). \label{deformation3dprinted}}
\end{figure}

The tensors $\bm\epsilon^{ini}$ and $\bm\epsilon^{cryst}$ are obtained using micromechanics. The results of the deformed tape are illustrated in figure \ref{deformation3dprinted}. The results show a shrinking along both $x$ and $y$ directions in the tape.

\subsection{Experimental validation}
\begin{figure}[!ht]
\centering
    \includegraphics[width=0.9\textwidth]{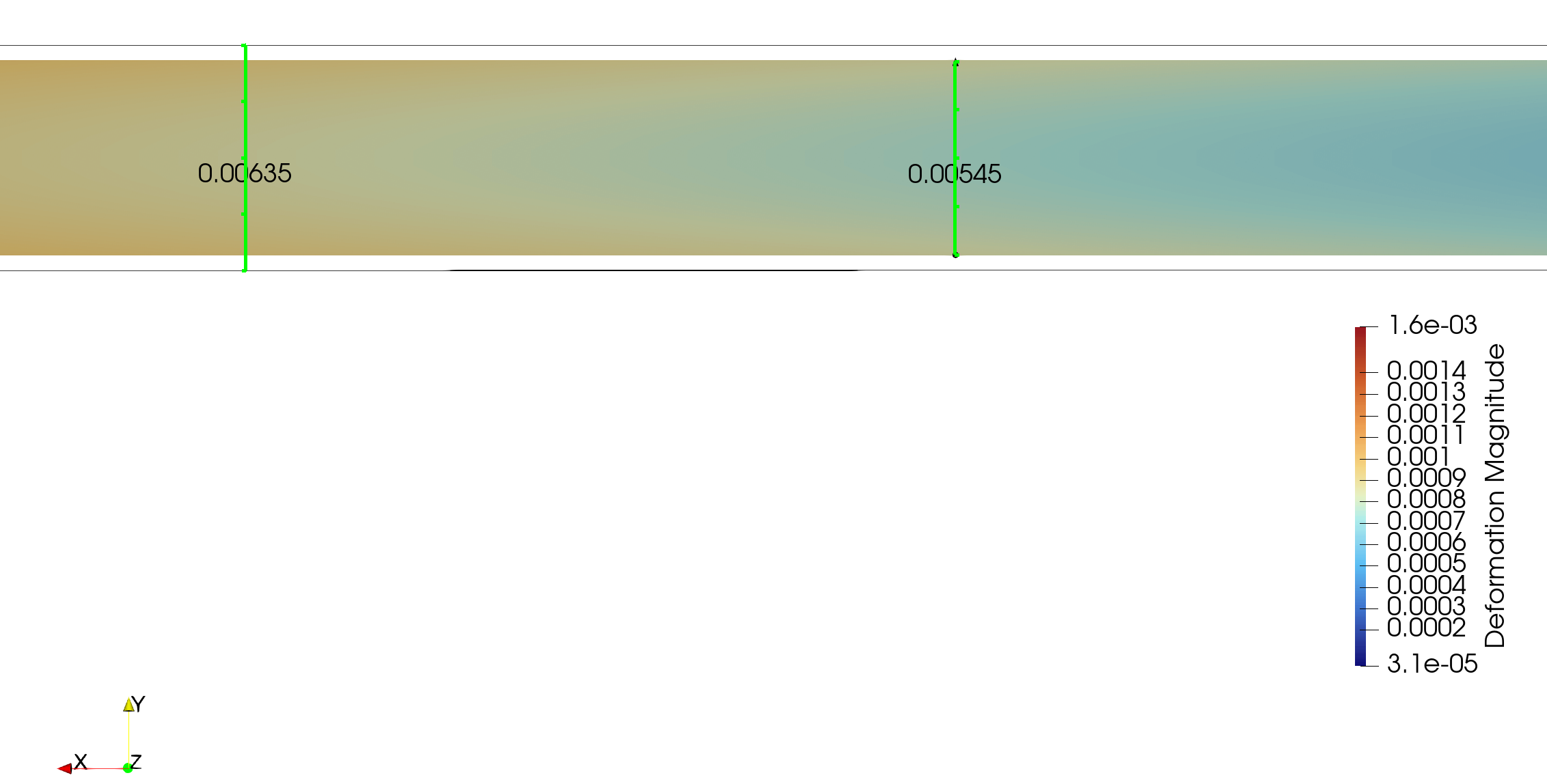}  
\caption{Lateral deformation (in meters) of the 3D printed tape as predicted by the simulation, with dimensions measured in meters, and the outline showcasing the original shape of $6.35mm$ width. \label{deformation3dprintedlateral}}
\end{figure}

To validate the simulation model, a tap with a length of 280 mm is deposited on the build platform. An experimental part is manufactured using the same operating conditions. The final lateral deformation from the simulation  is measured, and the lateral deformations are illustrated in Figure \ref{deformation3dprintedlateral}. The experimental tape is shown in figure \ref{fig:fig2akazem}, with its dimensions shown in Figure \ref{fig:fig2bkazem}.

The comparison of the experimental and simulation results shows the ability of the proposed framework to evaluate the deformation of the 3D printed tape using all the phenomena experienced during the deformation, with high fidelity. The difference between the simulation and experimental data  can be attributed to warping \cite{GHNATIOS2022102796}, and caliper measurement errors of a thin tape. Laser line scanning can be used in future for more accurate measurements of the tapes. 

\section{Conclusion}\label{comclusionsec}
In this work, we derived a Kelvin-Voigt viscoelastic model for low-melt poly-aryl ether ketone (LM PAEK) reinforced with high-strength continuous carbon fiber prepreg tapes. It is based on experimental results from Differential Scanning Calorimetry (DSC) and Dynamic Mechanical Analyzer (DMA) machines and numerical simulations. The model was derived from a DMA experiment with heating up to $120^\textrm{o}C$ and later was validated for a second experiment going up to $180^\textrm{o}C$, confirming its ability to reproduce the intrinsic material behavior. Moreover, an initial strain release model is created using a stabilized Neural ODE method. The residual strain release is again trained on the DMA thermal cycles with a maximum temperature of $120^\textrm{o}C$, and used for the second experiment with a maximum temperature of $180^\textrm{o}C$. The extrapolation showed the stability of the learned model, reaching a saturation intrinsically. The initial strain release showed an error in the strain of about 0.02\% at the saturation point in DMA-2, which is operating far beyond the training region. In fact, the model was shown only temperatures not exceeding $110^\textrm{o}C$, and used to predict deformations after heating up to $180^\textrm{o}C$. This shows the stability of the modeling framework and its ability to generalize beyond the training region while conserving a reasonable performance.

The crystallization-induced strains were modeled when the temperature went above the glass transition temperature for LM PAEK. Beyond this point, the modeling was performed again using the stabilized neural ODE, and showcased high stability and the ability to extrapolate beyond the learned time and temperature frames.

The previous three modeling efforts were combined into a holistic modeling effort to simulate the dimensional change of the prepreg tape after deposition by the robotic 3D printer. The width measurement from the simulation model and an actual deposited tape were compared. The results showed the ability of the proposed technique to predict the dimensional change of the prepreg tape for the 3D printing process parameter.

This work constitutes a cornerstone in the understanding of the feedstock deformation in robotic 3D printing of continuous fiber thermoplastic tapes. The work proposed a holistic approach for characterization, modeling, and integration of this effect into a digital-twin framework of the 3D printing process. The results from this study can be used to accurately place tapes on a build platform per their final dimensions, preventing defects like gaps and overlaps. The numerical model significantly reduces experimental trials required to find final tape dimensions for various 3D printing process parameters. The simulation efforts were completed for a tape deposited on a straight path. Robotic 3D printing can be used to 3D print tapes along a curvilinear path to tailor structural performance of final parts. This work can be expanded to predict deformation of tapes and provided guidelines for manufacturing complex parts.

\section*{Declarations}
\paragraph{Funding}
The Natural Sciences and Engineering Research Council of Canada (NSERC, RGPIN-2023-04091) supported this work.
\paragraph{Conflict of interest and competing interests}
Not applicable.
\paragraph{Consent for publication}
All authors read and agreed to the published version of the manuscript.
\paragraph{Availability of data and materials}
The raw data are available upon request.
\paragraph{Author Contributions}
Conceptualization, C.G., K.F. ; methodology, C.G. ; software, C.G. ; validation, C.G. and K.F.; formal analysis, C.G. and K.F.; investigation, C.G. ; resources, C.G., K.F.; data curation, K.F. ; original draft preparation (writing), C.G. ; review and editing (writing), C.G. and K.F. ; visualization, C.G. ; supervision, C.G. and K.F.

\paragraph{Acknowledgments}
Authors want to thank Teijin Carbon America for proving raw materials and Dr. Seyed Miri for 3D-printing efforts.\\

\appendix
\section{Micromechanical analysis of the tape lateral deformation}\label{micromechanicsappendix}
In this section, micromechanics assumptions in mechanics of materials are considered. Assuming the fibers are oriented along the $x$ direction and $v_f$ the fiber volume fraction, we can write at equilibrium in the absence of external loads:
\begin{equation}\label{microassumptions}
\begin{array}{l}
\epsilon_{xx}^{tape}=\epsilon_{xx}^{resin}=\epsilon_{xx}^{fiber}\\
\sigma_{xx}^{tape}=(1-v_f)\sigma_{xx}^{resin}+v_f\sigma_{xx}^{fiber}\\
\epsilon_{yy}^{tape}=(1-v_f)\epsilon_{yy}^{resin}+v_f\epsilon_{yy}^{fiber}\\
\sigma_{yy}^{tape}=\sigma_{yy}^{resin}=\sigma_{yy}^{fiber}=0\\
\end{array}
\end{equation}

writing that $\epsilon_{xx}^{tape}$ is known and equal to $Y$ (the one obtained from the DMA experiment), we can write that:
\begin{equation}\label{microstrains}
\begin{array}{l}
\epsilon_{xx}^{resin}=X+\frac{\sigma_{xx}^{resin}}{E_r}=Y\\
\sigma_{xx}^{fiber}=\epsilon_{xx}^{fiber} E_f=Y E_f,
\end{array}
\end{equation}

with $X$ being the intrinsic resin crystallization strains along $x$-direction, $E_f$ is the fibers' modulus of elasticity and $E_r$ the matrix one. The fibers are only subjected to the stress contribution to their deformation, as they do not exhibit crystallization deformation. Solving equations (\ref{microassumptions}) and (\ref{microstrains}) for $X$ we can find:
\begin{equation}
X=Y\left(1+\frac{E_f v_f}{E_r\left(1-v_f\right)}\right)
\end{equation}

To find the contribution of $X$ to the lateral direction of deformation, we can use the following.
\begin{equation}\label{crystlateral1}
\begin{array}{l}
\epsilon_{yy}^{tape}=(1-v_f)\epsilon_{yy}^{resin}+v_f\epsilon_{yy}^{fiber}\\
\epsilon_{yy}^{tape}=(1-v_f)\left(X-\frac{\nu_r \sigma_{xx}^{resin}}{E_r}\right)+v_f\left(-\frac{\sigma_{xx}^{fiber}}{E_f}\right)
\end{array}
\end{equation}

Equation (\ref{crystlateral1}) assumes the crystallization strains $X$ act on both longitudinal and lateral directions with the same amplitude, a logical assumption for a spherical crystal shape.

Finally, from equations (\ref{microassumptions}) and (\ref{microstrains}), one can also write:
\begin{equation}\label{crystlateral2}
\begin{array}{l}
\sigma_{xx}^{fiber}=\epsilon_{xx}^{fiber} E_f=Y E_f\\
\sigma_{xx}^{resin}=\frac{\sigma_{xx}^{tape}-v_f\sigma_{xx}^{fiber}}{1-v_f}
\end{array}
\end{equation}

Replacing equation (\ref{crystlateral2}) in (\ref{crystlateral1}) and considering no loads at equilibrium ($\sigma_{xx}^{tape}=0$), one can find the final crystallization strains along the lateral directions, equal for $y$ and $z$:
\begin{equation}
\epsilon_{yy}^{tape}=\epsilon_{zz}^{tape}
\end{equation}

Finally, we can note:
\begin{equation}\label{finalappaeq}
\bm\epsilon_0=\left(\begin{array}{c}
\epsilon_{xx}^{tape}\\
\epsilon_{yy}^{tape}\\
\epsilon_{zz}^{tape}
\end{array}\right)
\end{equation}

Equation (\ref{finalappaeq}) is substituted into (\ref{epsilonzerouse}) to solve the mechanical deformation of the 3D printed tape.

\bibliographystyle{bmc-mathphys} 
\bibliography{RefChady}

@manual{datasheettejin,
    key = {14186},
    title = {TEIJIN Datasheet T\_Prepreg; TPUD PAEK HTS45 P12 12K},
    year = {2022},
    month ={January},
}

@article{reviewmlinengineering, 
title={Data publishing in mechanics and dynamics: challenges, guidelines, and examples from engineering design}, 
volume={6}, 
DOI={10.1017/dce.2025.13}, 
journal={Data-Centric Engineering}, 
author={Ebel, Henrik and van Delden, Jan and Lüddecke, Timo and Borse, Aditya and Gulakala, Rutwik and Stoffel, Marcus and Yadav, Manish and Stender, Merten and Schindler, Leon and de Payrebrune, Kristin Miriam and et al.}, 
year={2025}, 
pages={e23}}

@article{KURTZ20074845,
title = {PEEK biomaterials in trauma, orthopedic, and spinal implants},
journal = {Biomaterials},
volume = {28},
number = {32},
pages = {4845-4869},
year = {2007},
issn = {0142-9612},
doi = {https://doi.org/10.1016/j.biomaterials.2007.07.013},
url = {https://www.sciencedirect.com/science/article/pii/S0142961207005467},
author = {Steven M. Kurtz and John N. Devine}
}

@article{GHNATIOS2024105542,
title = {A new methodology for anisotropic yield surface description using model order reduction techniques and invariant neural network},
journal = {Journal of the Mechanics and Physics of Solids},
volume = {184},
pages = {105542},
year = {2024},
issn = {0022-5096},
doi = {https://doi.org/10.1016/j.jmps.2024.105542},
url = {https://www.sciencedirect.com/science/article/pii/S0022509624000085},
author = {Chady Ghnatios and Oana Cazacu and Benoit Revil-Baudard and Francisco Chinesta}
}

@article{ATKINSON2002731,
title = {Enthalpic relaxation in semi-crystalline PEEK},
journal = {Polymer},
volume = {43},
number = {3},
pages = {731-735},
year = {2002},
note = {Mattice Special Issue},
issn = {0032-3861},
doi = {https://doi.org/10.1016/S0032-3861(01)00668-1},
url = {https://www.sciencedirect.com/science/article/pii/S0032386101006681},
author = {J.R. Atkinson and J.N. Hay and M.J. Jenkins}
}

@article{https://doi.org/10.1002/pen.26813,
author = {Ledesma, Rodolfo and Wohl, Christopher and Grimsley, Brian},
title = {Crystallization kinetics analysis and modeling of aerospace PAEK materials},
journal = {Polymer Engineering \& Science},
volume = {64},
number = {8},
pages = {3802-3816},
doi = {https://doi.org/10.1002/pen.26813},
url = {https://4spepublications.onlinelibrary.wiley.com/doi/abs/10.1002/pen.26813},
eprint = {https://4spepublications.onlinelibrary.wiley.com/doi/pdf/10.1002/pen.26813},
year = {2024}
}

@ARTICLE{Ghnatios2021-zq,
  title     = "Reduced order modeling of selective laser melting: from
               calibration to parametric part distortion",
  author    = "Ghnatios, Chady and Rai, Khalil El and Hascoet, Nicolas and
               Pires, Pierre-Adrien and Duval, Jean-Louis and Lambarri, Jon and
               Hascoet, Jean-Yves and Chinesta, Francisco",
  journal   = "Int. J. Mater. Form.",
  publisher = "Springer Science and Business Media LLC",
  volume    =  14,
  number    =  5,
  pages     = "973--986",
  month     =  sep,
  year      =  2021,
  copyright = "https://creativecommons.org/licenses/by/4.0",
  language  = "en"
}

@article{TUO2026113114,
title = {The defect quantification in additive manufacturing composites using fused infrared thermography and artificial intelligence},
journal = {Composites Part B: Engineering},
volume = {309},
pages = {113114},
year = {2026},
issn = {1359-8368},
doi = {https://doi.org/10.1016/j.compositesb.2025.113114},
url = {https://www.sciencedirect.com/science/article/pii/S1359836825010303},
author = {Hongliang Tuo and Liang Wang and Junqing Zhang and Xiaoyu Zhang and Shumeng Dong and Xinbo Li}
}

@article{DAGHIGH2024100600,
title = {Review of machine learning applications for defect detection in composite materials},
journal = {Machine Learning with Applications},
volume = {18},
pages = {100600},
year = {2024},
issn = {2666-8270},
doi = {https://doi.org/10.1016/j.mlwa.2024.100600},
url = {https://www.sciencedirect.com/science/article/pii/S2666827024000768},
author = {Vahid Daghigh and Hamid Daghigh and Thomas E. Lacy and Mohammad Naraghi}
}

@ARTICLE{Ghnatios2021-bd,
  title     = "A nonparametric probabilistic method to enhance {PGD} solutions
               with data-driven approach, application to the automated tape
               placement process",
  author    = "Ghnatios, Chady and Barasinski, Anais",
  journal   = "Adv. Model. Simul. Eng. Sci.",
  publisher = "Springer Science and Business Media LLC",
  volume    =  8,
  number    =  1,
  month     =  dec,
  year      =  2021,
  copyright = "https://creativecommons.org/licenses/by/4.0",
  language  = "en"
}

@article{OUYANG2025111209,
title = {Process modeling and deformation prediction of 3D printed continuous fiber-reinforced composites based on in-situ micro-scale measuring},
journal = {Composites Science and Technology},
volume = {267},
pages = {111209},
year = {2025},
issn = {0266-3538},
doi = {https://doi.org/10.1016/j.compscitech.2025.111209},
url = {https://www.sciencedirect.com/science/article/pii/S0266353825001770},
author = {Shiping Ouyang and Dongsheng Li and Weijun Zhu and Long Fu and Zhikun Zhang and Ning Wang and Quan Zhi}
}

@article{KABIR2020111476,
title = {A critical review on 3D printed continuous fiber-reinforced composites: History, mechanism, materials and properties},
journal = {Composite Structures},
volume = {232},
pages = {111476},
year = {2020},
issn = {0263-8223},
doi = {https://doi.org/10.1016/j.compstruct.2019.111476},
url = {https://www.sciencedirect.com/science/article/pii/S0263822319322706},
author = {S M Fijul Kabir and Kavita Mathur and Abdel-Fattah M. Seyam}
}

@article{https://doi.org/10.1002/pc.29895,
author = {Zhang, Yaru and Zheng, Wenkai and Wang, Yuzhong and Ma, Kaiyue and Feng, Xueming and Ji, Qianyu and Guo, Wenhua and Lu, Bingheng},
title = {A review of 3D printing continuous carbon fiber reinforced thermoplastic polymers: Materials, processes, performance enhancement, and failure analysis},
journal = {Polymer Composites},
volume = {46},
number = {14},
pages = {12619-12649},
doi = {https://doi.org/10.1002/pc.29895},
url = {https://4spepublications.onlinelibrary.wiley.com/doi/abs/10.1002/pc.29895},
eprint = {https://4spepublications.onlinelibrary.wiley.com/doi/pdf/10.1002/pc.29895},
year = {2025}
}

@ARTICLE{Zheng2025-jf,
  title     = "{3D} printing continuous fiber reinforced polymers: A review of
               material selection, process, and mechanics-function integration
               for targeted applications",
  author    = "Zheng, Haoyuan and Zhu, Shaowei and Chen, Liming and Wang,
               Lianchao and Zhang, Hanbo and Wang, Peixu and Sun, Kefan and
               Wang, Haorui and Liu, Chengtao",
  journal   = "Polymers (Basel)",
  publisher = "MDPI AG",
  volume    =  17,
  number    =  12,
  pages     = "1601",
  month     =  jun,
  year      =  2025,
  copyright = "https://creativecommons.org/licenses/by/4.0/",
  language  = "en"
}

@ARTICLE{El_Moumen2019-pl,
  title     = "Modelling of the temperature and residual stress fields during
               {3D} printing of polymer composites",
  author    = "El Moumen, A and Tarfaoui, M and Lafdi, K",
  journal   = "International Journal of Advanced Manufacturing Technology",
  publisher = "Springer Science and Business Media LLC",
  volume    =  104,
  number    = "5-8",
  pages     = "1661--1676",
  month     =  oct,
  year      =  2019,
  language  = "en"
}

@article{ZHANG2021101775,
title = {Fibre misalignment and breakage in 3D printing of continuous carbon fibre reinforced thermoplastic composites},
journal = {Additive Manufacturing},
volume = {38},
pages = {101775},
year = {2021},
issn = {2214-8604},
doi = {https://doi.org/10.1016/j.addma.2020.101775},
url = {https://www.sciencedirect.com/science/article/pii/S2214860420311477},
author = {Haoqi Zhang and Jiayun Chen and Dongmin Yang}
}

@article{Lu31122024,
author = {Lu Lu and Yongtang Yuan and Yongkang Xie and Shangqin Yuan and Jingwen Song and Han Luo and Yamin Li and Jihong Zhu and Weihong Zhang},
title = {Autonomous intelligent additive manufacturing of continuous fiber-reinforced composites: data-enhanced knowledgebase and multi-sensor fusion},
journal = {Virtual and Physical Prototyping},
volume = {19},
number = {1},
pages = {e2412192},
year = {2024},
publisher = {Taylor \& Francis},
doi = {10.1080/17452759.2024.2412192},
}

@article{MALAGUTTI2023117961,
title = {Effects of printed bead volume on thermal history, polymer degree of crystallinity and mechanical properties in large scale additive manufacturing},
journal = {Journal of Materials Processing Technology},
volume = {316},
pages = {117961},
year = {2023},
issn = {0924-0136},
doi = {https://doi.org/10.1016/j.jmatprotec.2023.117961},
url = {https://www.sciencedirect.com/science/article/pii/S0924013623001061},
author = {L. Malagutti and S. Charlon and V. Mazzanti and F. Mollica}
}

@article{WANG2026109310,
title = {Through-thickness crystallinity gradient controls warpage reduction in CF/PAEK via in-situ consolidation automated fiber placement},
journal = {Composites Part A: Applied Science and Manufacturing},
volume = {200},
pages = {109310},
year = {2026},
issn = {1359-835X},
doi = {https://doi.org/10.1016/j.compositesa.2025.109310},
url = {https://www.sciencedirect.com/science/article/pii/S1359835X25006049},
author = {Ye Wang and Zhibo Xin and Jie Yuan and Yugang Duan and Hong Xiao and Fanghong Yang and Daijun Zhang and Fuping Li}
}

@article{SREEJITH2023103789,
title = {A thermodynamic framework for the additive manufacturing of crystallizing polymers. Part I: A theory that accounts for phase change, shrinkage, warpage and residual stress},
journal = {International Journal of Engineering Science},
volume = {183},
pages = {103789},
year = {2023},
issn = {0020-7225},
doi = {https://doi.org/10.1016/j.ijengsci.2022.103789},
url = {https://www.sciencedirect.com/science/article/pii/S0020722522001550},
author = {P. Sreejith and K. Kannan and K.R. Rajagopal}
}

@article{YADAV2023107654,
title = {Review of in-process defect monitoring for automated tape laying},
journal = {Composites Part A: Applied Science and Manufacturing},
volume = {173},
pages = {107654},
year = {2023},
issn = {1359-835X},
doi = {https://doi.org/10.1016/j.compositesa.2023.107654},
url = {https://www.sciencedirect.com/science/article/pii/S1359835X23002300},
author = {Neha Yadav and Ralf Schledjewski}
}

@article{LEE2022100085,
title = {Effect of temperature history during additive manufacturing on crystalline morphology of PEEK},
journal = {Advances in Industrial and Manufacturing Engineering},
volume = {4},
pages = {100085},
year = {2022},
issn = {2666-9129},
doi = {https://doi.org/10.1016/j.aime.2022.100085},
url = {https://www.sciencedirect.com/science/article/pii/S2666912922000150},
author = {Austin Lee and Mathew Wynn and Liam Quigley and Marco Salviato and Navid Zobeiry}
}

@ARTICLE{Pourali2024-ny,
  title    = "Crystallization modeling of two semi-crystalline polyamides
              during material extrusion additive manufacturing",
  author   = "Pourali, Masoumeh and Adisa, Ahmed and Salunke, Shalmali and
              Peterson, Amy M",
  journal  = "Sci. Rep.",
  volume   =  14,
  number   =  1,
  pages    = "26297",
  month    =  nov,
  year     =  2024,
  language = "en"
}

@ARTICLE{Afanasev2026-jo,
  title     = "Development of a continuous fiber-reinforced {3D} printing
               process with a 6-axis robot arm: Process design and equipment",
  author    = "Afanasev, Anna and H{\"o}fer, Philipp and Holtmannsp{\"o}tter,
               Jens and Zimmer, Felix and Ehrlich, Ingo",
  journal   = "Int. J. Adv. Manuf. Technol.",
  publisher = "Springer Science and Business Media LLC",
  month     =  jan,
  year      =  2026,
  copyright = "https://creativecommons.org/licenses/by/4.0",
  language  = "en"
}

@article{KUMAR2026119819,
title = {A critical review of the past, present, and future of 3D printing for continuous and short fiber composites},
journal = {Composite Structures},
volume = {376},
pages = {119819},
year = {2026},
issn = {0263-8223},
doi = {https://doi.org/10.1016/j.compstruct.2025.119819},
url = {https://www.sciencedirect.com/science/article/pii/S0263822325009845},
author = {Sanjay Kumar and Dong-Hoon Yoo and Jun-Seop Song and Hak-Sung Kim}
}

@article{Jiang02012023,
author = {Bingnong Jiang and Yuan Chen and Lin Ye and Li Chang and Hang Dong},
title = {Residual stress and warpage of additively manufactured SCF/PLA composite parts},
journal = {Advanced Manufacturing: Polymer \& Composites Science},
volume = {9},
number = {1},
pages = {2171940},
year = {2023},
publisher = {Taylor \& Francis},
doi = {10.1080/20550340.2023.2171940}
}

@article{SAMY2021389,
title = {Finite element analysis of residual stress and warpage in a 3D printed semi-crystalline polymer: Effect of ambient temperature and nozzle speed},
journal = {Journal of Manufacturing Processes},
volume = {70},
pages = {389-399},
year = {2021},
issn = {1526-6125},
doi = {https://doi.org/10.1016/j.jmapro.2021.08.054},
url = {https://www.sciencedirect.com/science/article/pii/S1526612521006381},
author = {Anto Antony Samy and Atefeh Golbang and Eileen Harkin-Jones and Edward Archer and David Tormey and Alistair McIlhagger}
}

@ARTICLE{Miri2025-iu,
  title     = "Numerical and experimental study of the consolidation of
               continuous carbon fiber thermoplastics made by robotic {3D} printing",
  author    = "Miri, Seyed and Rana, Jash and Fayazbakhsh, Kazem and Ghnatios,
               Chady",
  journal   = "Int. J. Mater. Form.",
  publisher = "Springer Science and Business Media LLC",
  volume    =  18,
  number    =  1,
  month     =  mar,
  year      =  2025,
  copyright = "https://www.springernature.com/gp/researchers/text-and-data-mining",
  language  = "en"
}

@article{FEREIDOUNI2025108519,
title = {Transverse squeeze flow of thermoplastic composite tape during in-situ consolidation via automated fiber placement},
journal = {Composites Part A: Applied Science and Manufacturing},
volume = {188},
pages = {108519},
year = {2025},
issn = {1359-835X},
doi = {https://doi.org/10.1016/j.compositesa.2024.108519},
url = {https://www.sciencedirect.com/science/article/pii/S1359835X24005177},
author = {Mahmoud Fereidouni and Suong Van Hoa}
}

@article{DIAZ2025104749,
title = {Warpage of thin additively manufactured continuous fiber thermoset composites},
journal = {Additive Manufacturing},
volume = {102},
pages = {104749},
year = {2025},
issn = {2214-8604},
doi = {https://doi.org/10.1016/j.addma.2025.104749},
url = {https://www.sciencedirect.com/science/article/pii/S2214860425001137},
author = {Mateo Diaz and Jeffery W. Baur}
}

@article{GHNATIOS2022102796,
title = {Warping estimation of continuous fiber-reinforced composites made by robotic 3D printing},
journal = {Additive Manufacturing},
volume = {55},
pages = {102796},
year = {2022},
issn = {2214-8604},
doi = {https://doi.org/10.1016/j.addma.2022.102796},
url = {https://www.sciencedirect.com/science/article/pii/S221486042200197X},
author = {Chady Ghnatios and Kazem Fayazbakhsh}
}

@article{GHNATIOS2019769,
title = {A stabilized mixed formulation using the proper generalized decomposition for fluid problems},
journal = {Computer Methods in Applied Mechanics and Engineering},
volume = {346},
pages = {769-787},
year = {2019},
issn = {0045-7825},
doi = {https://doi.org/10.1016/j.cma.2018.09.030},
url = {https://www.sciencedirect.com/science/article/pii/S0045782518304821},
author = {Chady Ghnatios and Elie Hachem}
}

@BOOK{Donea2003-tn,
  title     = "Finite element methods for flow problems",
  author    = "Donea, Jean and Huerta, Antonio",
  publisher = "John Wiley \& Sons",
  month     =  apr,
  year      =  2003,
  address   = "Nashville, TN",
  language  = "en"
}

@phdthesis{ghnatios:tel-00867281,
  TITLE = {{Simulation avanc{\'e}e des probl{\`e}mes thermiques rencontr{\'e}s lors de la mise en forme des composites}},
  AUTHOR = {Ghnatios, Chady},
  URL = {https://theses.hal.science/tel-00867281},
  SCHOOL = {{Ecole Centrale de Nantes (ECN)}},
  YEAR = {2012},
  MONTH = Oct,
  TYPE = {Theses},
  HAL_ID = {tel-00867281},
  HAL_VERSION = {v1},
}

@article{KEMMISH1985905,
title = {The effect of physical ageing on the properties of amorphous PEEK},
journal = {Polymer},
volume = {26},
number = {6},
pages = {905-912},
year = {1985},
issn = {0032-3861},
doi = {https://doi.org/10.1016/0032-3861(85)90136-3},
url = {https://www.sciencedirect.com/science/article/pii/0032386185901363},
author = {D.J. Kemmish and J.N. Hay}
}

@Article{en16155790,
AUTHOR = {Ghnatios, Chady and Kestelyn, Xavier and Denis, Guillaume and Champaney, Victor and Chinesta, Francisco},
TITLE = {Learning Data-Driven Stable Corrections of Dynamical Systems—Application to the Simulation of the Top-Oil Temperature Evolution of a Power Transformer},
JOURNAL = {Energies},
VOLUME = {16},
YEAR = {2023},
NUMBER = {15},
ARTICLE-NUMBER = {5790},
URL = {https://www.mdpi.com/1996-1073/16/15/5790},
ISSN = {1996-1073},
DOI = {10.3390/en16155790}
}

@article{LINOT2023111838,
title = {Stabilized neural ordinary differential equations for long-time forecasting of dynamical systems},
journal = {Journal of Computational Physics},
volume = {474},
pages = {111838},
year = {2023},
issn = {0021-9991},
doi = {https://doi.org/10.1016/j.jcp.2022.111838},
url = {https://www.sciencedirect.com/science/article/pii/S0021999122009019},
author = {Alec J. Linot and Joshua W. Burby and Qi Tang and Prasanna Balaprakash and Michael D. Graham and Romit Maulik}
}

@INPROCEEDINGS{9959914,
  author={Kestelyn, Xavier and Denis, Guillaume and Champaney, Victor and Hascoet, Nicolas and Ghnatios, Chady and Chinesta, Fransisco},
  booktitle={2022 7th International Advanced Research Workshop on Transformers (ARWtr)}, 
  title={Towards a Hybrid Twin for Infrastructure Asset Management: Investigation on Power Transformer Asset Maintenance Management}, 
  year={2022},
  volume={},
  number={},
  pages={109-114},
  doi={10.23919/ARWtr54586.2022.9959914}}

@article{FERDOUSI2023110958,
title = {Investigation of 3D printed lightweight hybrid composites via theoretical modeling and machine learning},
journal = {Composites Part B: Engineering},
volume = {265},
pages = {110958},
year = {2023},
issn = {1359-8368},
doi = {https://doi.org/10.1016/j.compositesb.2023.110958},
url = {https://www.sciencedirect.com/science/article/pii/S1359836823004614},
author = {Sanjida Ferdousi and Rigoberto Advincula and Alexei P. Sokolov and Wonbong Choi and Yijie Jiang}
}

@Article{polym17182557,
AUTHOR = {Malashin, Ivan and Martysyuk, Dmitry and Tynchenko, Vadim and Gantimurov, Andrei and Nelyub, Vladimir and Borodulin, Aleksei},
TITLE = {Data-Driven Optimization of Discontinuous and Continuous Fiber Composite Processes Using Machine Learning: A Review},
JOURNAL = {Polymers},
VOLUME = {17},
YEAR = {2025},
NUMBER = {18},
ARTICLE-NUMBER = {2557},
URL = {https://www.mdpi.com/2073-4360/17/18/2557},
PubMedID = {41012320},
ISSN = {2073-4360},
DOI = {10.3390/polym17182557}
}

@Article{mynpmpgd,
  author =  {Chady Ghnatios and Anais Barasinski},
  title =   {A nonparametric probabilistic method to enhance PGD solutions with data-driven approach, application to the automated tape placement process},
  journal = {Advanced modeling and simulation in engineering sciences},
  year =    {2021},
  volume =  {submitted}
}

@Article{Archives-PGD,
  author =    {F. Chinesta and A. Ammar and E. Cueto},
  title =     {Recent Advances in the Use of the Proper Generalized Decomposition for Solving Multidimensional Models},
  journal =   {Archives of Computational Methods in Engineering},
  year =      {2010},
  volume =    {17},
  number =    {4},
  pages =     {327-350},
  comment =   {Explication de la PGD et quelques rÃ©sultats}
}

@Article{hybridfedatamodel2019,
  author =  {A. Clot and J.W.R. Meggitt and R.S. Langley and A.S. Elliott and A.T. Moorhouse},
  title =   {Development of a hybrid FE-SEA-experimental model},
  journal = {Journal of sound and vibration},
  year =    {2019},
  volume =  {452},
  pages =   {112-131},
  month =   {July}
}

@article{doi:10.1177/002199836900300419,
author = {J.C. Halpin},
title ={Stiffness and Expansion Estimates for Oriented Short Fiber Composites},

journal = {Journal of Composite Materials},
volume = {3},
number = {4},
pages = {732-734},
year = {1969},
doi = {10.1177/002199836900300419},
}

@inproceedings{Kulkarni2021,
  author    = {Kulkarni, S. S. and Khan, K. A. and Umer, R.},
  title     = {Quasi-Linear Viscoelastic Modelling of Uncured Prepregs under Compaction},
  booktitle = {Proceedings of the American Society for Composites – 36th Technical Conference},
  year      = {2021}
}

@inproceedings{Wei2023,
  author    = {Wei, Q. and Sun, Y. and Zhang, D. and Enos, R.},
  title     = {An Experimentally Validated Hyper-Viscoelastic Model for the Bending Responses of Pre-Impregnated Tapes Under Processing Conditions},
  booktitle = {Proceedings of the American Society for Composites – 38th Technical Conference},
  year      = {2023}
}

@mastersthesis{Chan2018,
  author    = {Chan, K. J.},
  title     = {Investigation of Processing Conditions and Viscoelastic Properties on Frictional Sliding Behavior of Unidirectional Carbon Fiber Epoxy Prepreg},
  school    = {Virginia Tech},
  year      = {2018}
}

@article{Wang2024,
  author    = {Wang, G. and et al.},
  title     = {Preparation and Mechanical Properties of Flexible Prepreg Resin with High Strength and Low Creep},
  journal   = {Polymers},
  volume    = {16},
  number    = {4},
  pages     = {558},
  year      = {2024}
}

@ARTICLE{Xiong2019-bd,
  title     = "Consolidation modeling during thermoforming of thermoplastic composite prepregs",
  author    = "Xiong, Hu and Hamila, Nahi{\`e}ne and Boisse, Philippe",
  journal   = "Materials",
  publisher = "MDPI AG",
  volume    =  12,
  number    =  18,
  pages     = "2853",
  month     =  sep,
  year      =  2019,
  copyright = "https://creativecommons.org/licenses/by/4.0/",
  language  = "en"
}

@Book{Jones1999,
  title =     {Mechanics of composite materials},
  publisher = {Taylor and Francis group},
  year =      {1999},
  author =    {Robert Jones},
  edition =   {2}
}

@InProceedings{ghnatios2017computational,
author = {Ghnatios, Chady and Barasinski, Anais and Villegas, IF and Palardy, G and Chinesta, Francisco},
title = {Computational vamedecum of the coupled mechanical/thermal behavior of composite materials during ultrasonic welding},
booktitle = {VII International Conference on Computational Methods for Coupled Problems in Science and Engineering, Coupled problems 2017},
year = {2017},
address = {Rhodes Island, Greece},
}

@article{GHNATIOS201229,
title = {Proper Generalized Decomposition based dynamic data-driven control of thermal processes},
journal = {Computer Methods in Applied Mechanics and Engineering},
volume = {213-216},
pages = {29-41},
year = {2012},
issn = {0045-7825},
doi = {https://doi.org/10.1016/j.cma.2011.11.018},
url = {https://www.sciencedirect.com/science/article/pii/S0045782511003641},
author = {Ch. Ghnatios and F. Masson and A. Huerta and A. Leygue and E. Cueto and F. Chinesta}
}

@Article{k,
  author =  {S.I. Abu-Eishah},
  title =   {Correlations for the Thermal Conductivity of Metals as a Function of Temperature},
  journal = {International Journal of Thermophysics},
  year =    {2001},
  volume =  {22},
  number =  {6},
  pages =   {1855-1868}
}

@InProceedings{neuralodeoriginal,
author = {R.T.Q. Chen and Y. Rubanova and J. Bettencourt and D. Duvenaud},
title = {Neural ordinary differential equations},
booktitle = {Conference on Neural Information Processing Systems (NeurIPS 2018)},
pages = {1-18},
year = {2018},
volume = {32},
address = {Montreal, Canada},
}

@article{10.1162/neco.1997.9.8.1735,
    author = {Hochreiter, Sepp and Schmidhuber, Jürgen},
    title = {Long Short-Term Memory},
    journal = {Neural Computation},
    volume = {9},
    number = {8},
    pages = {1735-1780},
    year = {1997},
    month = {11},
    issn = {0899-7667},
    doi = {10.1162/neco.1997.9.8.1735},
    url = {https://doi.org/10.1162/neco.1997.9.8.1735},
    eprint = {https://direct.mit.edu/neco/article-pdf/9/8/1735/813796/neco.1997.9.8.1735.pdf},
}

@article{SHAWLY20253033,
title = {A Neural ODE-Enhanced Deep Learning Framework for Accurate and Real-Time Epilepsy Detection},
journal = {CMES - Computer Modeling in Engineering and Sciences},
volume = {143},
number = {3},
pages = {3033-3064},
year = {2025},
issn = {1526-1492},
doi = {https://doi.org/10.32604/cmes.2025.065264},
url = {https://www.sciencedirect.com/science/article/pii/S1526149225001699},
author = {Tawfeeq Shawly and Ahmed A. Alsheikhy}
}

@INPROCEEDINGS{7780459,
  author={He, Kaiming and Zhang, Xiangyu and Ren, Shaoqing and Sun, Jian},
  booktitle={2016 IEEE Conference on Computer Vision and Pattern Recognition (CVPR)}, 
  title={Deep Residual Learning for Image Recognition}, 
  year={2016},
  volume={},
  number={},
  pages={770-778},
  doi={10.1109/CVPR.2016.90}}

@ARTICLE{mydigitaltwin,
  author = {C. Ghnatios},
  title = {A hybrid modeling combining the Proper Generalized Decomposition (PGD) approach to data-driven model learners, with application to non-linear biphasic materials},
  journal = {Comptes rendus m\'ecanique},
  year = {2021},
  volume = {In Press}
}
\end{document}